\documentclass[a4paper,11pt]{article}

\usepackage{jheppub}                
\usepackage[all]{hypcap}            
\usepackage{slashed}                
\usepackage[usenames,table]{xcolor} 
\usepackage{feyndiag}               
\usepackage{parskip}                
\usepackage[section]{placeins}      
\usepackage[utf8]{inputenc}         
\usepackage{bbm}                    

\newcommand{\irrep}[2][0]{\ensuremath{\mathrm{\textbf{#2}}}}
\newcolumntype{x}[1]{>{\centering\hspace{0pt}}p{#1}}
\newcommand{\tn}{\tabularnewline}

\graphicspath{{./plots/}}
\title{Cornering Colored Coannihilation}
\author[a,b]{Sonia~El~Hedri,}
\author[a]{Maikel~de~Vries}
\affiliation[a]{PRISMA Cluster of Excellence \& Mainz Institute for Theoretical Physics, Johannes Gutenberg University, 55099 Mainz, Germany}
\affiliation[b]{NIKHEF, Theory Group, Science Park 105, 1098 XG, Amsterdam, The Netherlands}
\emailAdd{soniaelh@nikhef.nl}
\emailAdd{mdt.maikel@gmail.com}
\preprint{MITP/18-048}

\abstract{In thermal dark matter models, allowing the dark matter candidate to coannihilate with another particle can considerably loosen the relic density constraints on the dark matter mass. In particular, introducing a single strongly interacting coannihilation partner in a dark matter model can bring the upper bound on the dark sector energy scale from a few TeV up to about $10$~TeV. While these energies are outside the LHC reach, a large part of the parameter space for such coannihilating models can be explored by future hadron colliders. In this context, it is essential to determine whether the current bounds on dark matter simplified models also hold in non-minimal scenarios. In this paper, we study extended models that include multiple coannihilation partners. We show that the relic density bounds on the dark matter mass in these scenarios are stronger than for the minimal models in most of the parameter space and that weakening these bounds requires sizable interactions between the different species of coannihilation partners. Furthermore, we discuss how these new interactions as well as the additional particles in the models  can lead to stronger collider bounds, notably in jets plus missing transverse energy searches. This study serves as a vital ingredient towards the determination of the highest possible energy scale for thermal dark matter models.
}

\begin{document}

\maketitle
\clearpage

\section{Introduction}
\label{sec:introduction}
The thermal hypothesis, or the assumption that the dark matter used to be in thermal and chemical equilibrium with the Standard Model (SM) in the early Universe, tightly links the dark matter relic density to the strength of its interactions with the SM particles. The associated models hence predict a plethora of experimental signatures from colliders, direct, and indirect detectors. For minimal models, where the dark matter is the only new particle, the relic density bounds on the masses of the dark sector particles typically lie around a few TeV and these theories are therefore under siege at the current experiments~\cite{Kahlhoefer:2015bea,Cirelli:2005uq,Cohen:2013ama,Cohen:2013kna}. One of the most efficient ways to loosen the existing limits is to introduce a new dark sector particle that will accelerate the dark matter depletion through the so-called coannihilation mechanism~\cite{Griest:1990kh}. If this new particle is strongly interacting, coannihilation would allow the dark matter to reach masses of up to $10$~TeV without overclosing the Universe~\cite{ElHedri:2017nny,Baker:2015qna,Buschmann:2016hkc,deSimone:2014pda,Low:2014cba,Ellis:2015vaa,Liew:2016hqo,Ibarra:2015nca,diCortona:2016fsn,Ellis:2015xba}. This scenario is however significantly constrained by the LHC, which can probe $\mathcal{O}(10\%)$ relative mass splittings between the dark matter and its coannihilation partner. While small mass splittings are particularly challenging to explore, a future $100$~TeV collider should be able to probe most of the remaining regions of the parameter space~\cite{Harris:2015kda,Low:2014cba,Arkani-Hamed:2015vfh,Golling:2016gvc}. Although these projections seem extremely encouraging, it is essential to keep in mind that they have been derived from simplified models involving only the dark matter and a single coannihilation partner. It is therefore crucial to determine whether the current limits on the masses of the dark sector particles still apply to more complex scenarios. 

A simplified model of dark matter coannihilation can be extended by adding either new mediators between the SM and the dark sector, new dark matter candidates, or new coannihilation partners. The first possibility is a straightforward way to loosen the constraints on the dark matter mass by adding new annihilation channels without increasing the new physics couplings. For these extended models, the constraints from perturbative unitarity, colliders, or relic density can be considerably weaker than the ones derived using simplified models, and are highly model-dependent~\cite{Chala:2015ama,Duerr:2016tmh}. We therefore do not study this configuration here. Conversely, as we will show in the rest of this paper, meaningful and generic constraints can still be derived for models with either multiple dark matter candidates or multiple coannihilation partners~\cite{Chialva:2012rq,Hur:2007ur,Baldi:2012kt,Baldi:2013axa}. Although both scenarios can be studied using similar approaches, the parameter space for models with multiple coannihilation partners is larger since there are fewer restrictions on the partners' quantum numbers. In what follows, we will therefore focus on these types of models, keeping in mind that our techniques can be straightforwardly applied to models with multiple dark matter candidates.

In this study, we focus on models involving one dark matter candidate (DM) and an arbitrary number of strongly interacting coannihilation partners X$_i$, close in mass to the dark matter. In this type of models, X$_i$ self-annihilates through strong interactions, thus causing an efficient depletion of the dark matter by shifting the DM-X$_i$ equilibrium. As shown in~\cite{ElHedri:2017nny,Ellis:2015vaa,Liew:2016hqo,deSimone:2014pda}, the bounds on the dark matter mass from these models are among the weakest for thermal dark matter models, and can reach up to  $\mathcal{O}(10)$~TeV. In the rest of this study, we build on the framework introduced in~\cite{ElHedri:2017nny}, where we assume that the dark matter relic density is entirely determined by the self-annihilation rate between the X$_i$ and all the processes involving the dark matter can be neglected. Note that in such scenarios the direct and indirect detection signals are suppressed and do not need to be considered. The dominant model-independent constraints on these type of models will therefore be the relic density requirement and the collider bounds on the X$_i$.

Requiring the dark matter relic density to agree with the current observations~\cite{Ade:2015xua} forces the dark matter mass into a narrow band whose central value highly depends on the mass splitting between the X$_i$ and the dark matter. For simplified models with a single coannihilation partner, the allowed dark matter mass ranges from $\mathcal{O}(100)$~GeV to about $10$~TeV for low DM-X$_i$ splittings. Since the dark matter relic density can be increased \textit{ad libitum} by introducing additional novel dark matter candidates, the lower bound on the dark matter mass set by relic density can be easily relaxed. The upper bound, however, is much more robust and can therefore be used to determine the range of energies that should be explored by the future experiments.  
Here, we show that the upper bounds on the dark matter mass derived for minimal coannihilating models are still valid when introducing new coannihilation partners, unless the ``mixed'' annihilation rates between different species of X$_i$ are particularly large. Moreover, we find that, in order to test the robustness of the simplified model constraints against introducing any number of X$_i$, it is sufficient to consider models with only two coannihilation partners. This result considerably simplifies the study of non-minimal thermal dark matter scenarios.

In what follows, we derive the upper bounds on the mixed annihilation rates X$_i$\,X$_j$\,$\to$\,SM\,SM up to which the simplified model constraints still apply. We study a representative set of models where the dark matter coannihilates with two strongly interacting particles X$_1$ and X$_2$. We model the  X$_1$\,X$_2$\,$\to$\,SM\,SM process using only renormalizable interactions, that can be mediated by either a SM particle, a new particle in the $s$-channel, or a new particle in the $t$-channel. This approach allows us to express the upper bounds on this mixed annihilation rate in terms of constraints on the couplings between the different X$_i$ and the masses of the new mediators. In particular, we characterize several specific kinematic configurations where introducing new coannihilation partners can considerably enhance the dark matter depletion rate even for moderate couplings. We finally explore how introducing the new X$_i$ and their associated vertices affects the collider searches at the LHC and at a future $100$~TeV accelerator.  

The applications of our results are manifold. First and foremost, demonstrating the robustness of the current simplified model constraints is a necessary condition for an extensive search program for coannihilating dark matter at current and future hadron colliders. On the theoretical side, exploring models with extended dark sectors is significantly simplified as, in most of the parameter space, only one or two coannihilation partners need to be considered at the same time. In particular, our results are directly applicable to SUSY models where the gluino and/or the different squark flavors coannihilate with the neutralino. Note that, for squark-neutralino coannihilation, since flavor constraints strongly limit the annihilation rates between different sfermion species, the existence of multiple degenerate squark or slepton flavors can lead to particularly tight relic density bounds~\cite{Profumo:2006bx,Chakraborti:2017dpu,Davidson:2017gxx}. Another major class of models with multiple coannihilation partners is Kaluza-Klein theories where the dark matter is the lightest Kaluza-Klein particle and can coannihilate with higher modes that are nearly degenerate with each other~\cite{Servant:2002aq}.  

Our study of models with multiple coannihilation partners is outlined as follows. In section~\ref{sec:relic:density}, we discuss how the relic density changes when we increase the number of coannihilation partners X$_i$ in a given model, and we derive a set of upper limits on the mixed X$_i$X$_j$ interactions. In the same section, we also introduce the simplified models of coannihilation that we are going to use in the rest of this work. We then describe all the possible tree-level mixed annihilation processes in more detail in section~\ref{sec:mixed:annihilation} and identify the regions of the parameter space for which adding new coannihilation partners can increase the dark matter depletion rate. Informed by these results, we discuss the collider constraints for these models in section~\ref{sec:collider:pheno}. Finally we conclude in section~\ref{sec:conclusions}.

\section{Relic density for multiple coannihilation partners}
\label{sec:relic:density}
In this section we discuss how increasing the number of coannihilation partners affects the dark matter relic density. In particular, we derive the conditions under which adding new coannihilation partners significantly increases the dark matter annihilation rate.  Since our goal is to determine how heavy the dark matter mass can be in thermal scenarios, we only consider strongly interacting coannihilation partners, that typically lead to the largest annihilation rates in the dark sector. We follow the approach described in~\cite{ElHedri:2017nny}, focusing on several representative models that allow to derive generic conclusions about strongly interacting coannihilating dark sectors. In what follows, we first discuss the dependence of the effective dark matter annihilation cross-section on the annihilation rates of the additional partners and what ingredients are necessary for these partners to loosen the relic density bounds on the dark matter mass in any given model. We then describe the simplified models that we will use throughout this paper and present a few examples of how introducing additional coannihilation partners can affect the dark matter relic density constraints. 

\subsection{Multiple partners and dark matter annihilation}
\label{sec:relic:density:annihilation}
Here, we study how introducing multiple coannihilation partners modifies the dark matter effective annihilation rate in generic coannihilating models. We consider a scenario with one dark matter candidate with $g_\mathrm{DM}$ degrees of freedom and $N$ coannihilation partners X$_i$ with $g_i$ degrees of freedom. We denote the relative mass splittings between the X$_i$ and the dark matter by $\Delta_i = \frac{m_{\mathrm{X}_i} - m_\mathrm{DM}}{m_\mathrm{DM}}$. Neglecting the dark matter self-annihilation rate, the effective annihilation rate of the dark matter~\cite{Griest:1990kh,Edsjo:1997bg} in the non-relativistic approximation is
\begin{equation}\label{eq:effective:annihilation:rate}
	\sigma_\mathrm{eff} = \sum_{ij} \sigma_{ij}(s) \frac{g_i g_j}{g_\mathrm{eff}^2} K_i K_j ,
\end{equation}
where $\sigma_{ij}$ is the X$_i$ X$_j$ annihilation rate and $g_\mathrm{eff}$ and $K_i$ are defined by
\begin{equation}
	\begin{aligned}
		g_\mathrm{eff} & \equiv g_\mathrm{DM} + \sum_i g_i K_i \\
		K_i & \equiv (1 + \Delta_i)^{3/2} e^{- x \Delta_i}.
	\end{aligned}
\end{equation}

In order to acquire an intuition on how the number of coannihilation partners affects the effective rate from equation~\eqref{eq:effective:annihilation:rate}, we first assume that all the X$_i$ are degenerate. We denote the mass splitting between the DM and all the X$_i$ as $\Delta$. If only X$_i$\,X$_i$ self-annihilation, with cross-section $\sigma_\mathrm{XX}$, is allowed, equation~\eqref{eq:effective:annihilation:rate} becomes
\begin{equation} \label{eq:effective:annihilation:rate:only:self}
	\sigma_\mathrm{eff} = \frac{N \, g_\mathrm{X}^2 (1+\Delta)^3 e^{-2x\Delta}}{(g_\mathrm{DM}+ N g_\mathrm{X} (1+\Delta)^{3/2}e^{-x\Delta})^2}\, \sigma_\mathrm{XX}.    
\end{equation}
When $\Delta$ approaches zero, which usually corresponds to the largest effective annihilation rates, this equation simplifies to
\begin{equation}
	\sigma_\mathrm{eff} = \frac{N \, g_\mathrm{X}^2}{(g_\mathrm{DM}+ N g_\mathrm{X})^2}\, \sigma_\mathrm{XX}   
\end{equation}
which decreases as $1/N$ in the large $N$ or small $g_\mathrm{DM}$ limit. In the absence of mixed interactions between different species of dark sector particles, introducing additional copies of the coannihilation partner in any given model thus tightens the relic density constraints at low $\Delta$.  This rather counter-intuitive behavior has already been pointed out in the context of flavor violation in the squark sector~\cite{Herrmann:2011xe}.

At large $\Delta$, the dynamics of coannihilation drastically changes. Although increasing the number of coannihilation partners still makes X$_i$\,X$_i$ self-annihilation more difficult, the limiting process is now the conversion of DM into X$_i$. Since adding new coannihilation partners increases the number of possible final states for this process, the effective dark matter annihilation rate will also increase. This phenomenon can be observed analytically by taking the large $\Delta$ limit in equation~\eqref{eq:effective:annihilation:rate:only:self}, which gives
\begin{equation}\label{eq:large:delta}
	\sigma_\mathrm{eff} = N \, \frac{g_\mathrm{X}^2 (1+\Delta)^3 e^{-2x\Delta}}{g_\mathrm{DM}^2}\, \sigma_\mathrm{XX} .    
\end{equation}
In this regime, the increase of the effective annihilation rate is thus linear with the number of coannihilation partners.

Let us now introduce mixed X$_i$\,X$_j\,\to\,$SM\,SM annihilation processes with cross-sections all equal to a given $\sigma_\mathrm{mix}$. Now, equation~\eqref{eq:effective:annihilation:rate:only:self} becomes
\begin{equation}
	\sigma_\mathrm{eff} = \frac{N^2\,g_\mathrm{X}^2 (1+\Delta)^3 e^{-2x\Delta}}{(g_\mathrm{DM}+ N g_\mathrm{X} (1+\Delta)^{3/2}e^{-x\Delta})^2}\, \sigma_\mathrm{XX} \left[\frac{1}{N} + \left(1 - \frac{1}{N}\right)\frac{\sigma_\mathrm{mix}}{\sigma_\mathrm{XX}}\right].    
\end{equation}
Since allowing for mixed annihilation does not modify the DM-X conversion rate, $\sigma_\mathrm{eff}$ still increases with $N$ at large $\Delta$. In the small $\Delta$ limit, however, the total annihilation rate is now nearly independent of $N$. In particular, if the X$_i$ annihilate with each other indifferently ($\sigma_\mathrm{mix} = \sigma_\mathrm{XX}$), at $\Delta = 0$, the effective dark matter annihilation cross-section is the same as in a model with only one coannihilation partner. Conversely, when the mixed annihilation dominates over the self-annihilation ($\sigma_\mathrm{mix} > \sigma_\mathrm{XX}$), we observe an increase of $\sigma_\mathrm{eff}$ compared to a single-partner model. Interestingly, as long as there is more than one coannihilation partner in the model, this increase will depend only weakly on the actual value of $N$. This behavior indicates that focusing on models with only two coannihilation partners could be an efficient way to estimate how much $\sigma_\mathrm{eff}$ can increase in generic coannihilating models. 

As found in~\cite{ElHedri:2017nny}, the current LHC bounds on $m_\mathrm{DM}$ and $\Delta$ combined with the relic density constraints exclude values of $\Delta$ down to about $10$\% for simplified models with a single coannihilation partner. Understanding the behavior of the dark matter annihilation rate in this low $\Delta$ region is therefore crucial to inform the future collider search program. In the rest of this work, we thus focus on the $\Delta \approx 0$ region and derive a set of sufficient conditions for the simplified model constraints on the dark matter mass to still hold in scenarios with multiple coannihilation partners X$_i$. We will briefly comment on the effect of a non-zero $\Delta$ on models with two different coannihilation partners in section~\ref{sec:model:coannihilation}. As already found in the simple case study above, we establish that the only way to invalidate the current simplified model constraints at low $\Delta$ is to introduce sizable mixed annihilation rates between the X$_i$. We will thus derive a set of upper bounds on these mixed rates for three different scenarios: $N$ identical coannihilation partners, two different coannihilation partners, and finally $N$ different X$_i$. The latter is the completely general case and hence serves as our main result.

\subsubsection{Multiple equal species}
\label{subsubsec:multiple:equal:species}
We first consider a simple scenario similar to the one discussed above, with one dark matter candidate that does not self-annihilate, and $N$ identical coannihilation partners X$_i$ with the same numbers of degrees of freedom $g_\mathrm{X}$ and self-annihilation cross-sections $\sigma_\mathrm{X}$. The mixed annihilation cross-sections between the different X$_i$ are assumed to be all identical and equal to $\sigma^\mathrm{mix}_\mathrm{XX}$. The effective dark matter annihilation rate in the small $\Delta$ limit then reads
\begin{equation}
	\sigma^\mathrm{eff}_N = \frac{g_\mathrm{X}^2}{(g_\mathrm{DM} + N g_\mathrm{X})^2} \left[ N \sigma_\mathrm{X} + N (N - 1) \sigma^\mathrm{mix}_{\mathrm{X} \mathrm{X}} \right] ,
\end{equation}
This rate needs to be compared to the effective annihilation rate of a model with only one species, which in this case is defined by
\begin{equation}
	\sigma^\mathrm{eff}_\mathrm{X} = \frac{g_\mathrm{X}^2}{(g_\mathrm{DM} + g_\mathrm{X})^2} \sigma_\mathrm{X} .
\end{equation}
In order for $\sigma^\mathrm{eff}_N$ to be smaller than $\sigma^\mathrm{eff}_\mathrm{X}$ for $N \geq 2$ we therefore need
\begin{equation} \label{eq:sigma:mix:limit:same:species}
	\sigma^\mathrm{mix}_{\mathrm{X} \mathrm{X}} \leq \frac{N g_\mathrm{X}^2 - g_\mathrm{DM}^2}{N (g_\mathrm{DM} + g_\mathrm{X})^2} \, \sigma_\mathrm{X} .
\end{equation}
In our models, the highest number of degrees of freedom for the dark matter is $g_\mathrm{DM} = 6$ for a complex vector boson, while the lowest possible number of degrees of freedom for a strongly interacting X is $g_\mathrm{X} = 6$ for a complex scalar triplet. The upper bound on the mixed annihilation rate will then range from $\frac{\sigma_\mathrm{X}}{8}$ for a complex vector dark matter candidate, a complex scalar triplet X, and $N = 2$, up to values close to $\sigma_\mathrm{X}$ for large N and $g_\mathrm{X} \gg g_\mathrm{DM}$. Since the X$_i$ can self-annihilate through the strong interaction, obtaining a large enough $\sigma^\mathrm{mix}_{\mathrm{X} \mathrm{X}}$ to break condition~\eqref{eq:sigma:mix:limit:same:species} requires either particularly large couplings or specific kinematic features such as resonances, interferences, or a suppressed (e.g. $p$-wave) self-annihilation cross-section.

\subsubsection{Two different species}
We now consider a model with one dark matter candidate and two coannihilation partners X$_1$ and X$_2$ that can have different properties. The effective annihilation cross-section for this model is
\begin{equation}
	\sigma^\mathrm{eff}_{\mathrm{X}_1 \mathrm{X}_2} = \frac{1}{(g_\mathrm{DM} + g_{\mathrm{X}_1} + g_{\mathrm{X}_2})^2} \left( g_{\mathrm{X}_1}^2 \sigma_{\mathrm{X}_1} + g_{\mathrm{X}_2}^2 \sigma_{\mathrm{X}_2} + 2 g_{\mathrm{X}_1} g_{\mathrm{X}_2} \sigma^\mathrm{mix}_{\mathrm{X}_1 \mathrm{X}_2} \right) .
\end{equation}
This cross-section needs to be smaller than the single effective annihilation cross-section for a single-partner model with either DM and X$_1$ or DM and X$_2$, given by
\begin{equation} \label{eq:single:partner:model}
	\sigma^\mathrm{eff}_{\mathrm{X}_i} = \frac{g_{\mathrm{X}_i}^2}{(g_\mathrm{DM} + g_{\mathrm{X}_i})^2} \sigma_{\mathrm{X}_i} .
\end{equation}
Now we can freely assume $\sigma^\mathrm{eff}_{\mathrm{X}_1} > \sigma^\mathrm{eff}_{\mathrm{X}_2}$, which leads to a condition of the form
\begin{equation} \label{eq:sigma:mix:limit:two:species}
	\sigma^\mathrm{mix}_{\mathrm{X}_1 \mathrm{X}_2} \leq \frac{1}{2} \left[ \frac{g_{\mathrm{X}_1} ( 2 g_\mathrm{DM} + 2 g_{\mathrm{X}_1} + g_{\mathrm{X}_2} )}{(g_\mathrm{DM} + g_{\mathrm{X}_1})^2} \sigma_{\mathrm{X}_1} - \frac{g_{\mathrm{X}_2}}{g_{\mathrm{X}_1}} \sigma_{\mathrm{X}_2} \right] .
\end{equation}
Note that this equation reduces to the constraint in equation~\eqref{eq:sigma:mix:limit:same:species} for $N=2$ and equal $\mathrm{X}_1$ and $\mathrm{X}_2$, as expected. 

\subsubsection{Multiple different species}
\label{sec:multiple:different:species}
We finally consider the most general scenario, with one dark matter candidate and $N$ coannihilating partners that are allowed to have different properties from each other. We can rewrite equation~\eqref{eq:effective:annihilation:rate} as
\begin{equation} \label{eq:general:coannihilation}
	\sigma^\mathrm{eff}_{\mathrm{X}_1 \cdots \mathrm{X}_N} = \frac{1}{(g_\mathrm{DM} + \sum_i g_{\mathrm{X}_i})^2} \left( \sum_i g_{\mathrm{X}_i}^2 \sigma_{\mathrm{X}_i} + 2 \sum_{i < j} g_{\mathrm{X}_i} g_{\mathrm{X}_j} \sigma^\mathrm{mix}_{\mathrm{X}_i \mathrm{X}_j} \right).
\end{equation}
Here, the second term sums over all mixed annihilation cross-sections between the coannihilation partners $\mathrm{X}_i$ and $\mathrm{X}_j$. This time, $\sigma^\mathrm{eff}_{\mathrm{X}_1 \cdots \mathrm{X}_N}$ may not exceed $\mathrm{max} \left( \sigma^\mathrm{eff}_{\mathrm{X}_1}, \ldots, \sigma^\mathrm{eff}_{\mathrm{X}_N} \right)$, with $\sigma^\mathrm{eff}_{\mathrm{X}_i}$ the effective annihilation cross-section for a model with only the dark matter and X$_i$ defined in~\eqref{eq:single:partner:model}. Although this constraint naively translates into heavily model-dependent bounds on the different mixed rates, it is also automatically satisfied when all the submodels of the form DM + X$_i$ + X$_j$ satisfy equation~\eqref{eq:sigma:mix:limit:two:species}. The detailed derivation of this property is shown in Appendix~\ref{sec:general:case:calculation}. As a consequence, all the possible mechanisms that could loosen the existing simplified model bounds in models with multiple coannihilation partners already appear in models with two X particles. This is our main result and considerably simplifies our analysis, since it allows us to focus exclusively on such models for the rest of this work.

We now apply the requirements derived here to specific dark matter models. The formalism used throughout this paper, in particular the different models that we are going to study, is detailed in the next section. We highlight the importance of the mixed X$_i$\,X$_j$ interactions through several examples and briefly comment on the cases where the mass splittings $\Delta_i$ are different from each other.

\subsection{Models of the coannihilating dark sector}
\label{sec:model:coannihilation}
In this work we extend the scope of the previous colored dark sector studies~\cite{ElHedri:2017nny,deSimone:2014pda,Ellis:2015vaa,Liew:2016hqo,Ibarra:2015nca} to include multiple coannihilation partners. We follow the methodology described in~\cite{ElHedri:2017nny} and consider simplified models where the dark matter is a pure SM singlet that does not self-annihilate. The dark matter is in thermal and chemical equilibrium with at least two coannihilation partners X$_i$ that are strongly interacting. As shown in~\cite{ElHedri:2017nny}, in such models, the self-annihilation of the X$_i$ through the strong interaction significantly contribute to the dark matter depletion. These processes alone can in fact loosen the upper bound on the dark matter mass from a few TeV up to more than 10~TeV. In this study, in addition to the SM couplings, we will also introduce interactions that allow for \emph{mixed} annihilation channels of the form X$_i$\,X$_j$\,$\to$\,SM\,SM. We however neglect the influence of the X$_i$\,DM\,$\to$\,SM\,SM coannihilation channels (see reference~\cite{Baker:2015qna}) on the final relic density. Since these processes are still necessary to ensure that the X$_i$ decay and are in equilibrium with the dark matter, we introduce them as effective operators of the form DM\,X$_i$\,SM\,SM. The final Lagrangian for a given model will then be of the form
\begin{equation}
	\mathcal{L} = \mathcal{L}_\mathrm{DM} + \sum_i \mathcal{L}_{\mathrm{X}_i} + \sum_i\mathcal{L}_{\mathrm{DM} + \mathrm{X}_i} + \sum_{i\neq j} \mathcal{L}_{\mathrm{X}_i + \mathrm{X}_j} .
\end{equation}

The kinetic and mass terms of X and DM, $\mathcal{L}_\mathrm{X}$ and $\mathcal{L}_\mathrm{DM}$ as well as the effective operator describing the DM-X$_i$ interactions are taken from~\cite{ElHedri:2017nny}. In what follows, we will allow the X$_i$ to be either complex scalars or Dirac fermions, in the triplet, sextet or octet representation of $SU(3)$. The corresponding $\mathcal{L}_{\mathrm{X}_i}$ for X being either a scalar $S$ or a fermion $\psi$ are of the form
\begin{equation} \label{eq:lagrangians:x}
	\begin{aligned}
		\mathcal{L}_S & = \left[ D_{\mu,ij}^{\vphantom{\mu}} S_j\right]^\dagger \left[D_{ij}^\mu S_j \right] - m_S^2 \, S_i^\dagger S_i \\
		\mathcal{L}_\psi & = \bar{\psi}_i i \slashed{D}_{ij} \psi_j - m_\psi \bar{\psi}_i \psi_i ,
	\end{aligned}
\end{equation}
where $i, j$ are color indices and the $T^a_{\irrep{R}}$ matrices are the generators for the color representation $\irrep{R}$ of X. The covariant derivatives are given by
\begin{equation}
	D_{\mu, ij} = \partial_\mu \delta_{ij} - i g_s G_\mu^a (T^a_{\irrep{R}})_{ij} .
\end{equation}
We assume that the dark matter field as well as X$_i$ are protected by a $\mathbb{Z}_2$ symmetry. Since the dark matter quantum numbers will only enter in our final results through the effective number of degrees of freedom, we put no restriction on the spin of the DM.

While, as in~\cite{ElHedri:2017nny}, the DM-X$_i$ interactions are modeled by a suppressed effective operator, this formalism is not suitable for the X$_i$\,X$_j$\,SM\,SM interactions since in our study the associated annihilation rates can be large. In what follows, we will therefore fully resolve these mixed interactions by introducing a mediator that can be either a SM particle, a new physics $s$-channel, or a new physics $t$-channel particle. In the last two cases, we allow the mediator to be either a scalar, a fermion, or a vector, in the triplet, sextet, or octet representation of $SU(3)$, and choose the color and the spin that leads to the largest dark matter effective annihilation cross-section for each model. We introduce the $\mathcal{L}_{\mathrm{X}_i + \mathrm{X}_j}$ associated to the three possible mediator configurations in section~\ref{sec:mixed:models}. We provide all the models discussed in this work and in~\cite{ElHedri:2017nny}, including the mixed interactions, in the updated version of our \texttt{FeynRules v2.3.24}~\cite{Christensen:2008py, Alloul:2013bka} model package~\cite{deVries:2017xyz}. Using a \texttt{Mathematica} notebook each of the specific models can be generated in both \texttt{UFO}~\cite{Degrande:2011ua} and \texttt{CalcHEP v3.6.25}~\cite{Belyaev:2012qa} format and then be further used to do collider studies or calculate the thermal relic abundance.

\begin{figure}[!t]
	\centering
	\includegraphics[width=0.49\textwidth]{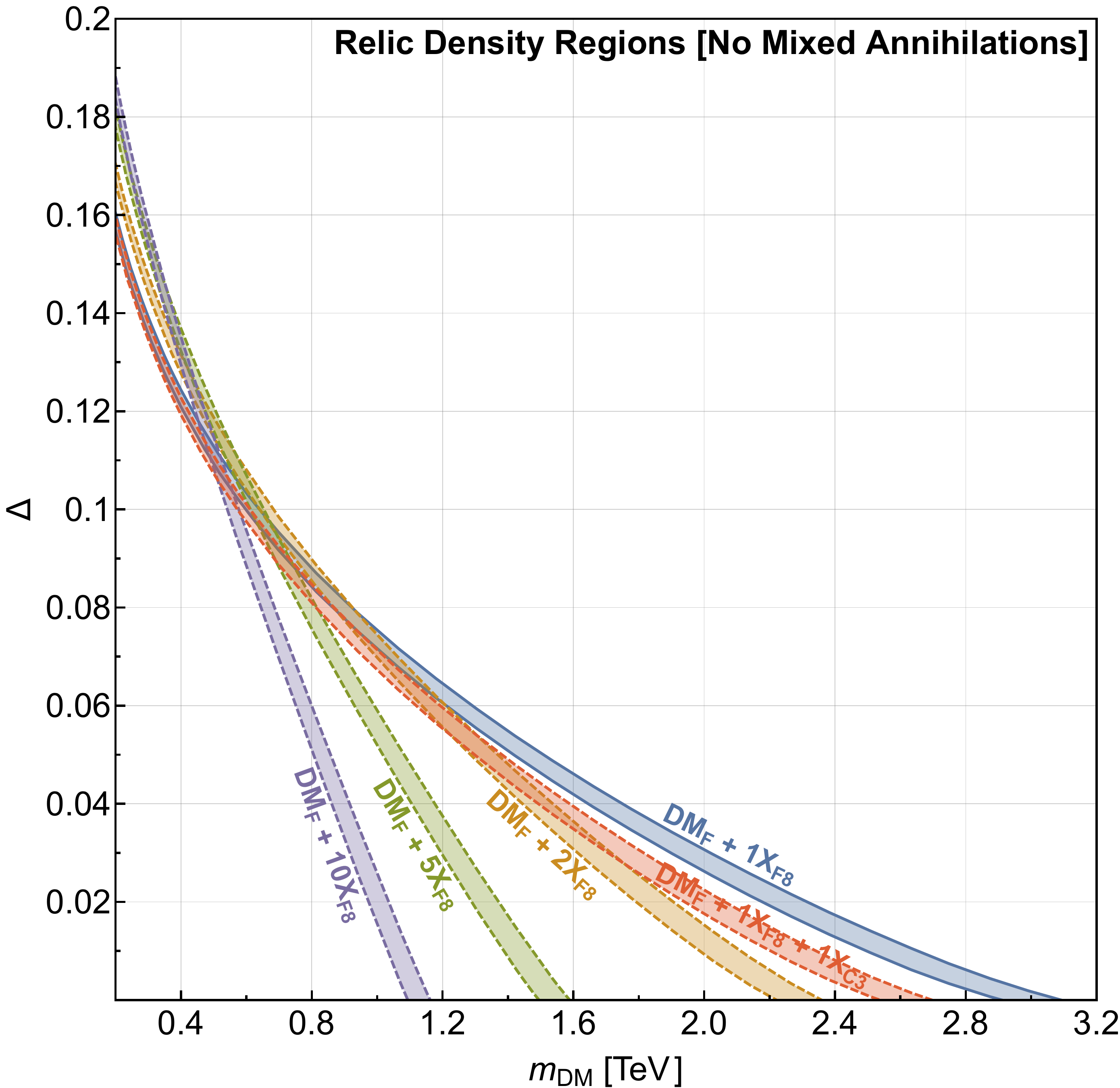}
	\includegraphics[width=0.49\textwidth]{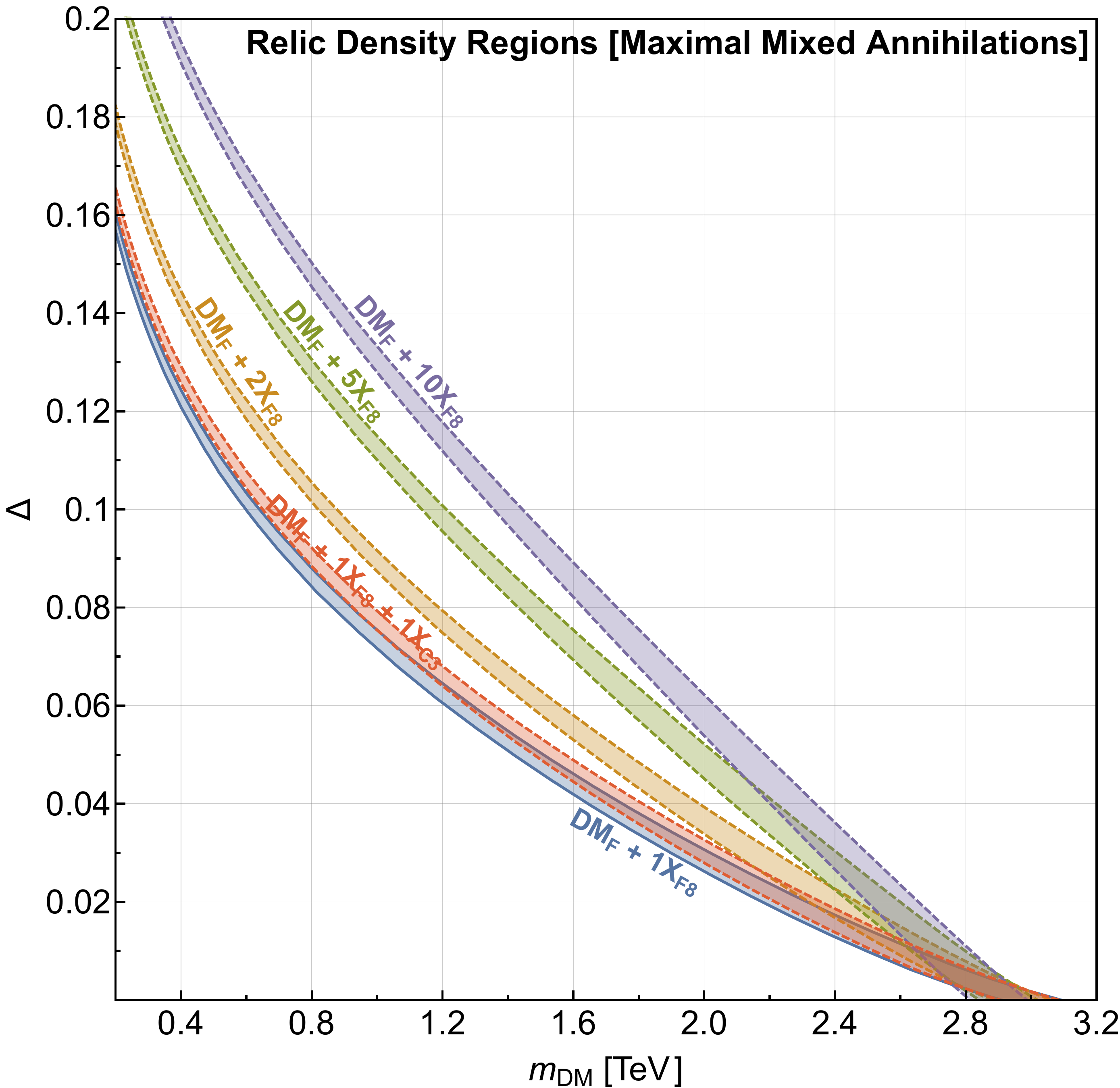}
	\caption{Relic density constraints in the $\Delta$ versus $m_\mathrm{DM}$ plane for models where DM is a Dirac fermion. We show the parameter space regions that agree with the relic abundance measured by Planck. We compare models with multiple coannihilation partners against the benchmark model with a single partner $\mathrm{X}_\mathrm{F8}$ (solid blue line). We either increase the multiplicity of $\mathrm{X}_\mathrm{F8}$ or introduce an additional partner $\mathrm{X}_\mathrm{C3}$. The left figure shows the contours in the absence of mixed annihilation channels, while the right figure shows the models with the mixed annihilation cross-section saturating condition~\eqref{eq:sigma:mix:limit:two:species}.}
	\label{fig:relic:density:multiplicity}
\end{figure}

In figure~\ref{fig:relic:density:multiplicity}, we show the effect of additional coannihilation partners, with and without mixed interactions, on the relic density constraints for a few of the models described in equation~\eqref{eq:lagrangians:x}. Here, we start from a model involving one fermionic dark matter candidate and one fermionic color octet X$_\mathrm{F8}$ (blue band) and add new coannihilation partners either neglecting mixed interactions (left figure) or saturating condition~\eqref{eq:sigma:mix:limit:two:species} (right figure). For all models we compute the relic density using \texttt{micrOMEGAs v4.3.2}~\cite{Belanger:2014vza}, directly inputting the cross-sections for the mixed processes into the code without assuming a specific type of interaction. As discussed in section~\ref{sec:relic:density:annihilation}, introducing extra particles, whether with the same or different quantum numbers always lowers the allowed dark matter masses when $\Delta$ is small and the mixed condition~\eqref{eq:sigma:mix:limit:two:species} is satisfied. However, for $\Delta \gtrsim 7\%$ in the absence of mixed interactions there is a crossover and introducing new coannihilation partners increases the allowed dark matter mass, as expected from equation~\eqref{eq:large:delta}. Finally, when saturating the condition~\eqref{eq:sigma:mix:limit:two:species}, the allowed dark matter masses become independent on the number of coannihilation partners at $\Delta \approx 0$ and increase with the number of X particles for larger $\Delta$. Importantly, for any model the largest allowed dark matter masses are still obtained at $\Delta = 0$. 

\begin{figure}[!t]
	\centering
	\includegraphics[width=0.49\textwidth]{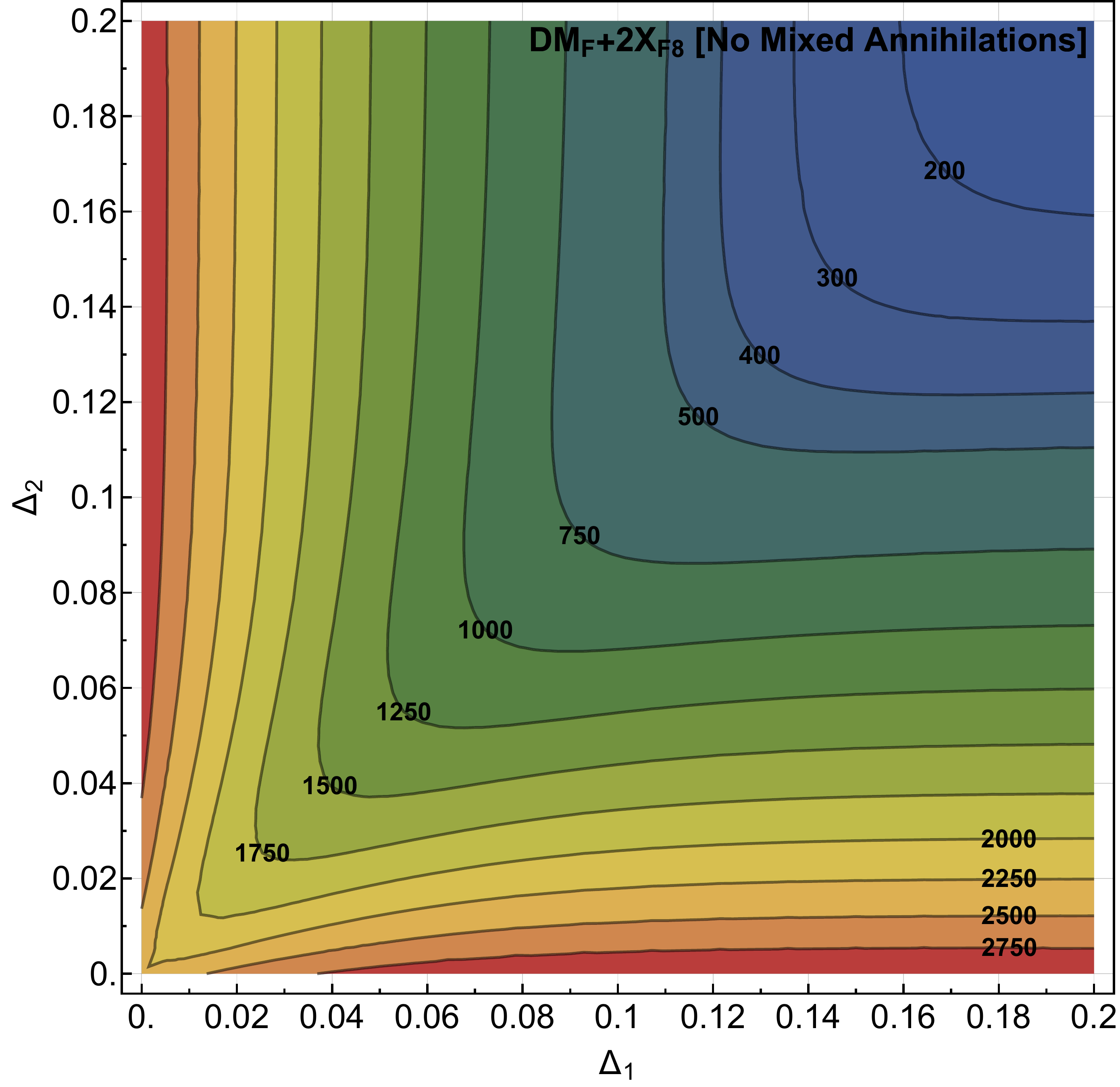}
	\includegraphics[width=0.49\textwidth]{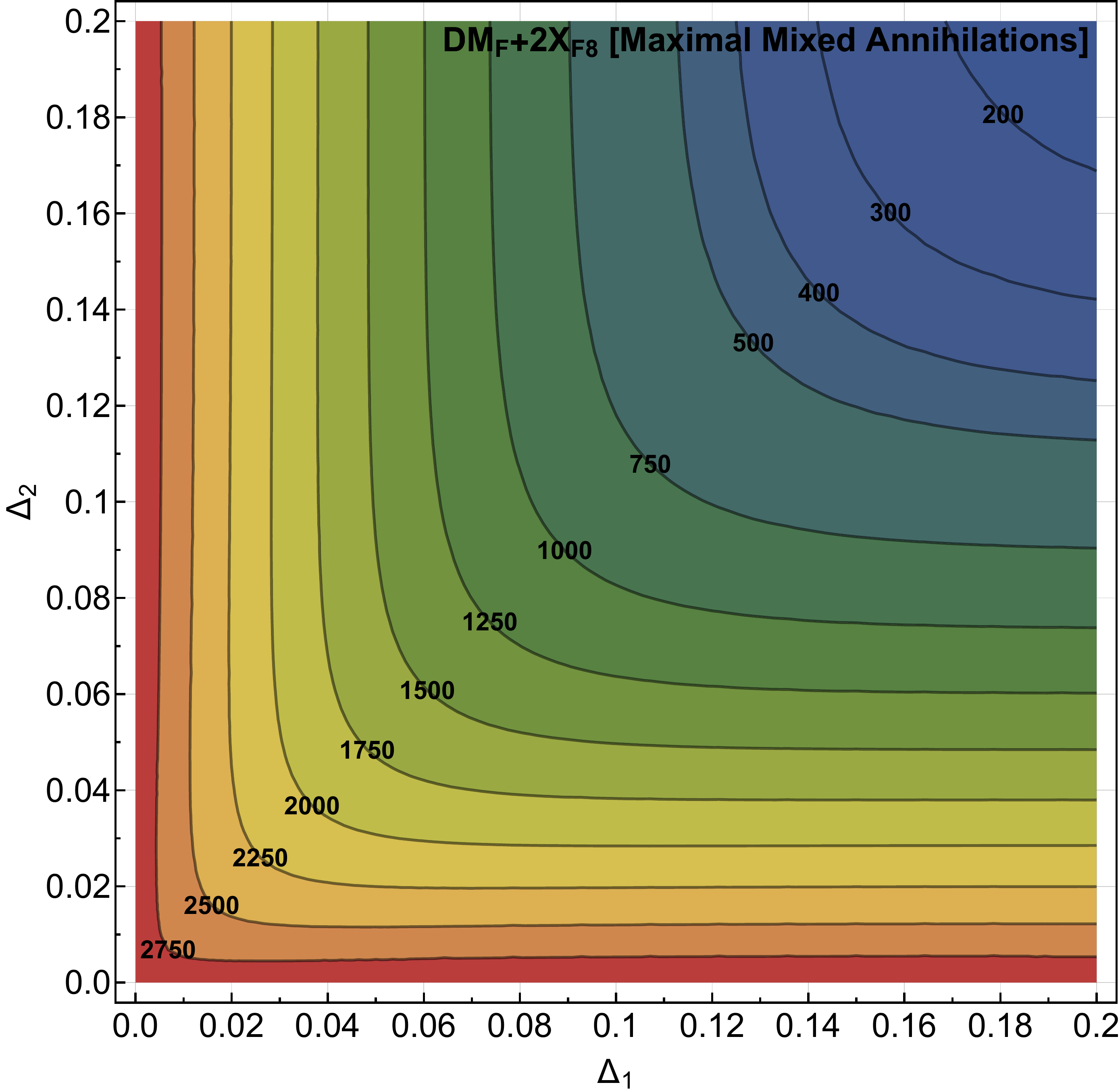}
	\caption{Contours of dark matter masses in GeV required to obtain the relic abundance measured by Planck as a function of the two mass splittings, $\Delta_1$ and $\Delta_2$. The dark matter is a Dirac fermion, while both coannihilation partners are fermionic color octets. The left figure shows the case with no mixed annihilation, whereas in the right figure the mixed annihilation rate is maximally saturating condition~\eqref{eq:sigma:mix:limit:two:species}.}
	\label{fig:relic:density:different:delta}
\end{figure}

For the former analysis as well as for the rest of this paper, we restrict our study to the case of $\Delta_1 \simeq \Delta_2$. In order to illustrate how relaxing this assumption changes the bounds on the DM masses, we compute the relic density constraints on a model with two fermionic color octet coannihilation partners, X$_\mathrm{F8}^{(1)}$ and X$_\mathrm{F8}^{(2)}$, with mass splittings $\Delta_1$ and $\Delta_2$ with the dark matter. The dark matter masses leading to the correct relic density are shown in figure~\ref{fig:relic:density:different:delta} as a function of $\Delta_1$ and $\Delta_2$. In this figure the straight horizontal and vertical lines indicate the regions where one of the coannihilation partners decouples. In the absence of any X$_1$\,X$_2$ annihilation process and for $\Delta_{1,2} \lesssim 10$\%, bringing the $\Delta_i$ closer to each other leads to a decrease in the DM mass, and thus to tighter relic density constraints. For $\Delta_{1,2} \gtrsim 10$\%, the opposite behavior occurs. These results are a direct generalization of the behavior observed on the left panel of figure~\ref{fig:relic:density:multiplicity} since the $\Delta_i \gg \Delta_j$ and the $\Delta_1 = \Delta_2$ cases correspond to the DM$_\mathrm{F}$ + 1X$_\mathrm{F8}$ and DM$_\mathrm{F}$ + 2X$_\mathrm{F8}$ models respectively. A similar reasoning can be applied to the scenario where the X$_1$\,X$_2$ annihilation rate saturates the condition~\eqref{eq:sigma:mix:limit:two:species}, shown on the right panel of figure~\ref{fig:relic:density:different:delta}. Here, by construction, for low values of $\Delta_{1,2}$ the allowed dark matter mass for $\Delta_1 = \Delta_2$ is similar to the one obtained when one of the coannihilation partners decouples. For larger $\Delta_i$, however, an increase in the allowed dark matter mass of up to about $100$~GeV can be observed in the $\Delta_1 \approx \Delta_2$ region. The diagonal symmetry observed for both the left and right panels of figure~\ref{fig:relic:density:different:delta} stems from the fact that the two coannihilation partners in our model have identical quantum numbers. For models where this is not the case, the diagonal would be shifted but the general features of figure~\ref{fig:relic:density:different:delta} are conserved. The results obtained throughout this section for identical $\Delta_i$ are therefore directly applicable to scenarios with different $\Delta_i$ and we can safely assume identical mass splittings for the rest of this study.

Note that, since we present figures~\ref{fig:relic:density:multiplicity} and~\ref{fig:relic:density:different:delta} for illustrative purposes, we computed only perturbative annihilation rates. In principle, these rates get modified by non-perturbative correction factors from the Sommerfeld effect~\cite{Sommerfeld:1931aa,ElHedri:2016onc} and bound state formation processes~\cite{Liew:2016hqo,Mitridate:2017izz}. As discussed in detail for models with a single coannihilation partner in~\cite{ElHedri:2017nny}, these effects generically lead to an increase in the dark matter mass allowed by the relic density requirement. This shift in the dark matter mass, however, will occur for both the single and multiple partner models so the relative positions of the relic density bands shown in figure~\ref{fig:relic:density:multiplicity} will not be significantly modified. Moreover, the results derived in section~\ref{sec:relic:density:annihilation} do not depend on how the annihilation cross-sections are computed and are thus valid whether or not non-perturbative effects are taken into account. We confirm this result through a quantitative study in section~\ref{sec:mixed:mediator:sm}.

As discussed in section~\ref{sec:multiple:different:species}, in order to study the validity of the usual simplified model constraints for models with multiple coannihilation partners, it is sufficient to consider models with only two X particles. In the next section we will consider all possible models with one dark matter candidate and two coannihilation partners X$_1$ and X$_2$ that are charged under $SU(3)$. Besides the gauge interactions, we introduce additional vertices and particle content in order to allow for the mixed X$_1$\,X$_2$\,$\to$\,SM\,SM interaction. The corresponding models will therefore be characterized by the quantum numbers of X$_1$ and X$_2$, the dark matter mass $m_\mathrm{DM}$, the  relative mass splittings $\Delta_1$ and $\Delta_2$ between the dark matter and X$_{1}$, X$_{2}$, and the parameters associated with the mixed interactions. In what follows, as discussed in section~\ref{sec:relic:density:annihilation}, since we are interested in the region of parameter space giving the loosest limits on the dark matter mass, we assume that $\Delta_1 = \Delta_2 = 0$. As derived in section~\ref{sec:relic:density:annihilation}, the maximum value on the mixed annihilation rate up to which the simplified model constraints are valid is then
\begin{equation} \label{eq:sigma:mix:limit}
	\sigma^\mathrm{mix}_{\mathrm{X}_1 \mathrm{X}_2} \leq \frac{(g_\mathrm{DM} \! + \! g_{\mathrm{X}_1} \! + \! g_{\mathrm{X}_2} \!)^2}{2 g_{\mathrm{X}_1} g_{\mathrm{X}_2}} \! \left[ \mathrm{max} \! \left( \!\! \frac{g_{\mathrm{X}_1}^2 \sigma_{\mathrm{X}_1}}{(g_\mathrm{DM} \! + \! g_{\mathrm{X}_1} \!)^2} , \frac{g_{\mathrm{X}_2}^2 \sigma_{\mathrm{X}_2}}{(g_\mathrm{DM} \! + \! g_{\mathrm{X}_2} \!)^2} \!\! \right) \!\! - \! \frac{g_{\mathrm{X}_1}^2 \sigma_{\mathrm{X}_1} \!\! + \! g_{\mathrm{X}_2}^2 \sigma_{\mathrm{X}_2}}{(g_\mathrm{DM} \! + \! g_{\mathrm{X}_1} \! + \! g_{\mathrm{X}_2} \!)^2} \right] \!\! .
\end{equation}
This condition is the main result of our work and serves to ensure that at low $\Delta_i$ the introduction of additional coannihilation partners does not increase the maximal dark matter mass allowed by the relic density constraint. 

In what follows, we fully resolve the structure of the  X$_1$\,X$_2$\,$\to$\,SM\,SM interaction and investigate the impact of condition~\eqref{eq:sigma:mix:limit} on the corresponding parameters. Since, in order to saturate this condition, the mixed annihilation rate needs to be large, we will consider only tree-level and renormalizable interactions. This restriction leaves us with three possible topologies: annihilation via a SM mediator, $s$-channel annihilation via a new physics mediator, and $t$-channel annihilation via a new physics mediator. The mixed interaction will therefore be characterized by at most two coupling constants as well as the mass and the quantum numbers of the mediator. A detailed study of these three scenarios and the associated constraints is presented in the next section.

\section{Mixed annihilation}
\label{sec:mixed:annihilation}
In the previous section, we derived a sufficient condition for the constraints from simplified models with a single coannihilation partner to still apply to models with multiple coannihilation partners. This condition can be written as an upper bound on the annihilation rates for ``mixed'' processes of the form X$_i$\,X$_j$\,$\to$\,SM\,SM. Moreover, we showed that to understand whether the coannihilating simplified model constraints still apply to a model with an arbitrary number of coannihilation partners, it is sufficient to apply this requirement to each X$_i$\,X$_j$ pair individually. In what follows, we explore how this condition on the mixed annihilation rates constrains the masses and couplings in the dark sector for models with two coannihilation partners. To this end, we first construct all the possible tree-level renormalizable interactions that can lead to mixed coannihilation processes, introducing new particles and new couplings if necessary. For each of these processes and for representative choices of quantum numbers for the X$_i$ and the dark matter, we then derive the bounds on these new masses and couplings within which the coannihilating simplified model limits on the dark matter mass are still valid. 

\subsection{Modelling the mixed interaction}
\label{sec:mixed:models}
Adding coannihilation partners to a given model opens new annihilation channels that will modify the dark matter depletion rate. In particular, aside from the self-annihilation channels X$_i$\,X$_i$\,$\to$\,SM\,SM studied in~\cite{ElHedri:2017nny}, processes of the form X$_i$\,X$_j$\,$\to$\,SM\,SM can now take place. As outlined in section~\ref{sec:relic:density:annihilation}, in order to establish whether introducing additional coannihilation partners can significantly loosen the bounds on the dark matter mass, it is essential to model these mixed processes and estimate their strength. 

In the rest of this section, we focus on all the possible X$_i$\,X$_j$ interactions that could significantly modify the dark matter annihilation cross-section. As discussed in section~\ref{sec:relic:density:annihilation}, we consider models with  two coannihilation partners X$_1$ and X$_2$ and choose both of these partners to be strongly interacting since we focus on the models typically associated to the loosest relic density bounds. Following the approach described in~\cite{ElHedri:2017nny}, we assume that the thermal and chemical equilibrium between a given X$_i$ and the dark matter is ensured by an effective operator of the form X$_i$\,DM\,SM\,SM, with the DM being an SM singlet. In the rest of this work, we will assume that this operator is suppressed and can be neglected compared to the other annihilation channels. The Lagrangian for a given model can then be written as
\begin{align}
    \mathcal{L} &= \mathcal{L}_\mathrm{DM} + \mathcal{L}_{\mathrm{X}_1} + \mathcal{L}_{\mathrm{X}_2} + \mathcal{L}_{\mathrm{DM-X}} + \mathcal{L}_{\mathrm{X}_1-\mathrm{X}_2} ,
\end{align}
where $\mathcal{L}_\mathrm{DM}$, $\mathcal{L}_{\mathrm{X}_1}$, $\mathcal{L}_{\mathrm{X}_2}$, $\mathcal{L}_{\mathrm{DM-X}}$ are given in section~\ref{sec:model:coannihilation}. We have implemented several examples for $\mathcal{L}_{\mathrm{X}_1-\mathrm{X}_2}$ in the Mathematica package~\cite{deVries:2017xyz} introduced in~\cite{ElHedri:2017nny} for single partner models. As seen in section~\ref{sec:relic:density:annihilation}, the rates of the mixed and self-annihilation processes need to be comparable to lead to an increase of the DM annihilation cross-section. In what follows, we therefore ignore possible non-renormalizable or loop-level interactions between X$_1$ and X$_2$ and focus on the dimension four tree-level interactions susceptible to lead to X$_1$\,X$_2$\,$\to$\,SM\,SM processes. These interactions are of three types: X$_1$\,X$_2$\,SM, X$_1$\,X$_2$\,M$_s$ with M$_s$ being a new physics $s$-channel mediator, and X$_{1,2}$\,M$_t$\,SM with M$_t$ being a $t$-channel mediator interacting with both X$_1$ and X$_2$. In the rest of this work, we study these three possible scenarios separately for various color representations of X$_1$ and X$_2$. Since our goal is to establish what the loosest constraints on the dark sector can be in these models, we will consider only the Standard Model particles giving the weakest LHC bounds, i.e. quarks and gluons. We would like to emphasize, however, that the procedure outlined in this section is generic and is expected to lead to similar results for a wide range of models involving weakly interacting particles.

For two-to-two annihilation processes into quark or gluons, the allowed color representations for the new physics particles naively range from $\irrep{1}$ to $\irrep{27}$. In fact, since two different particles cannot annihilate into two gluons at tree-level, the only allowed $SU(3)$ representations for X$_1$ and X$_2$ are $\irrep{1}$, $\irrep{3}$, $\irrep{6}$ and $\irrep{8}$. In what follows, we generally choose the X partners to be either triplets or octets of $SU(3)$ and the mediators to be either singlets, triplets, or octets. Although sextets are commonly encountered in new physics models~\cite{Bauer:2009cc,Ma:1998pi}, for most processes, their associated color factors are similar to the ones for octets. We therefore consider color sextets only for processes where color octets are forbidden by gauge invariance, such as $s$-channel annihilations to $q\,q$ or $\bar{q}\,\bar{q}$. We do not impose any particular restriction on the spin of the two X particles, that can be either scalars, fermions, or vectors. For the sake of clarity, we do not consider all the possible spin configurations for the new physics particles and, for a given mixed process, just choose the spin and coupling structure that give the least suppressed amplitude. Finally, note that, in our framework, the dark matter does not self-annihilate and its quantum numbers contribute to the effective annihilation cross-section only through the number of degrees of freedom $g_\mathrm{DM}$. Since according to equation~\eqref{eq:sigma:mix:limit} a larger $g_\mathrm{DM}$ enhances the contribution of the additional X$_j$ to the effective annihilation cross-section, we choose DM to be a complex vector, which corresponds to $g_{\mathrm{DM}}= 6$.

Although we neglect processes involving weak interactions, it is important to notice that specific UV completions of the X$_1$-X$_2$-SM-SM interaction can lead to constraints on the $SU(2)$ quantum numbers of X$_1$ and X$_2$. These constraints notably arise when X$_1$ and X$_2$ annihilate to $q\,\bar{q}$ via an $s$-channel mediator M$_s$ since the $SU(2)$ quantum numbers of the quarks depend on their chiralities. After EWSB, scenarios where the coannihilation partners are $SU(2)$ multiplets can involve a large number of coannihilation partners and channels and need to be treated with care~\cite{Cirelli:2009uv,Cirelli:2007xd,Cirelli:2005uq}. Since the dark sector particles in our models typically have multi-TeV masses, we can however safely treat all the components of a given $SU(2)$ multiplet as identical copies of the same particle that --- since we neglect the effects of the weak interaction --- do not mix with each other. In this approximation, according to the results from~\ref{sec:multiple:different:species}, scenarios where X$_1$ and X$_2$ are $SU(2)$ multiplets can be studied by considering only a single component of each multiplet. Hence, in what follows, in models where X$_1$ and X$_2$ are required to be charged under $SU(2)$, we will only study the annihilation of the electrically neutral components of X$_1$ and X$_2$ through the neutral component of the mediator, effectively treating these particles as $SU(2)$ singlets. 

We now introduce the different types of X$_1$\,X$_2$\,$\to$\,SM\,SM processes, classifying them by their mediator: SM particle, new physics particle in the $s$-channel, or new physics particle in the $t$-channel. For each category, we present the Lagrangians for the models we study and discuss in detail all the associated annihilation processes.

\subsubsection{SM mediator}
\label{sec:mixed:model:sm}
\begin{figure}[!t]
	\centering
	\begin{tikzpicture}[line width=1.4pt, scale=1]
	\draw[gluon] (0.8,0.8)--(0.5,0.0);
	\draw[fermionbar] (0.8,-0.8)--(0.5,0.0);
	\draw[fermionna] (-0.8,0.8)--(-0.5,0);
	\draw[fermionna] (-0.8,-0.8)--(-0.5,0);
	\draw[fermion] (-0.5,0)--(0.5,0);
	
	\node at (-1.05,0.8) {X$_1$};
	\node at (-1.05,-0.8) {X$_2$};
	\node at (0,0.225) {$q$};
	\node at (1.05,0.8) {$g$};
	\node at (1.05,-0.8) {$q$};
\end{tikzpicture}
	\hspace{4mm}
	\begin{tikzpicture}[line width=1.4pt, scale=1]
	\draw[gluon] (0.8,0.8)--(0,0.5);
	\draw[fermionbar] (0.8,-0.8)--(0,-0.5);
	\draw[fermionna] (-0.8,0.8)--(0,0.5);
	\draw[fermionna] (-0.8,-0.8)--(0,-0.5);
	\draw[fermionna] (0,0.5)--(0,-0.5);
	
	\node at (-1.05,0.8) {X$_1$};
	\node at (-1.05,-0.8) {X$_2$};
	\node at (-0.3,0) {X$_1$};
	\node at (1.05,0.8) {$g$};
	\node at (1.05,-0.8) {$q$};
\end{tikzpicture}
	\hspace{4mm}
	\begin{tikzpicture}[line width=1.4pt, scale=1]
	\draw[fermionbar] (0.8,0.8)--(0,0.5);
	\draw[gluon] (0.8,-0.8)--(0,-0.5);
	\draw[fermionna] (-0.8,0.8)--(0,0.5);
	\draw[fermionna] (-0.8,-0.8)--(0,-0.5);
	\draw[fermionna] (0,0.5)--(0,-0.5);
	
	\node at (-1.05,0.8) {X$_1$};
	\node at (-1.05,-0.8) {X$_2$};
	\node at (-0.3,0) {X$_2$};
	\node at (1.05,0.8) {$q$};
	\node at (1.05,-0.8) {$g$};
\end{tikzpicture}
	\hspace{12mm}
	\begin{tikzpicture}[line width=1.4pt, scale=1]
	\draw[fermion] (0.8,0.8)--(0,0.5);
	\draw[fermionbar] (0.8,-0.8)--(0,-0.5);
	\draw[fermionna] (-0.8,0.8)--(0,0.5);
	\draw[fermionna] (-0.8,-0.8)--(0,-0.5);
	\draw[fermionna] (0,0.5)--(0,-0.5);
	
	\node at (-1.2,0.8) {X$_{1/2}$};
	\node at (-1.2,-0.8) {X$_{1/2}$};
	\node at (-0.4,0) {X$_{2/1}$};
	\node at (1.05,0.8) {$\bar{q}$};
	\node at (1.05,-0.8) {$q$};
\end{tikzpicture}
	\caption{The first three diagrams show the mixed annihilation processes when a coupling between the two X's and a SM quark is present. This coupling also introduces a new contribution to the self-annihilation rates of X$_1$ and X$_2$ as shown in the fourth diagram.}
	\label{fig:mixed:channels:sm:mediator}
\end{figure}
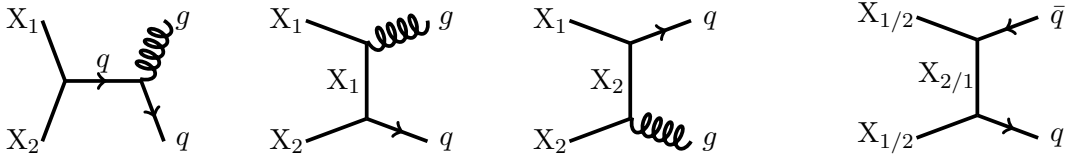

A common new physics scenario is the existence of a coupling between two dark sector particles, X$_1$ and X$_2$, and a Standard Model field. For the models studied here, since we are considering only annihilations into quarks or gluons, this Standard Model field can only be a quark. The $SU(3)$ representations for X$_1$ and X$_2$ allowed by gauge invariance are therefore $(\irrep{3},\irrep{3}$), ($\irrep{3}, \overline{\irrep{6}}$), ($\irrep{3}, \irrep{8}$) and ($\irrep{6}, \irrep{8}$) as well as their conjugates. Similarly, the only allowed spin configurations are the ones where one of the X is bosonic and the other one is fermionic. For these choices of quantum numbers, introducing the new X$_1$-X$_2$-quark interaction leads to the additional coannihilation processes shown in figure~\ref{fig:mixed:channels:sm:mediator}. Note that a new $t$-channel X$_i$\,X$_i$\,$\to$\,SM\,SM annihilation diagrams appears, which will either increase or decrease the self-annihilation rate of the coannihilation partners, depending on the quantum numbers of the X$_i$.

For this study, we choose X$_1$ to be a complex scalar and X$_2$ to be either a Dirac or a Majorana fermion. Note that, if X$_2$ is a Majorana fermion, either of the new annihilation channels X$_2$\,X$_2\,\to\,q\,q,\bar{q}\,\bar{q}$ opens. The existence of this new channel leads to an increase in the X$_2$ self-annihilation rate that makes is more difficult for the mixed annihilation rate to dominate compared to the Dirac fermion scenario. We focus on a few representative models for which the X$_1$-X$_2$ interaction Lagrangian is shown below 
\begin{equation} \label{eq:lagrangians:mixed:sm}
	\begin{aligned}
		\mathcal{L}_{\mathrm{X}_\mathrm{C3} + \mathrm{X}_\mathrm{F3}} & = y_\mathrm{NP} \, \epsilon_{ijk} \, \overline{\mathrm{X}}_{\mathrm{C3}, k} \, \overline{\mathrm{X}}_{\mathrm{F3}, i} \, q^c_{R,j} + \mathrm{h.c.} \\
		\mathcal{L}_{\mathrm{X}_\mathrm{C3} + \mathrm{X}_\mathrm{F8}} & = y_\mathrm{NP} \, T^a_{ij} \, \mathrm{X}_{\mathrm{C3}, i} \, \overline{\mathrm{X}}_\mathrm{F8}^a \, q_{R,j} + \mathrm{h.c.} \\
		\mathcal{L}_{\mathrm{X}_\mathrm{C3} + \mathrm{X}_\mathrm{M8}} & = y_\mathrm{NP} \, T^a_{ij} \, \mathrm{X}_{\mathrm{C3}, i} \, \overline{\mathrm{X}}_\mathrm{M8}^a \, q_{R,j} + \mathrm{h.c.} \\
		\mathcal{L}_{\mathrm{X}_\mathrm{C8} + \mathrm{X}_\mathrm{F3}} & = y_\mathrm{NP} \, T^a_{ij} \, \mathrm{X}_\mathrm{C8}^a \, \overline{\mathrm{X}}_{\mathrm{F3},i} \, q_{R,j} + \mathrm{h.c.} \\
		\mathcal{L}_{\mathrm{X}_\mathrm{C3} + \mathrm{X}_\mathrm{F6}} & = y_\mathrm{NP} \, K^u_{ij} \, \mathrm{X}_{\mathrm{C3}, i} \, \overline{\mathrm{X}}_\mathrm{F6}^u \, q_{R,j} + \mathrm{h.c.} \\
		\mathcal{L}_{\mathrm{X}_\mathrm{C6} + \mathrm{X}_\mathrm{F3}} & = y_\mathrm{NP} \, \overline{K}^u_{ij} \, \mathrm{X}_\mathrm{C6}^u \, \overline{\mathrm{X}}_{\mathrm{F3}, i} \, q^c_{R,j} + \mathrm{h.c.} \\
		\mathcal{L}_{\mathrm{X}_\mathrm{C6} + \mathrm{X}_\mathrm{F8}} & = y_\mathrm{NP} \, \mathcal{K}^{aui} \, \mathrm{X}_\mathrm{C6}^u \, \overline{\mathrm{X}}_\mathrm{F8}^a \, q_{R,i} + \mathrm{h.c.} \\
		\mathcal{L}_{\mathrm{X}_\mathrm{C8} + \mathrm{X}_\mathrm{F6}} & = y_\mathrm{NP} \, \mathcal{K}^{aui} \, \mathrm{X}_\mathrm{C8}^a \, \overline{\mathrm{X}}_\mathrm{F6}^u \, q^c_{R,i} + \mathrm{h.c.}
	\end{aligned}
\end{equation}
In these Lagrangians the $K^u_{ij}$ and their properties are defined in~\cite{Han:2009ya}, and the color factor $\mathcal{K}^{aui}$ for the $\irrep{3}-\irrep{6}-\irrep{8}$ interaction is
\begin{equation}
	\mathcal{K}^{aui} = (T^a)_j^l \, \epsilon^{ijk} K^u_{kl} ,
\end{equation}
where $j$, $k$, and $l$ run from one to three and $i$, $u$, $a$ are the indices of the external triplet, sextet, and octet particles respectively. Here, $q$ can be either an up or a down-type quark depending on the electromagnetic charges of X$_1$ and X$_2$. Besides the particle masses, the only new physics parameter for these models is the coupling $y_\mathrm{NP}$, which will typically be of order one as all particles involved are colored.

\subsubsection{New physics mediator ($s$-channel)}
\label{sec:mixed:model:schannel}
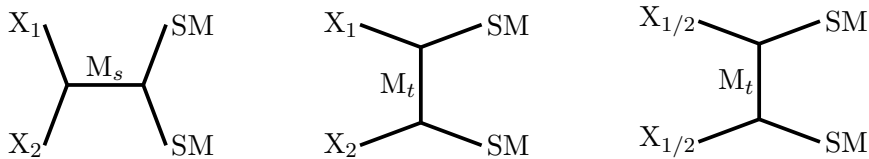
\begin{figure}[!t]
	\centering
	\begin{tikzpicture}[line width=1.4pt, scale=1]
	\draw[fermionna] (0.8,0.8)--(0.5,0.0);
	\draw[fermionna] (0.8,-0.8)--(0.5,0.0);
	\draw[fermionna] (-0.8,0.8)--(-0.5,0);
	\draw[fermionna] (-0.8,-0.8)--(-0.5,0);
	\draw[fermionna] (-0.5,0)--(0.5,0);
	
	\node at (-1.05,0.8) {X$_1$};
	\node at (-1.05,-0.8) {X$_2$};
	\node at (0,0.225) {M$_s$};
	\node at (1.15,0.8) {SM};
	\node at (1.15,-0.8) {SM};
\end{tikzpicture}
	\hspace{8mm}
	\begin{tikzpicture}[line width=1.4pt, scale=1]
	\draw[fermionna] (0.8,0.8)--(0,0.5);
	\draw[fermionna] (0.8,-0.8)--(0,-0.5);
	\draw[fermionna] (-0.8,0.8)--(0,0.5);
	\draw[fermionna] (-0.8,-0.8)--(0,-0.5);
	\draw[fermionna] (0,0.5)--(0,-0.5);
	
	\node at (-1.05,0.8) {X$_1$};
	\node at (-1.05,-0.8) {X$_2$};
	\node at (-0.3,0) {M$_t$};
	\node at (1.15,0.8) {SM};
	\node at (1.15,-0.8) {SM};
\end{tikzpicture}
	\hspace{8mm}
	\begin{tikzpicture}[line width=1.4pt, scale=1]
	\draw[fermionna] (0.8,0.8)--(0,0.5);
	\draw[fermionna] (0.8,-0.8)--(0,-0.5);
	\draw[fermionna] (-0.8,0.8)--(0,0.5);
	\draw[fermionna] (-0.8,-0.8)--(0,-0.5);
	\draw[fermionna] (0,0.5)--(0,-0.5);
	
	\node at (-1.2,0.8) {X$_{1/2}$};
	\node at (-1.2,-0.8) {X$_{1/2}$};
	\node at (-0.3,0) {M$_t$};
	\node at (1.15,0.8) {SM};
	\node at (1.15,-0.8) {SM};
\end{tikzpicture}
	\caption{The mixed annihilation diagram when a new physics mediator is present. The left diagram describes the case of an $s$-channel mediator whereas the middle diagram describes $t$-channel mediators. The right diagram denotes the modification of the self-annihilation rates of the coannihilation partners for a $t$-channel mediator.}
	\label{fig:mixed:channels:np:mediator}
\end{figure}

Instead of directly coupling to a SM particle, X$_1$ and X$_2$ can also interact with the SM through a new physics $s$-channel mediator M$_s$. The corresponding interaction requires introducing two new trilinear couplings, for the M$_s$-X$_1$-X$_2$ and the M$_s$-SM-SM vertices, thus constraining M$_s$ to be $\mathbb{Z}_2$ even. An interesting aspect of this model is that the new interactions open the mixed annihilation channel shown in figure~\ref{fig:mixed:channels:np:mediator} without introducing any new self-annihilation diagrams. Hence, such $s$-channel models could potentially lead to significant increases in the total dark matter annihilation cross-section with respect to models with a single partner, especially near the resonant $m_{\mathrm{M}} \approx m_{\mathrm{X}_1} + m_{\mathrm{X}_2}$ region. The increase of the annihilation rate near the resonance, however, is primarily due to the introduction of M$_s$ in the model and could therefore also occur in a scenario with only one coannihilation partner that self-annihilates through a new physics mediator. In order to disentangle the effects of adding a new coannihilation partner from the effect of having a new mediator, our two-partner model therefore needs to be compared to a one-partner model with a new mediator that has similar properties as M$_s$.

Since, for a new physics mediator, M$_s$-SM-SM interactions involving gluons are forbidden at tree-level, we focus only on scenarios where X$_1$ and X$_2$ annihilate into quarks. These annihilation processes can occur if the mediator is either a scalar or a vector, and a singlet, triplet, sextet or octet of $SU(3)$. Here, as mentioned at the beginning of this section, we consider only triplet and octet mediators and study models for which the mixed annihilation cross-section has a sizable $s$-wave component. For each X$_1$\,X$_2$ pair, we choose the spin of the mediator that leads to the largest X$_1$-X$_2$ annihilation cross-section. We therefore always use vector mediators for models with fermionic X$_1$ and X$_2$ and scalar mediators for models where the coannihilation partners are scalar fields. Mediators in the self-conjugate $\irrep{1}$ and $\irrep{8}$ representations of $SU(3)$ are taken to be real while triplet mediators are constrained to be complex. These choices lead us to select the following representative models 
\begin{equation} \label{eq:lagrangians:mixed:schannel}
	\begin{aligned}
    \mathcal{L}_{\mathrm{X}_\mathrm{C}\,\overline{\mathrm{X}}_\mathrm{C}\,\mathrm{M}_\mathrm{C1}} &= y_{X} m_S \,\mathrm{M}_\mathrm{C1}\,\mathrm{X}_\mathrm{C}\,\overline{\mathrm{X}}_\mathrm{C} + y_{SM}\,\mathrm{M}_\mathrm{C1}\,Q_L\,\overline{u}_R + h.c.\\
    \mathcal{L}_{\mathrm{X}_\mathrm{C}\,\overline{\mathrm{X}}_\mathrm{C}\,\mathrm{M}_\mathrm{C8}} &= y_{X} m_S \,(T^a_{R})_{ij}\,\mathrm{M}_\mathrm{C8}^a\,\mathrm{X}_{\mathrm{C}i}\,\overline{\mathrm{X}}_{\mathrm{C}j} + y_{SM}\,(T^a_{R})_{ij}\,\mathrm{M}_\mathrm{C8}^a\,Q_{Li}\,\overline{u}_{Rj} + h.c.\\
    \mathcal{L}_{\mathrm{X}_\mathrm{C3}\,\overline{\mathrm{X}}_\mathrm{C8}\,\mathrm{M}_\mathrm{C3}} &= y_{X} m_S \,\mathrm{M}_\mathrm{C3,i} T^a_{ij}\,\mathrm{X}_\mathrm{C3,j}\,\overline{\mathrm{X}}_\mathrm{C8}^a + y_{SM}\,\epsilon^{ijk}\mathrm{M}_\mathrm{C3,k}\,u_{R,i}\,u^C_{R,j} + h.c.\\
    \mathcal{L}_{\mathrm{X}_\mathrm{F} \, \overline{\mathrm{X}}_\mathrm{F} \, \mathrm{M}_\mathrm{V1}} & = y_\mathrm{X}\, \mathrm{M}_\mathrm{V1}^\mu \overline{\mathrm{X}}_\mathrm{F} \gamma_\mu \mathrm{X}_\mathrm{F} + y_\mathrm{SM} \mathrm{M}_\mathrm{V1}^\mu \bar{u}_{R} \gamma^\mu u_{R} + \mathrm{h.c.} \\
    \mathcal{L}_{\mathrm{X}_\mathrm{F} \, \overline{\mathrm{X}}_\mathrm{F} \, \mathrm{M}_\mathrm{W8}} & = y_\mathrm{X}\, (T^a_{R})_{ij}\,\mathrm{M}_\mathrm{W8}^{a\mu} \,\overline{\mathrm{X}}_{\mathrm{F}i} \gamma_\mu \mathrm{X}_{\mathrm{F}j} + y_\mathrm{SM}\, (T^a_{R})_{ij}\, \mathrm{M}_\mathrm{W8}^{a\mu} \,\bar{u}_{R,i} \gamma^\mu u_{R,j} + \mathrm{h.c.} \\
    \mathcal{L}_{\mathrm{X}_\mathrm{F3} \, \overline{\mathrm{X}}_\mathrm{F8} \, \mathrm{M}_\mathrm{W3}} & = y_\mathrm{X}\, (T^a_3)_{ij} \mathrm{M}_{\mathrm{W3},i}^\mu\,\overline{\mathrm{X}}_{\mathrm{F8}}^a\,\gamma_\mu\,\mathrm{X}_{\mathrm{F3},j} + y_\mathrm{SM}\epsilon^{ijk} \mathrm{M}_{\mathrm{W3},i}^\mu\, \overline{u}_{R, j} \gamma_\mu\, Q^c_{L, k} + \mathrm{h.c.}
	\end{aligned}
\end{equation}

Additionally, for scalar X$_1$ and X$_2$, all the M$_s$-X$_1$-X$_2$ interactions are parameterized by a dimensionful trilinear coupling that we write as $A_X = y_X\,m_S$. This notation highlights the fact that perturbative unitarity bounds prevent the trilinear coupling to be larger than a few times the mass of the mediator~\cite{Betre:2014fva,Schuessler:2007av}. 

As mentioned at the beginning of this section, in order to isolate the effects of introducing a new coannihilation partner for a given $s$-channel model, we need to compare this model to a scenario with only one coannihilation partner X that annihilates to $q\,\bar{q}$ via a mediator M$_s^\prime$. In order for the two models to be comparable, M$_s^\prime$ needs to be as similar as possible to the mediator of the mixed coannihilation process M$_s$. For models where this process is mediated by either a singlet or an octet, the quantum numbers of M$_s$ and M$_s^\prime$ can be taken to be exactly identical. For models where the X$_1$\,X$_2$ annihilation is mediated by a triplet, however, M$_s$ cannot be reused to mediate the X\,$\bar{\mathrm{X}}$ self-annihilation due to gauge invariance. In these cases, we consider both the cases of a singlet and an octet M$_s^\prime$, that has the same spin as M$_s$. In order to further ensure a fair comparison between the two- and one-partner scenarios, the couplings $y'_\mathrm{X}$ and $y'_\mathrm{SM}$ of M$_s^\prime$ to X\,$\bar{\mathrm{X}}$ and the SM respectively as well as the spin structure of the corresponding vertices are taken to be the same as for their mixed counterparts. For example, a model where X$_1$ and X$_2$ are both scalar triplets can be compared to a model with only one scalar triplet X and a mediator M$_\mathrm{C1}$, where the new physics interactions are parameterized by the Lagrangian $\mathcal{L}_{\mathrm{X}_\mathrm{C}\,\overline{\mathrm{X}}_\mathrm{C}\,\mathrm{M}_\mathrm{C1}}$ shown in equation~\eqref{eq:lagrangians:mixed:schannel}. 

\subsubsection{New physics mediator ($t$-channel)}
\label{sec:mixed:model:tchannel}
In this scenario, two new vertices M$_t$-X$_1$-SM and M$_t$-X$_2$-SM need to be introduced in order for X$_1$ and X$_2$ to annihilate to SM particles in the $t$-channel. Here, M is required to be $\mathbb{Z}_2$ odd and part of the dark sector. The mediator should therefore be heavier than both the dark matter and X for our choice of parameters. We will in particular focus on regions of parameter space where the splitting between M$_t$ and X$_1$, X$_2$ is sufficiently large to avoid chemical or thermal equilibrium between M$_t$ and the dark matter before and during freeze-out.

The introduction of the mediator and X$_1$, X$_2$ with their associated vertices opens the new annihilation channels shown in figure~\ref{fig:mixed:channels:np:mediator}. Interestingly, in this class of models, it is impossible to increase the X$_1$\,X$_2$ annihilation rate without also increasing the self-annihilation rates of X$_1$ and X$_2$.
Consequently, even for large couplings, any increase in the dark matter annihilation cross-section in the two-partner model compared to models with only either X$_1$ or X$_2$ is primarily due to the introduction of a new mediator. Since, as mentioned in the introduction, studying the effect of extra mediators on the dark matter depletion rate is beyond the scope of this work, we carefully factor out any mediator contribution to the dark matter annihilation rate in our subsequent study, as for the $s$-channel models studied in section~\ref{sec:mixed:model:schannel}. We describe the associated procedure in more details in section~\ref{sec:mixed:mediator:tchannel}.

As in the previous scenarios, annihilation of X$_1$ and X$_2$ into two gluons is forbidden by gauge invariance and we focus on X$_1$\,X$_2$ annihilation into either $q \, \overline{q}$ or $q \, q$. We therefore consider scenarios where either X$_1$ and X$_2$ are both complex scalars with a fermionic mediator, or X$_1$ and X$_2$ are both Dirac fermions with a complex scalar mediator. In this scenario, whether the fields are self-conjugate does not change the coupling strength for the interactions involved in the annihilation processes and we therefore made arbitrary choices. For all our models, we constrain X$_1$, X$_2$, and the mediator to be $SU(2)$ singlets, their hypercharges being defined by gauge invariance for each model. Finally, as for the models with an $s$-channel mediator, we choose the new physics particles to be either color triplet or octets. The Lagrangians describing the X$_1$ and X$_2$ interactions with the SM for our choice of models are 
\begin{equation} \label{eq:lagrangians:mixed:tchannel:scalarx}
	\begin{aligned}
		\mathcal{L}_{\mathrm{X}_\mathrm{C3} + \mathrm{M}_\mathrm{F1}} & = y_\mathrm{NP} \, \overline{\mathrm{X}}_{\mathrm{C3}, i} \, \overline{\mathrm{M}}_\mathrm{F1} \, q_{R,i} + \mathrm{h.c.} \\
		\mathcal{L}_{\mathrm{X}_\mathrm{C3} + \mathrm{M}_\mathrm{F8}} & = y_\mathrm{NP} \, T^a_{ij} \, \overline{\mathrm{X}}_{\mathrm{C3}, i} \, \overline{\mathrm{M}}_\mathrm{F8}^a \, q_{R,j} + \mathrm{h.c.} \\
		\mathcal{L}_{\mathrm{X}_\mathrm{C3} + \mathrm{M}_\mathrm{F3}} & = y_\mathrm{NP} \, \epsilon_{ijk} \, \overline{\mathrm{X}}_{\mathrm{C3}, k} \, \overline{\mathrm{M}}_{\mathrm{F3}, i} \, q_{R,j} + \mathrm{h.c.} \\
		\mathcal{L}_{\mathrm{X}_\mathrm{C8} + \mathrm{M}_\mathrm{F3}} & = y_\mathrm{NP} \, T^a_{ij} \, \overline{\mathrm{X}}_\mathrm{C8}^a \, \overline{\mathrm{M}}_{\mathrm{F3}, i} \, q_{R,j} + \mathrm{h.c.}
	\end{aligned}
\end{equation}
for scalar coannihilation partners and
\begin{equation} \label{eq:lagrangians:mixed:tchannel:fermionx}
	\begin{aligned}
		\mathcal{L}_{\mathrm{X}_\mathrm{F3} + \mathrm{M}_\mathrm{S1}} & = y_\mathrm{NP} \, \mathrm{M}_\mathrm{S1} \, \overline{\mathrm{X}}_{\mathrm{F3}, i} \, q_{R,i} + \mathrm{h.c.} \\
		\mathcal{L}_{\mathrm{X}_\mathrm{F3} + \mathrm{M}_\mathrm{S8}} & = y_\mathrm{NP} \, T^a_{ij} \, \mathrm{M}_\mathrm{S8}^a \, \overline{\mathrm{X}}_{\mathrm{F3}, i} \, q_{R,j} + \mathrm{h.c.} \\
		\mathcal{L}_{\mathrm{X}_\mathrm{F3} + \mathrm{M}_\mathrm{C3}} & = y_\mathrm{NP} \, \epsilon_{ijk} \, \overline{\mathrm{M}}_{\mathrm{C3}, k} \, \overline{\mathrm{X}}_{\mathrm{F3}, i} \, q^c_{R,j} + \mathrm{h.c.} \\
		\mathcal{L}_{\mathrm{X}_\mathrm{F8} + \mathrm{M}_\mathrm{C3}} & = y_\mathrm{NP} \, T^a_{ij} \, \overline{\mathrm{M}}_{\mathrm{C3}, i} \, \overline{\mathrm{X}}_\mathrm{F8}^a \, q_{R,j} + \mathrm{h.c.}
	\end{aligned}
\end{equation}
for fermionic coannihilation partners. A model for the mixed annihilation is constructed by combining either two different Lagrangians with a same mediator or two copies of a same Lagrangian with different coupling strengths if X$_1$ and X$_2$ have the same quantum numbers. 

\subsubsection{Our procedure}
\label{sec:mixed:model:condition:ratio}
In what follows, we describe how the effective annihilation cross-section changes when introducing new coannihilation partners for each of the models described above. More specifically, in addition to determining when condition~\eqref{eq:sigma:mix:limit} is verified, we also estimate how large the annihilation rate can become for models outside its range of validity. To this end, for each model with X$_1$, X$_2$, and possibly a mediator, we evaluate the ratio of the effective annihilation cross-section in the complete model over the value that this cross-section would take if either X$_1$ or X$_2$ is removed. Thus, in models where the X$_1$-X$_2$ annihilation process requires a new physics mediator, any possible effect of this mediator on the self-annihilation cross-section of the X$_i$ will appear both in the numerator and the denominator. Our procedure therefore allows to factor out the effects of the mediator on the self-annihilation cross-section in the $t$-channel models studied in section~\ref{sec:mixed:model:tchannel}. 

For fixed particle momenta, the ratio of the effective annihilation rates, denoted by $r$, can be written as follows
\begin{equation} \label{eq:sigma:mix:ratio:nonaveraged}
	r \equiv \frac{1}{(g_\mathrm{DM} + g_{\mathrm{X}_1} + g_{\mathrm{X}_2})^2} \frac{g_{\mathrm{X}_1}^2 \sigma_{\mathrm{X}_1} + g_{\mathrm{X}_2}^2 \sigma_{\mathrm{X}_2} + 2 g_{\mathrm{X}_1} g_{\mathrm{X}_2} \sigma^\mathrm{mix}_{\mathrm{X}_1 \mathrm{X}_2}}{\mathrm{max} \left( \frac{g_{\mathrm{X}_1}^2 \sigma_{\mathrm{X}_1}}{(g_\mathrm{DM} \! + \! g_{\mathrm{X}_1})^2} , \frac{g_{\mathrm{X}_2}^2 \sigma_{\mathrm{X}_2}}{(g_\mathrm{DM} \! + \! g_{\mathrm{X}_2})^2} \right)} .
\end{equation}

For a given choice of X$_1$ and X$_2$, $r$ depends on the masses and couplings of the new particles as well as on the quantum numbers of the dark matter and the mediator. Condition~\eqref{eq:sigma:mix:limit} now corresponds to $r \leq 1$ and will translate into constraints on all these parameters. For models with more than two coannihilation partners, the range of validity of condition~\eqref{eq:sigma:mix:limit} is simply the intersection of the constraints corresponding to all the possible X$_i$-X$_j$ combinations. The actual values of $r$, however, cannot be readily extrapolated from models with two coannihilation partners up to models with an arbitrary number of X$_i$. Estimating these values for two-partner models nevertheless provides a resonable indication of how large annihilation rates can become in regions where~\eqref{eq:sigma:mix:limit} is violated.

Since $r$ heavily depends on the momenta of the particles involved in the annihilation processes, in the rest of this study, we will consider a slightly modified version of this ratio in order to select the particle velocities most relevant around freeze-out times. More precisely, we average all the annihilation cross-sections over a Maxwell-Boltzmann velocity distribution, and define a new ratio $R$ as follows
\begin{equation} \label{eq:sigma:mix:ratio}
	R \equiv \frac{1}{(g_\mathrm{DM} + g_{\mathrm{X}_1} + g_{\mathrm{X}_2})^2} \frac{g_{\mathrm{X}_1}^2 \langle\sigma_{\mathrm{X}_1}v\rangle + g_{\mathrm{X}_2}^2 \langle\sigma_{\mathrm{X}_2}v\rangle + 2 g_{\mathrm{X}_1} g_{\mathrm{X}_2} \langle\sigma^\mathrm{mix}_{\mathrm{X}_1 \mathrm{X}_2}v\rangle}{\mathrm{max} \left( \frac{g_{\mathrm{X}_1}^2 \langle\sigma_{\mathrm{X}_1}v\rangle}{(g_\mathrm{DM} \! + \! g_{\mathrm{X}_1})^2} , \frac{g_{\mathrm{X}_2}^2 \langle\sigma_{\mathrm{X}_2}v\rangle}{(g_\mathrm{DM} \! + \! g_{\mathrm{X}_2})^2} \right)} ,
\end{equation}
where
\begin{equation} \label{eq:maxwell:boltzmann:distribution}
	\begin{aligned}
		\langle \sigma v \rangle & = \int_0^\infty \sigma v \left( \frac{x}{\pi} \right)^{3/2} 8 \pi v^2 e^{-x v^2} \mathrm{d} v ,
	\end{aligned}
\end{equation}
with $v$ being the velocity of an incoming particle in the center of mass frame and $x\equiv m_\mathrm{DM}/T$. Since our main goal is to investigate the relic density constraints on the dark sector, we evaluate $R$ at $x = m_\mathrm{DM}/T = 25$, which is a typical value for the freeze-out temperature. As mentioned in section~\ref{subsubsec:multiple:equal:species}, since $R$ increases with $g_\mathrm{DM}$, we choose the dark matter to be a complex vector in order to obtain the weakest possible relic density bounds. Now, naively, $R$ is expected to only depend on the masses of the X$_i$ and the mediator, as well as on the new physics couplings. For $\Delta_1 = \Delta_2 = 0$, however, at perturbative level, $R$ will depend on the masses of the new particles only through the ratio of the mediator mass $m_\mathrm{S}$ over the mass $m_\mathrm{X}$ of the X$_i$. In the rest of this work, we will therefore study the dependence of $R$ in $m_\mathrm{S}/m_\mathrm{X}$ and in the new physics couplings. In scenarios where we include Sommerfeld corrections, since the value of $\alpha_s$ in the QCD potential depends on the mass of X, we fix $m_\mathrm{X} = 1$~TeV and check that varying $m_\mathrm{X}$ in the range $500 \, \mathrm{GeV} \leq m_\mathrm{X} \leq 10 \, \mathrm{TeV}$ does not quantitatively change our results.

In what follows, we will compute $R$ for each of the models described in sections~\ref{sec:mixed:model:sm}, \ref{sec:mixed:model:schannel}, and~\ref{sec:mixed:model:tchannel} as a function of the new physics couplings as well as the ratio of the mediator mass over $m_\mathrm{X}$. We consider these three scenarios separately and, for each of them, discuss in detail the regions of parameter space where condition~\eqref{eq:sigma:mix:limit} is violated. For most scenarios, we consider only color triplets and octets and ignore non-perturbative effects. In order to assess the validity of these restrictions, we perform a complete study of the scenarios where the X$_1$-X$_2$ annihilation is mediated by a SM quark, including both Sommerfeld corrections and models involving color sextets. We show that the Sommerfeld corrections do not qualitatively modify the constraints on the new physics parameters from condition~\ref{eq:sigma:mix:limit}. Moreover, for any given model involving a color sextet, for couplings of order $\alpha_s$ and above, the values of $R$ are always smaller than the ones obtained when the sextet is replaced by either an octet or a triplet, as long as these other color configurations are allowed. For the models involving a new physics mediator, these color configurations are usually permitted and these models will therefore not qualitatively change our results. We thus consider scenarios with color sextets only for the models with SM mediators shown in the next section.

\subsection{Standard Model mediators}
\label{sec:mixed:mediator:sm}
We first consider the case of mixed annihilation processes mediated by a SM quark. These processes require introducing a X$_1$-X$_2$-$q$ interaction, that will give rise to three new mixed annihilation channels, depicted in figure~\ref{fig:mixed:channels:sm:mediator}. Interestingly, this new vertex also allows for new self-annihilation diagrams for X$_1$ and X$_2$, as shown in figure~\ref{fig:mixed:channels:sm:mediator}. Since these new self-annihilation modes only appear when the mixed process is introduced, their impact on the self-annihilation cross-sections is not taken into account in condition~\eqref{eq:sigma:mix:limit}. For the models studied in this section, we therefore slighly modify equation~\eqref{eq:sigma:mix:limit} to account for the new self-annihilation rates, as follows
\begin{equation} \label{eq:sigma:mix:limit:modified}
	\!\sigma^\mathrm{mix}_{\mathrm{X}_1 \mathrm{X}_2} \! \leq \! \frac{(g_\mathrm{DM} \! + \! g_{\mathrm{X}_1} \! + \! g_{\mathrm{X}_2} \!)^2}{2 g_{\mathrm{X}_1} g_{\mathrm{X}_2}} \! \! \left[ \! \mathrm{max} \! \left( \!\! \frac{g_{\mathrm{X}_1}^2 \sigma_{\mathrm{X}_1}}{(g_\mathrm{DM} \! + \! g_{\mathrm{X}_1} \!)^2} ,\! \frac{g_{\mathrm{X}_2}^2 \sigma_{\mathrm{X}_2}}{(g_\mathrm{DM} \! + \! g_{\mathrm{X}_2}\!)^2} \!\! \right) \! - \! \frac{g_{\mathrm{X}_1}^2 \sigma_{\mathrm{X}_1}^\mathrm{mod} \! + \! g_{\mathrm{X}_2}^2 \sigma_{\mathrm{X}_2}^\mathrm{mod}}{(g_\mathrm{DM} \! + \! g_{\mathrm{X}_1} \! + \! g_{\mathrm{X}_2}\!)^2} \! \right] \! ,
\end{equation}
and the ratio $R$ in equation~\eqref{eq:sigma:mix:ratio} is modified accordingly. Here $\sigma_{\mathrm{X}_i}^\mathrm{mod}$ is the modified self-annihilation rate obtained when including the fourth diagram in figure~\ref{fig:mixed:channels:sm:mediator}. 

\begin{table}[!ht]
	\centering
	\begin{tabular}{| c | c | c | c | c | c |}
		\hline
		X$_1$ & X$_2$ &$g_{\mathrm{X}_1}$ & $g_{\mathrm{X}_2}$ & $\sigma^\mathrm{mix}_{\mathrm{X}_1 \mathrm{X}_2}$ condition & $\sigma^\mathrm{mix}_{\mathrm{X}_1 \mathrm{X}_2}$ Sommerfeld \\
		\hline
		X$_\mathrm{C3}$ & X$_\mathrm{F3}$ & $6$ & $12$ & $\alpha_\mathrm{NP} \lesssim 1.5 \, \alpha_s$ & $\alpha_\mathrm{NP} \lesssim 1.3 \, \alpha_s$ \\
		X$_\mathrm{C3}$ & X$_\mathrm{F6}$ & $6$ & $24$ & $\alpha_\mathrm{NP} \lesssim 3.8 \, \alpha_s$ & $\alpha_\mathrm{NP} \lesssim 4.9 \, \alpha_s$ \\
		X$_\mathrm{C3}$ & X$_\mathrm{F8}$ & $6$ & $32$ & $\alpha_\mathrm{NP} \lesssim 5.3 \, \alpha_s$ & $\alpha_\mathrm{NP} \lesssim 7.0 \, \alpha_s$ \\
		X$_\mathrm{C6}$ & X$_\mathrm{F3}$ & $12$ & $12$ & $\alpha_\mathrm{NP} \lesssim 2.1 \, \alpha_s$ & $\alpha_\mathrm{NP} \lesssim 3.9 \, \alpha_s$ \\
		X$_\mathrm{C6}$ & X$_\mathrm{F8}$ & $12$ & $32$ & $\alpha_\mathrm{NP} \lesssim 3.7 \, \alpha_s$ & $\alpha_\mathrm{NP} \lesssim 3.9 \, \alpha_s$ \\
		X$_\mathrm{C8}$ & X$_\mathrm{F3}$ & $16$ & $12$ & $\alpha_\mathrm{NP} \lesssim 2.6 \, \alpha_s$ & $\alpha_\mathrm{NP} \lesssim 6.3 \, \alpha_s$ \\
		X$_\mathrm{C8}$ & X$_\mathrm{F6}$ & $16$ & $24$ & $\alpha_\mathrm{NP} \lesssim 6.9 \, \alpha_s$ & $\alpha_\mathrm{NP} \lesssim 7.6 \, \alpha_s$ \\
		\hline
	\end{tabular}
	\caption{Values of the new physics coupling $\alpha_\mathrm{NP}$ that saturate the mixed condition~\eqref{eq:sigma:mix:limit:modified} for new physics models where the mixed annihilation is mediated by an interaction with SM quarks. We marginalize over the other relevant parameters $g_\mathrm{DM}$, $m_\mathrm{X}$ and $v$ as described in the main text.}
	\label{tab:mixed:conditions:sm}
\end{table}

Since the models studied here do not include a new physics mediator, the ratio of the rates, $R$, only depends on the X$_1$-X$_2$-$q$ coupling, $\alpha_\mathrm{NP} \equiv \frac{y_\mathrm{NP}^2}{4 \pi}$. For each of the models described by Lagrangian~\eqref{eq:lagrangians:mixed:sm} we compute $R$ as a function of $\alpha_{\mathrm{NP}}$ and find the maximal value of $\alpha_\mathrm{NP}$ for which condition~\eqref{eq:sigma:mix:limit:modified} is verified, or, equivalently, $R \leq 1$. The results for the different models are summarized in table~\ref{tab:mixed:conditions:sm}, both with and without Sommerfeld corrections. Details about our calculation of the Sommerfeld effect can be found in a previous work~\cite{ElHedri:2016onc} and accompanying Mathematica package~\cite{ElHedri:2016pac}, as well as in appendix~\ref{sec:sommerfeld:corrections}.

\begin{figure}[!t]
	\centering
	\includegraphics[width=0.49\textwidth]{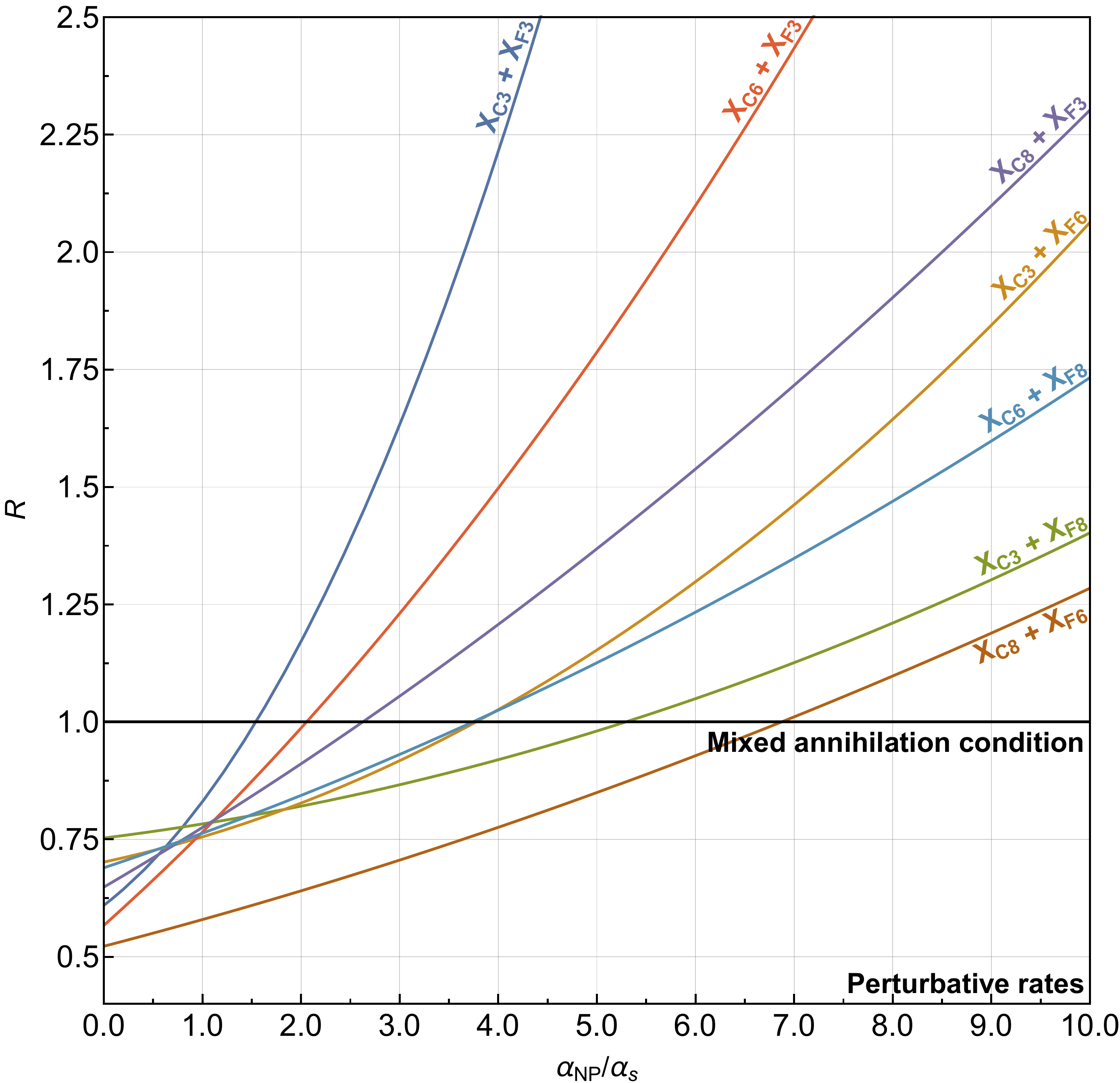}
	\includegraphics[width=0.49\textwidth]{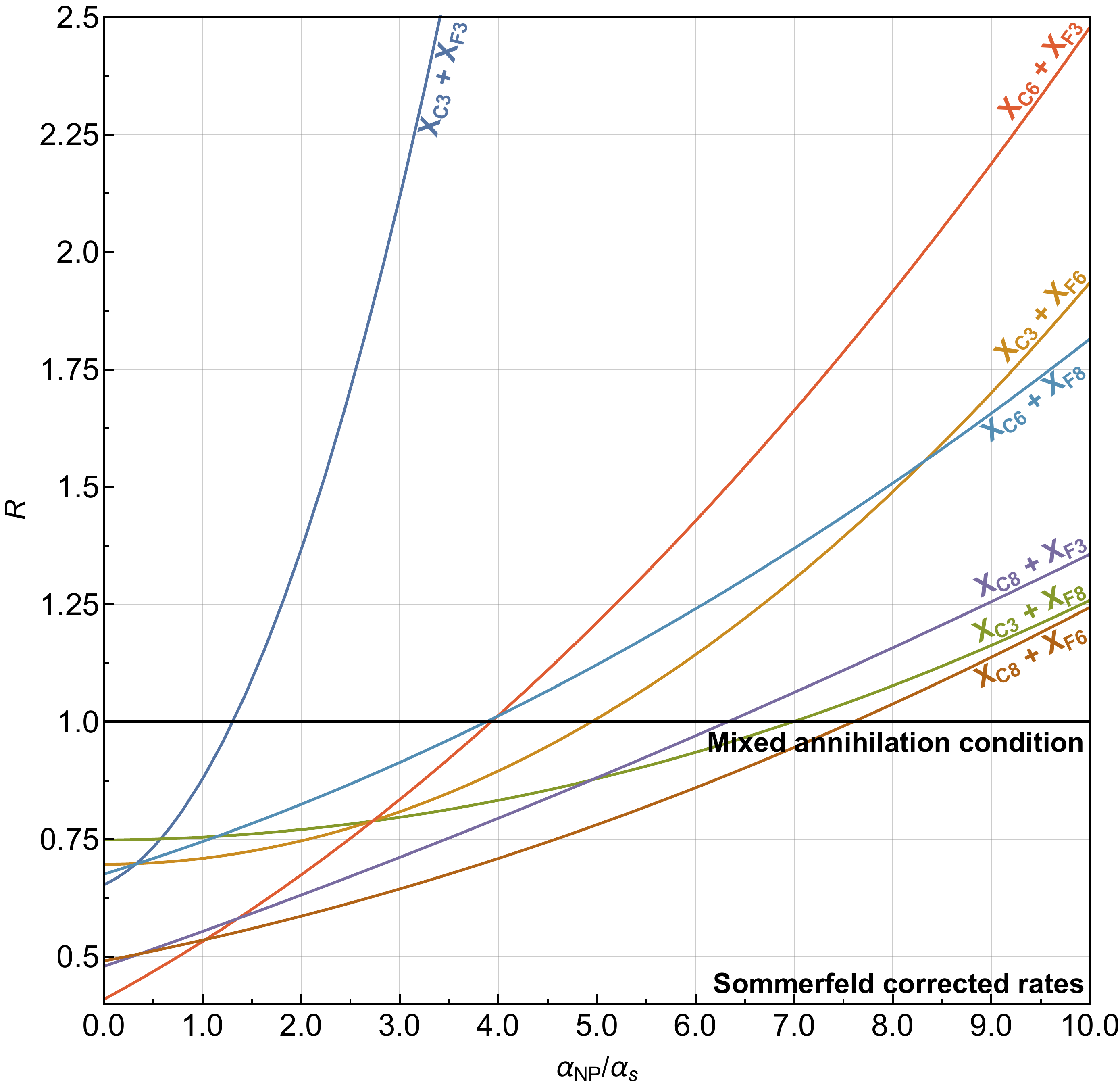}
	\caption{Values of the ratio $R$ of the effective annihilation cross-section for two coannihilation partners over the maximum of the effective rate with only either particle present, as defined in equation~\ref{eq:sigma:mix:limit}, as a function of $\alpha_\mathrm{NP}/\alpha_s$. We marginalize over the other relevant parameters $g_\mathrm{DM}$, $m_\mathrm{X}$ and $v$ as described in the main text.}
	\label{fig:mixed:conditions:sm}
\end{figure}

For all the models listed in table~\ref{tab:mixed:conditions:sm}, in order to significantly increase the dark matter depletion rate compared to single-partner models, we need either new physics couplings $\alpha_\mathrm{NP}$ much larger than the strong coupling or couplings of X$_1$ and X$_2$ to multiple quark flavors at a time. Since the second option is typically strongly constrained by flavor measurements, we conclude that for models where the mixed X$_1$-X$_2$ annihilation occurs via a SM quark, increasing the number of coannihilation partners typically leads to stronger relic density bounds on the DM mass. In order to estimate the effect of $\alpha_\mathrm{NP}$ on the DM annihilation for theories with very large couplings, we also show $R$ as a function of $\alpha_\mathrm{NP}$ in figure~\ref{fig:mixed:conditions:sm}. We can see that, for large values of $\alpha_\mathrm{NP}$, models involving color triplets are associated with the largest values of $R$, followed by models with sextets and octets. These results confirm our previous hypothesis about the dependence of $R$ on the colors of the annihilating particles. Scenarios where mixed annihilations occur via new physics mediators also exhibit this feature, which is due to the fact that the X particles in the lowest $SU(3)$ representations are also associated with the lowest annihilation rates and hence the tightest relic density bounds. Even for color triplets, however, adding new coannihilation partners loosens these bounds only for large couplings $\alpha_\mathrm{NP}\gtrsim \alpha_s$. For such couplings, the pair-production of X at colliders can be significantly enhanced, leading to tighter collider bounds. We discuss this last point in section~\ref{sec:collider:pheno}.

\subsection{\texorpdfstring{$s$}{s}-channel mediators}
\label{sec:mixed:mediator:schannel}
Here, we evaluate the contribution of the mixed annihilation rate to the effective DM annihilation cross-section for models where the X$_1$\,X$_2$\,$\to$\,SM\,SM interaction occurs through a new physics $s$-channel mediator M$_s$. We focus on a set of representative models, whose Lagrangians are shown in~\eqref{eq:lagrangians:mixed:schannel}. As outlined in section~\ref{sec:mixed:model:schannel}, in order to factor out the effect of introducing the $s$-channel mediator M$_s$, we compare the effective annihilation rates for a model with X$_1$, X$_2$, and M$_s$ to a model with either X$_1$ or X$_2$ and a mediator M$_s^\prime$ that couples to $q \, \bar{q}$ and X$_i \, \overline{\mathrm{X}}_i$. The mass of this mediator as well as the values of its couplings to the SM and the dark sector are taken to be the same as for M$_s$. Besides the masses and degrees of freedom of the dark matter and the X partners, the DM depletion rate for these models is thus characterized by four parameters: $m_\mathrm{M}$, $m_\mathrm{X}$, $\alpha_\mathrm{SM}$, and $\alpha_\mathrm{X}$. Since, as shown in section~\ref{sec:mixed:mediator:sm}, the Sommerfeld corrections should not significantly influence the final results, we consider only the perturbative (co)annihilation cross-sections, that depend on $\alpha_\mathrm{SM}$, $\alpha_\mathrm{X}$, and $\rho= \frac{m_\mathrm{M}}{m_\mathrm{X}}$. In order to facilitate the interpretation of our results, we consider the ratio $\tilde{R}(r, \alpha_\mathrm{max})$, defined by 
\begin{equation}
	\tilde{R}(\rho, \alpha_\mathrm{max}) \equiv \mathrm{max}_{\{\alpha_\mathrm{SM}, \alpha_\mathrm{X} \leq \alpha_\mathrm{max}\}}R(\rho, \alpha_\mathrm{SM}, \alpha_\mathrm{X}) ,
	\label{eq:rtilde}
\end{equation}
with $R$ given in equation~\eqref{eq:sigma:mix:ratio}.

In order to understand how the mixed X$_1$\,X$_2$ $s$-channel interactions modify the DM annihilation rate, we rewrite the annihilation cross-sections corresponding to the X$_i$\,X$_j$\,$\to$\,M$_s^{(\prime)}\,\to\,q\,\bar{q}$ processes as
\begin{equation}
	\sigma_{ij} = \frac{C_{ij}\alpha_\mathrm{X}\alpha_\mathrm{SM}S_{ij}\hat{\sigma}}{(s - m_{\mathrm{M}}^2)^2 + m_\mathrm{M}^2\Gamma_{ij}^2} ,
\end{equation}
where $C_{ij}$ is the color factor corresponding to the interaction, $\Gamma_{ij}$ is the total decay width of the mediator, $S_{ij}$ is a symmetry factor that is equal to $\tfrac{1}{2}$ if exactly one of the initial states is self-conjugate and to one otherwise, and $\hat{\sigma}$ is the part of the cross-section that is the same for the one- and two-partner models. The decay width $\Gamma_{ij}$ depends on the new physics couplings in the following way
\begin{equation}
	\Gamma_{ij} = \alpha_\mathrm{SM}\Gamma_{ij, \mathrm{SM}} + \alpha_\mathrm{X}\Gamma_{ij, \mathrm{X}} ,
\end{equation}
where the $\Gamma_{ij, \mathrm{SM}}$ and the $\Gamma_{ij, \mathrm{X}}$ depend only on the spins and the color charges of the mediator and the X$_i$, as well as on the masses of the dark sector particles. In what follows, since we are considering only perturbative theories, we assume that this width is narrow, enforcing the loose requirement that $\Gamma_{ii}, \Gamma_{ij} < \tfrac{1}{4} m_\mathrm{M}$, with $\Gamma_{ii}$ and $\Gamma_{ij}$ the decay widths of the mediator in the one-partner model leading to the largest annihilation rate and in the two-partner model respectively. Here, contrary to the SM mediator case studied in section~\ref{sec:mixed:mediator:sm}, the new physics couplings appear both in the numerator and the denominator of $\sigma_{ij}$. Consequently, just increasing these couplings does not necessarily lead to an increase in the cross-section for the mixed annihilation processes. This cross-section is in fact large only in the following regimes:
\begin{itemize}
	\item The resonance region $m_\mathrm{M}\sim 2 m_\mathrm{X}$: for non-relativistic dark sector particles, the annihilation cross-section reduces to
		\begin{equation*}
			\sigma_{ij} \sim \frac{C_{ij}\alpha_\mathrm{X}\alpha_\mathrm{SM}S_{ij}\hat{\sigma}}{m_\mathrm{M}^2\Gamma_{ij}^2} .
		\end{equation*}
		For a reasonably narrow width $\Gamma_{ij} < \tfrac{1}{4} m_\mathrm{M}$, this gives
		\begin{equation*}
				\sigma_{ij} \gtrsim 4 \frac{C_{ij}\alpha_\mathrm{X}\alpha_\mathrm{SM}S_{ij}\hat{\sigma}}{m_\mathrm{M}^2 m_\mathrm{X}^2} .
		\end{equation*}
	\item The large coupling limit $\alpha_\mathrm{X}\gg 1$ with $m_\mathrm{M} < 2 m_\mathrm{X}$: in this scenario, $\Gamma_\mathrm{X} = 0$ so the denominator of the cross-section does not depend on $\alpha_\mathrm{X}$. $\sigma_{ij}$ can therefore be rewritten as 
		\begin{equation*}
			\sigma_{ij} = \frac{C_{ij} \alpha_\mathrm{X}\alpha_\mathrm{SM}S_{ij}\hat{\sigma}}{(s - m_{\mathrm{M}}^2)^2 + \alpha_\mathrm{SM}^2 m_\mathrm{M}^2\Gamma_{\mathrm{SM}}^2} ,
		\end{equation*}
		and grows linearly with $\alpha_\mathrm{X}$.
	\item The large coupling limit $\alpha_\mathrm{X} \gg 1$ ($\alpha_\mathrm{SM}\gg 1$) when $\Gamma_\mathrm{X}$ ($\Gamma_\mathrm{SM}$) is small: in this case, it is possible to show that, away from the resonance and in the narrow width approximation, the cross-section is maximized when the total width is equal to its maximal allowed value $\Gamma = \tfrac{1}{4} m_\mathrm{M}$. Optimizing over $\alpha_\mathrm{SM}$, this value can be obtained for
		\begin{equation} \label{eq:alpha}
			\alpha_\mathrm{SM,max} = \frac{1}{\Gamma_\mathrm{SM}} \left(\frac{m_\mathrm{M}}{4} - \alpha_\mathrm{X} \Gamma_\mathrm{X}\right) ,
		\end{equation}
		and the cross-section is then
		\begin{equation*}
			\sigma_{ij} = \frac{C_{ij} \alpha_\mathrm{X}S_{ij} \hat{\sigma} (m_\mathrm{M}/4 - \alpha_\mathrm{X}\Gamma_\mathrm{X})}{\Gamma_{\mathrm{SM}}\left[(s - m_{\mathrm{M}}^2)^2 + \frac{m_\mathrm{M}^4}{16}\right]}. 
		\end{equation*}
		An equivalent formula can be obtained when optimizing over $\alpha_\mathrm{X}$, by replacing $\alpha_\mathrm{SM}\leftrightarrow\alpha_\mathrm{X}$ and $\Gamma_\mathrm{SM}\leftrightarrow\Gamma_\mathrm{X}$. The maximal coupling given in equation~\eqref{eq:alpha} can reach large values if $\Gamma_\mathrm{SM}$ is small, which can in turn lead to large values for the cross-section. In this case, the latter would be only bounded by perturbativity.
\end{itemize}
In all these scenarios, for $\alpha_\mathrm{X} \gtrsim \alpha_s$, the cross-sections for the new physics processes will dominate over the ones for the QCD processes. The ratio of the rates $\tilde{R}$ can then be approximated as
\begin{equation} \label{eq:ratiosimp}
	\tilde{R} \sim \mathrm{min}_{i=1,2}\left\{2\frac{g_{\mathrm{X}_1}g_{\mathrm{X}_2}}{g_{\mathrm{X}_i}^2}\left(\frac{g_\mathrm{DM} + g_{\mathrm{X}_i}}{g_\mathrm{DM} + g_{\mathrm{X}_1} + g_{\mathrm{X}_2}}\right)^2\,S_{12}\,\frac{C_{12}}{C_{ii}}\frac{(s - m_\mathrm{M}^2)^2 + \Gamma_{ii}^2}{(s - m_\mathrm{M}^2)^2 + \Gamma_{12}^2}\right\}.
\end{equation}
This ratio reaches a maximum at the resonance, where it can be approximated by
\begin{equation} \label{eq:res:ratio}
	\tilde{R} \sim \mathrm{min}_{i=1,2} \left\{2\frac{g_{\mathrm{X}_1}g_{\mathrm{X}_2}}{g_{\mathrm{X}_i}^2}\left(\frac{g_\mathrm{DM} + g_{\mathrm{X}_i}}{g_\mathrm{DM} + g_{\mathrm{X}_1} + g_{\mathrm{X}_2}}\right)^2\frac{C_{12}}{C_{ii}}\left(\frac{\Gamma_{ii}}{\Gamma_{12}}\right)^2\right\}.
\end{equation}

Our estimates of $\tilde{R}$ in the regions where the new physics annihilation processes dominate thus show that $\tilde{R}$ is always less than one for the models where X$_1$ and X$_2$ have the same quantum numbers, as for these models, $\Gamma_{ii} = \Gamma_{12}$ and $C_{ii} = C_{12}$ and $\tilde{R}$ can therefore be approximated as
\begin{equation}
	\tilde{R} = 2 \left(\frac{g_\mathrm{DM} + g_\mathrm{X}}{g_\mathrm{DM} + 2g_\mathrm{X}}\right)^2 \leq \frac{8}{9}.
\end{equation}

\begin{figure}[!t]
	\centering
	\includegraphics[width=0.6\textwidth]{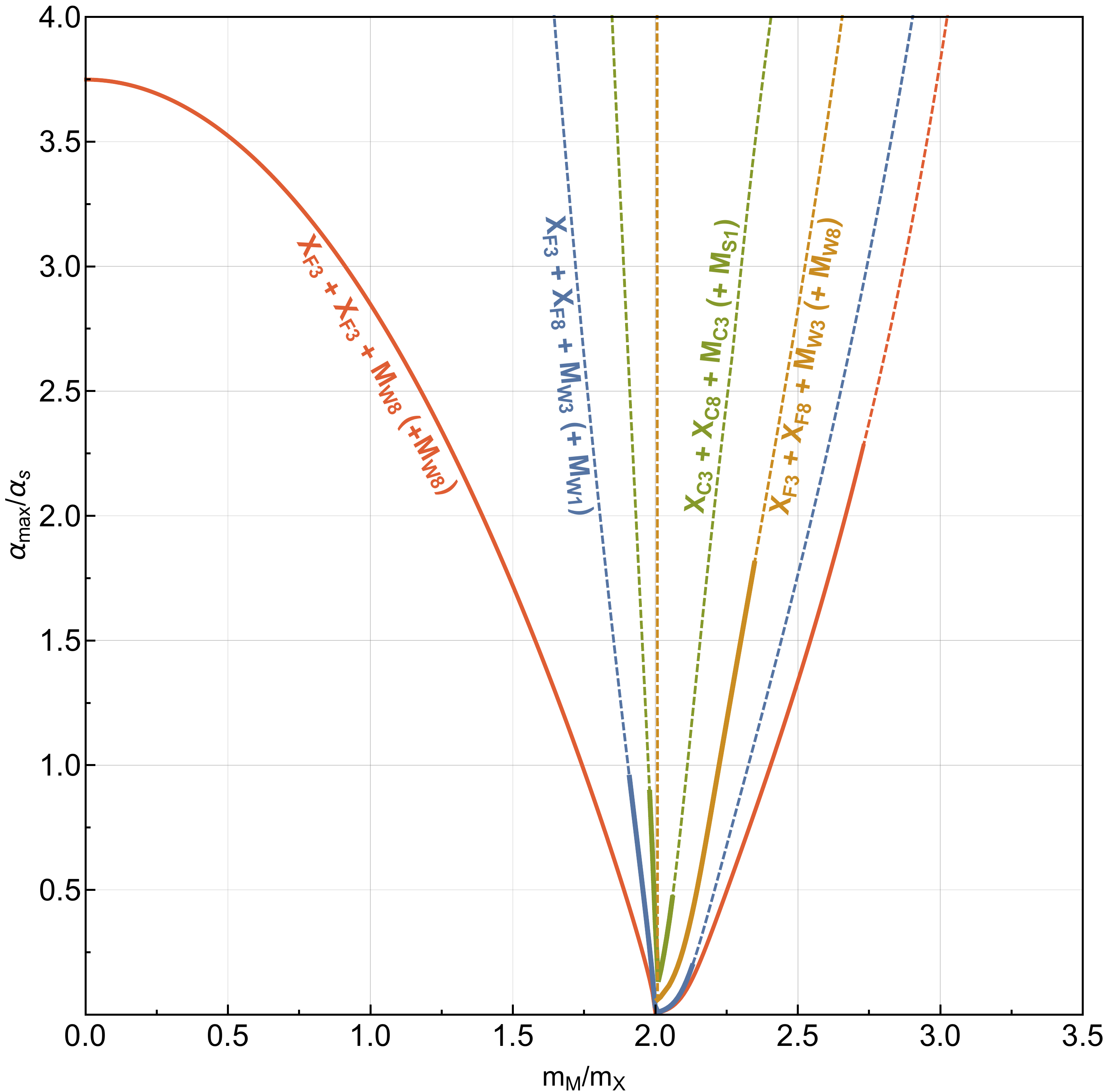}
	\caption{$\tilde{R} = 1$ contours in the $(\alpha_\mathrm{max}, \frac{m_\mathrm{M}}{m_\mathrm{X}})$ plane, with $\tilde{R}$ defined in equation~\eqref{eq:rtilde}. The thick solid lines show the regions for which the width of the mediators are smaller than $25$\% of their mass. The particle content of each model is shown as a label next to the corresponding line and is of the form X$_1$ + X$_2$ + M$_\textrm{2-partner}$ (+ M$_\textrm{1-partner}$) where M$_\textrm{2-partner}$ and M$_\textrm{1-partner}$ are the mediators of the two-partner and one-partner models respectively. For the X$_\mathrm{F3}$ + X$_\mathrm{F3}$ + M$_\mathrm{W8}$ (+ M$_\mathrm{W8}$) model, the two-partner mediator mediates both the mixed and self annihilation of the X$_i$.}
	\label{fig:mixed:conditions:schannel}
\end{figure}

For models where X$_1$ and X$_2$ have different quantum numbers, $\tilde{R}$ can be larger than one and is expected to be maximal in the resonant $m_\mathrm{M} \approx 2 \, m_\mathrm{X}$ region. We verify this hypothesis for the X$_\mathrm{F3}$ + X$_\mathrm{F8}$ + M$_\mathrm{W3}$ and the X$_\mathrm{C3}$ + X$_\mathrm{C8}$ + M$_\mathrm{C3}$ models, for which the $\tilde{R} = 1$ contours in the $\alpha_\mathrm{max}$ versus $m_\mathrm{M}$ plane are shown in figure~\ref{fig:mixed:conditions:schannel}. Since we are exploring regions of the parameter space where the new physics couplings are large, the decay width of the mediators in these models can be comparable to their mass, which would indicate an underlying non-perturbative dynamics. In figure~\ref{fig:mixed:conditions:schannel} we therefore also indicate the regions for which the decay width of all the mediators in our models are less than $25$\% of their mass. Applying this narrow width requirement constrains the $\tilde{R} > 1$ regions for all our models to be within about $30$\% of the resonance. Even when this criterium is relaxed, the $\tilde{R} > 1$ domain remains relatively narrow, with the mediator mass remaining within $50$\% of the resonant region for $\alpha_\mathrm{max} < 2 \, \alpha_s$.

\begin{figure}[!t]
	\centering
	\includegraphics[width=0.6\textwidth]{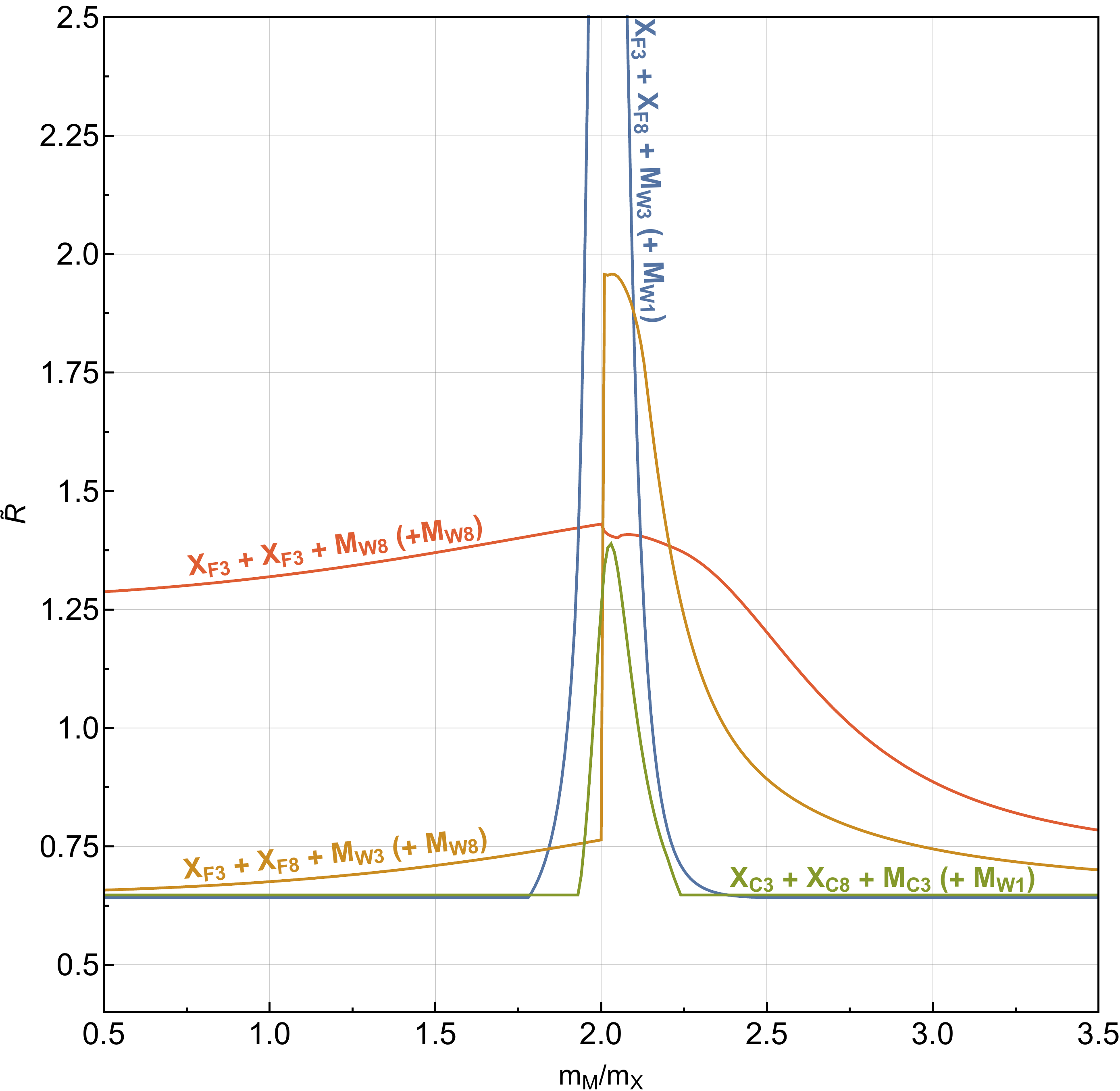}
	\caption{Maximal value of $\tilde{R}$ for which all the mediators verify $\Gamma_\mathrm{M} < \tfrac{1}{4} m_\mathrm{M}$ as a function of $\frac{m_\mathrm{M}}{m_\mathrm{X}}$. The particle content of each model is written as X$_1$ + X$_2$ + M$_\textrm{2-partner}$ (+ M$_\textrm{1-partner}$) where M$_\textrm{2-partner}$ and M$_\textrm{1-partner}$ are the mediators of the two-partner and one-partner models respectively. For the X$_\mathrm{F3}$ + X$_\mathrm{F3}$ + M$_\mathrm{W8}$ (+ M$_\mathrm{W8}$) model, the two-partner mediator mediates both the mixed and self annihilation of the X$_i$.}
	\label{fig:mixed:max:ratio:schannel}
\end{figure}

In order to estimate how much our two-partner models can enhance the dark matter annihilation rate, we now fix $\alpha_\mathrm{max}$ so that it saturates the narrow width requirement $\Gamma_\mathrm{M} < \tfrac{1}{4} m_\mathrm{M}$, and plot $\tilde{R}$ as a function of the mediator mass in figure~\ref{fig:mixed:max:ratio:schannel}. Here again, $\tilde{R}$ sharply peaks near the resonance and its maximal values are close to ones predicted from equation~\eqref{eq:res:ratio}. For most of our models, however, $\tilde{R}$ remains less than two, which corresponds to an order one increase in the relic density bound on the dark matter mass. Significantly loosening this bound without adding new mediators would thus require introducing a large number of coannihilation partners. If these coannihilation partners also annihilate via an $s$-channel interaction with a new physics mediator they would all need to verify $m_\mathrm{X} \approx \tfrac{1}{2} m_\mathrm{M}$, making the model particularly contrived.

Finally, we consider the possibility that, for models where X$_1$ and X$_2$ have the same quantum numbers, M$_s$ can couple not only to X$_1$-X$_2$ but also to X$_1$-X$_1$ and X$_2$-X$_2$. In this case, the total DM annihilation rate for the two-partner models can be further enhanced with respect to the one-partner model, even far from the resonant region.  More specifically, $\tilde{R}$ can be written as
\begin{equation}
	\tilde{R} = 2\left(\frac{g_\mathrm{DM} + g_\mathrm{X}}{g_\mathrm{DM} + 2g_\mathrm{X}}\right)^2 \,\left(1 + \frac{\sigma_{\mathrm{X}_1\mathrm{X}_2}}{\sigma_{\mathrm{X}_i\mathrm{X}_i}}\right) ,
\end{equation}
where $i$ can be either $1$ or $2$. For $s$-channel models, the QCD and the new physics processes interfere positively for the X$_i$\,X$_i\,\to\,q \, \bar{q}$ annihilation so $\sigma_{\mathrm{X}_i\mathrm{X}_i} \geq\sigma_{\mathrm{X}_1\mathrm{X}_2}$. Hence, we always have
\begin{equation} \label{eq:rtilde:max}
	\tilde{R} \leq 4\left(\frac{g_\mathrm{DM} + g_\mathrm{X}}{g_\mathrm{DM} + 2g_\mathrm{X}}\right)^2 \leq \frac{16}{9} ,
\end{equation}
which always lead to values smaller than two for colored coannihilation partners. We confirm these results by showing the $\tilde{R} = 1$ contours for this model in figure~\ref{fig:mixed:conditions:schannel} as well as the maximal values for $\tilde{R}$ as a function of the mediator mass in figure~\ref{fig:mixed:max:ratio:schannel}. As shown in figure~\ref{fig:mixed:conditions:schannel}, $\tilde{R}$ can now be larger than one further away from the resonance than in the X$_\mathrm{F3}$ + X$_\mathrm{F8}$ + M$_\mathrm{W3}$ and the X$_\mathrm{C3}$ + X$_\mathrm{C8}$ + M$_\mathrm{C3}$ models. Figure~\ref{fig:mixed:max:ratio:schannel} shows however that the increase in the dark matter annihilation rate for this model is less than about $40$\%, which is compatible with the maximal value predicted in equation~\ref{eq:rtilde:max}.  

In summary, although introducing new coannihilation partners and $s$-channel mixed annihilation processes can in principle enhance the dark matter annihilation rate, the concrete possibilities to do so are very limited. For two-partner models, increases of the rate by more than $100$\% can only be obtained by introducing mediators that have a particularly narrow width, can couple only to different species of coannihilation partners, and annihilate resonantly. Aside from this particularly contrived scenario, it is also possible to simply increase the number of mediator interactions with the dark sector but, as shown in equation~\ref{eq:rtilde:max}, the resulting enhancement of the dark matter depletion rate will be extremely limited.

\subsection{\texorpdfstring{$t$}{t}-channel mediators}
\label{sec:mixed:mediator:tchannel}
As described in section~\ref{sec:mixed:models}, introducing a new $t$-channel interaction between X$_1$, X$_2$, and the SM requires the existence of a mediator M$_t$ that is also part of the dark sector, and of two new vertices M$_t$\,X$_1$\,SM$_1$ and M$_t$\,X$_2$\,SM$_2$. The existence of these new vertices will not only enable the mixed X$_1$\,X$_2$\,$\to$\,SM$_1$\,SM$_2$ interaction but also open new self-annihilation channels for both X$_1$ and X$_2$ as shown in figure~\ref{fig:mixed:channels:np:mediator}. In order to isolate the effects of the introduction of a new coannihilation partner from the effects related to the mediator itself, we compare the annihilation rates in models with X$_1$, M$_t$, and X$_2$ to the rates in models with either X$_1$ or X$_2$, and the mediator M$_t$.

\begin{figure}[!t]
	\centering
	\includegraphics[width=0.49\textwidth]{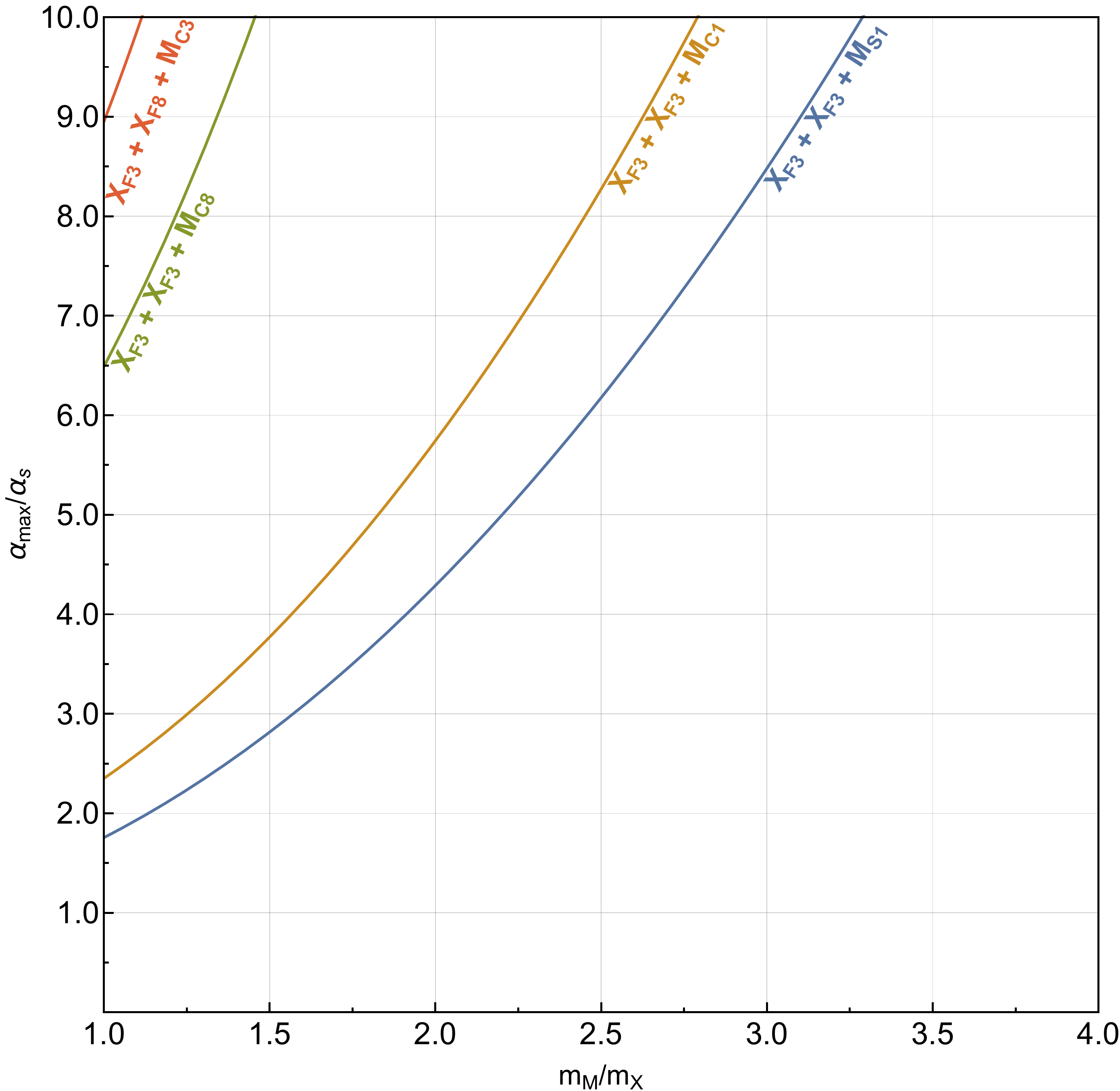}
	\includegraphics[width=0.49\textwidth]{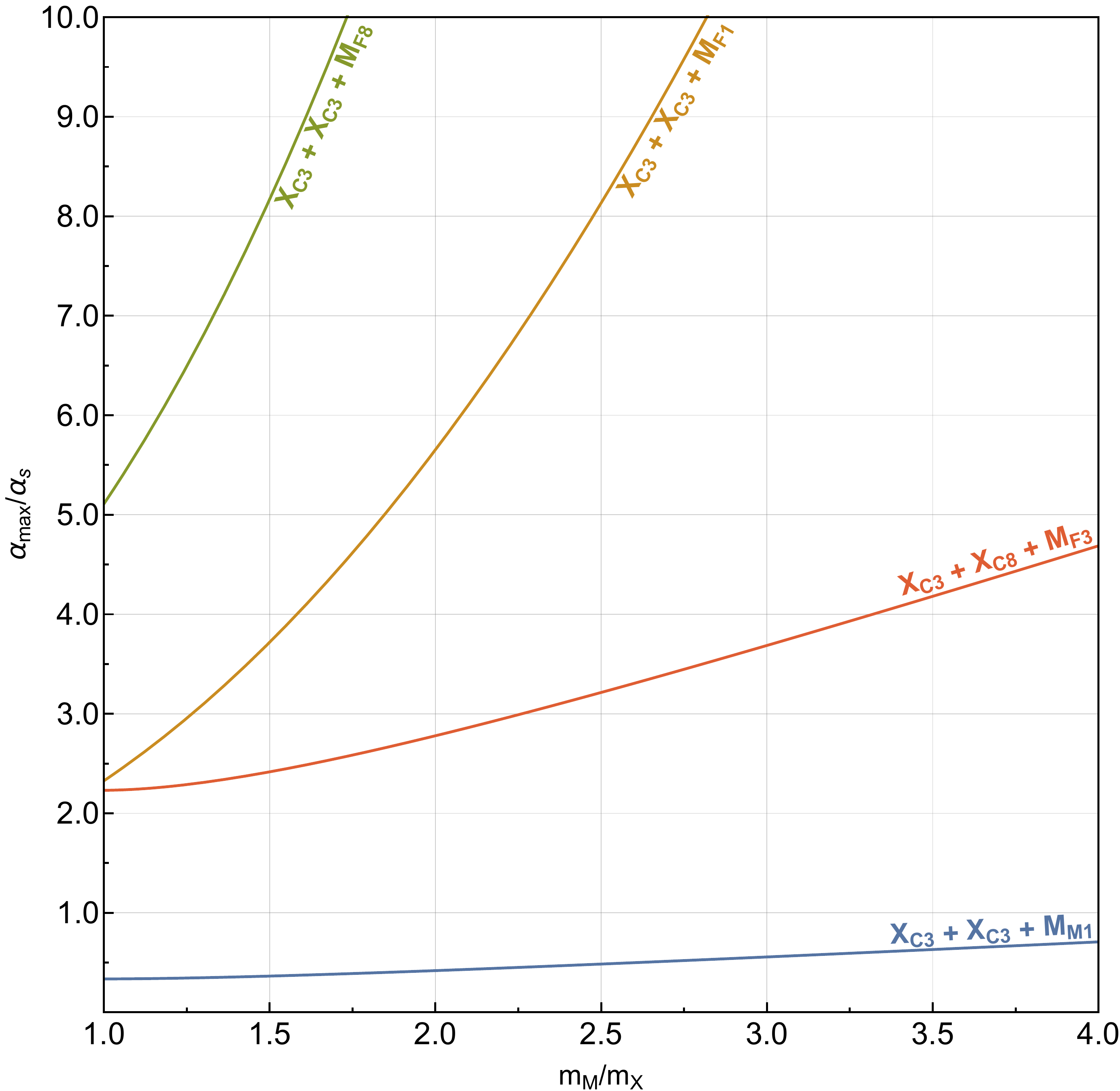}
	\caption{$\tilde{R} = 1$ contours for models where the mixed annihilation occurs via a  $t$-channel mediator. The regions below the contours in the $\alpha_\mathrm{max}$ versus $m_\mathrm{M} / m_\mathrm{X}$ plane satisfy the mixed condition~\eqref{eq:sigma:mix:limit}.}
	\label{fig:mixed:conditions:tchannel}
\end{figure}

Since the DM and the X particles all have equal masses in the $\Delta \approx 0$ region that we are studying here, the $t$-channel models are described by four parameters: $m_\mathrm{M}$, $m_\mathrm{X}$, $\alpha_{1}$, and $\alpha_{2}$. For perturbative annihilation cross-sections, the mass dependence is entirely contained in the ratio $\rho = \frac{m_\mathrm{M}}{m_\mathrm{X}}$. Similarly to the $s$-channel procedure described in section~\ref{sec:mixed:mediator:schannel}, we compute $\tilde{R}(\rho, \alpha_\mathrm{max})$ defined as
\begin{equation}
	\tilde{R}(\rho, \alpha_\mathrm{max}) \equiv \mathrm{max}_{\alpha_1, \alpha_2 \leq \alpha_\mathrm{max}} R(\rho, \alpha_1, \alpha_2) ,
\end{equation}
with $R$ given in equation~\eqref{eq:sigma:mix:ratio}. We show the regions of the $(\rho, \alpha_\mathrm{max})$ space where the mixed condition~\eqref{eq:sigma:mix:limit} is saturated ($\tilde{R} = 1$) in figure~\ref{fig:mixed:conditions:tchannel} for a subset of the models constructed from equations~\eqref{eq:lagrangians:mixed:tchannel:scalarx} and~\eqref{eq:lagrangians:mixed:tchannel:fermionx}. These models have been chosen so that they illustrate all the possible types of behavior encountered in $t$-channels scenarios. As shown in figure~\ref{fig:mixed:conditions:tchannel}, while a few models require large couplings $\alpha_\mathrm{max}\gtrsim 5 \, \alpha_s$ to reach $\tilde{R} > 1$, in other models, an enhancement of the dark matter annihilation rate is possible even for moderate couplings. For all the scenarios studied here, $\tilde{R}$ is maximal for $m_\mathrm{M} = m_\mathrm{X}$ and the mediator cannot be lighter than m$_\mathrm{X}$ in order for the dark matter to be stable. Interestingly, for $m_\mathrm{M}\sim m_\mathrm{X}$ the mediator also becomes a coannihilation partner. For the simplified models considered here, however, this particular scenario requires a sizable fine-tuning\footnote{This fine-tuning can be avoided if the X$_i$ and the mediator all belong to a same gauge multiplet of a spontaneously broken symmetry, such as $SU(2)$ as shown in~\cite{Cirelli:2009uv}. For groups like $SU(2)$ where the symmetry breaking scale is $\mathcal{O}(100)$~GeV, however, it is possible to reasonably approximate the relic density bounds on the dark matter mass by remaining in the unbroken phase. In this case, the constraints on the dark matter mass can be derived by following an approach similar to the one from this paper and from~\cite{ElHedri:2017nny}.} of the masses of the dark sector particles and we will thus not consider the effects of the mediator coannihilation in the rest of this section.
 
\begin{figure}[!t]
	\centering
	\includegraphics[width=0.49\textwidth]{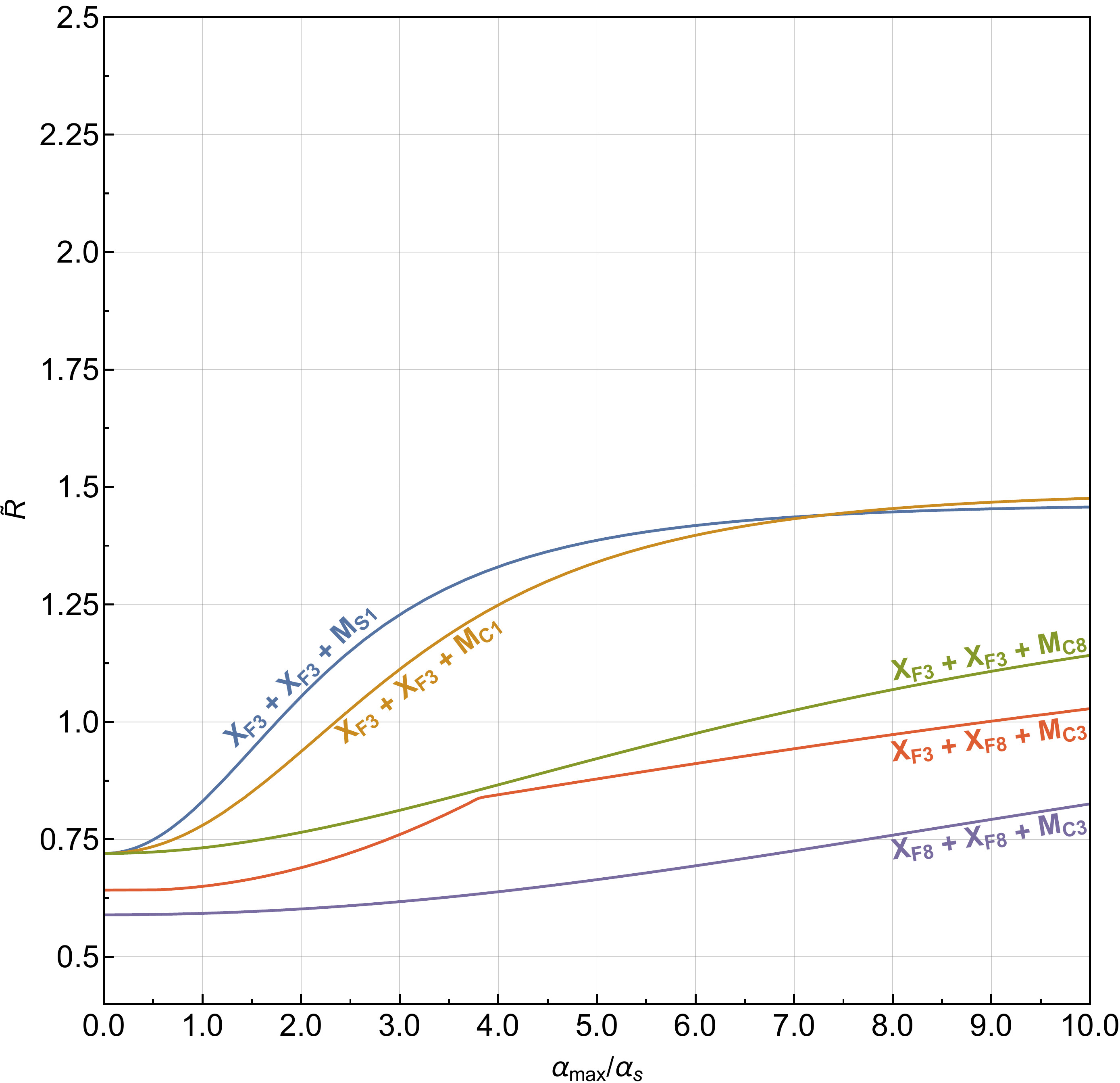}
	\includegraphics[width=0.49\textwidth]{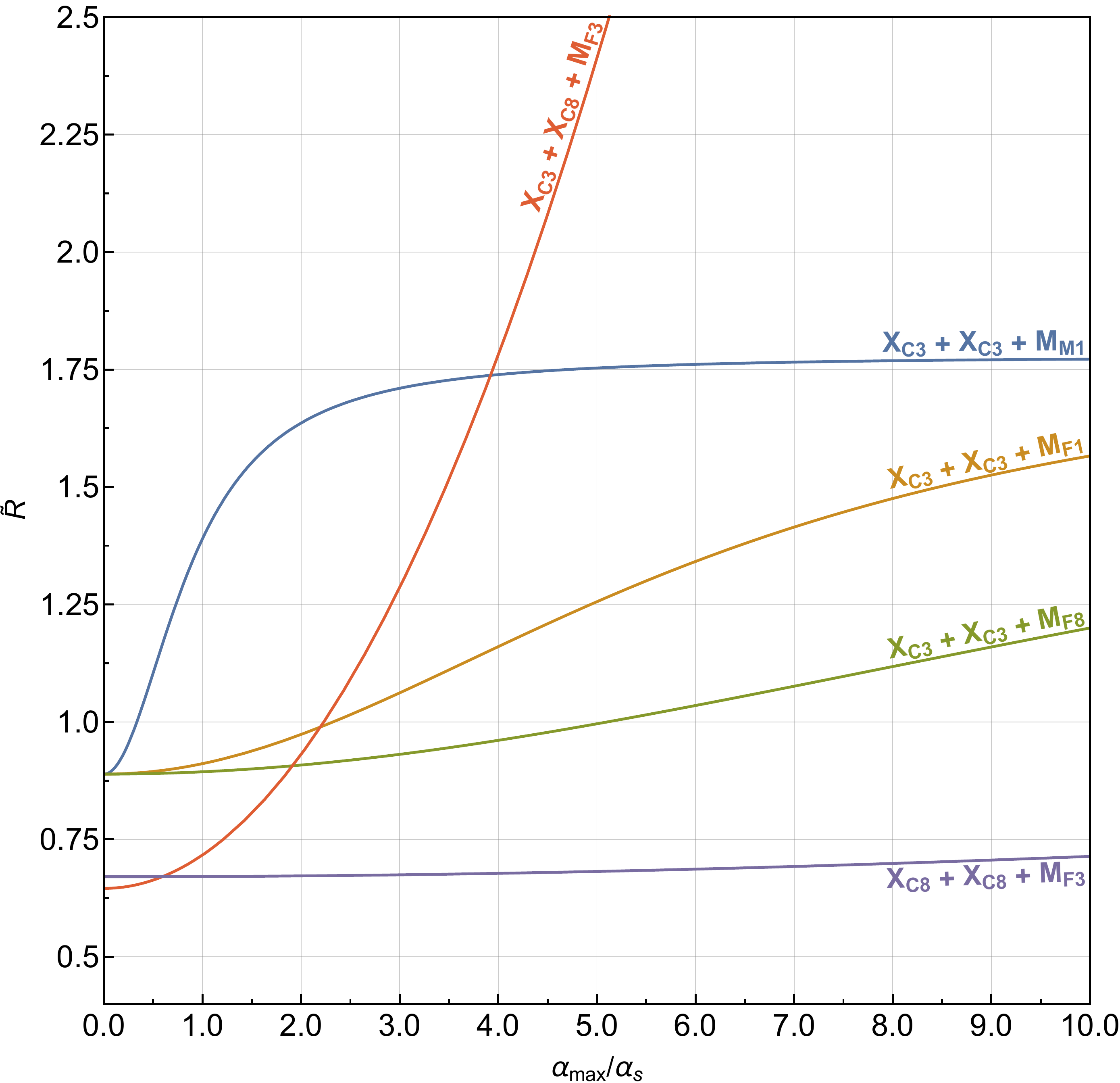}
	\caption{Ratio of the mixed annihilation rate over the self-annihilation rate $\tilde{R}$ for $t$-channel models as a function of $\alpha_\mathrm{max}$ for $m_\mathrm{M} = m_\mathrm{X}$. The scalar and fermion $\mathrm{X}_\mathrm{8} + \mathrm{X}_\mathrm{8} + \mathrm{M}_\mathrm{3}$ models (purple lines) do not appear in figure~\ref{fig:mixed:conditions:tchannel} since $\tilde{R}$ is always less than one in these scenarios.}
	\label{fig:mixed:max:ratio:tchannel}
\end{figure}

We now investigate how the ratio $\tilde{R}$ changes as $\alpha_\mathrm{max}$ increases to arbitrarily high values. In figure~\ref{fig:mixed:max:ratio:tchannel} we show the ratio $\tilde{R}$ as a function of $\alpha_\mathrm{max}$ for $m_\mathrm{M} = m_\mathrm{X}$, the value of the mediator mass for which $\tilde{R}$ is maximal at fixed couplings. We first observe that all $t$-channel coannihilation models asymptote to a maximum $\tilde{R}$ when $\alpha_\mathrm{max}$ becomes large. This is rooted in the fact that both the self-annihilation rate induced by the new mediator in the one-partner model and the mixed annihilation rate in the two-partner model scale as $\alpha_\mathrm{max}^4$ so the coupling dependence vanishes at large $\alpha_\mathrm{max}$. The speed at which this asymptotic value is reached is highly model-dependent. In particular, when X$_1$ and X$_2$ have different quantum numbers, as for the X$_\mathrm{F3}$-X$_\mathrm{F8}$-M$_\mathrm{C3}$ model, a sharp change in the slope of $\tilde{R}$ can sometimes be observed for a given $m_\mathrm{M}$. This feature indicates a change in the one-partner model giving the largest X self-annihilation rate, and hence in the denominator of $\tilde{R}$ as shown in equation~\eqref{eq:sigma:mix:ratio}. Another important observation is that, for the $\mathrm{X}_\mathrm{C3} + \mathrm{X}_\mathrm{C8} + \mathrm{M}_\mathrm{F3}$ model, $\tilde{R}$ can increase to particularly large values. This behavior is due to the fact that the self-annihilation rates in the one-partner models are $p$-wave suppressed while the mixed annihilation rate has a $s$-wave component. Hence, $\tilde{R}$ asymptotes only at values of order $50$ for this model. Even for this scenario, however, $\tilde{R}$ remains smaller than two for $\alpha_\mathrm{max} < 4 \, \alpha_s$. Moreover, even for large couplings, although the dark matter annihilation rates for the two-partner model will be much larger than for the $p$-wave suppressed one-partner models, they will remain comparable to the ones obtained in other one-partner models where $s$-wave self-annihilation is allowed. Hence, we do not expect this particular scenario to challenge the $10$~TeV bound found in~\cite{ElHedri:2017nny}.

We conclude from this analysis that, although models with $t$-channel mixed annihilation can lead to increased dark matter annihilation rates compared to single-partner coannihilating models, this increase is extremely limited. As in the case of mixed annihilation with a SM mediator explored in section~\ref{sec:mixed:mediator:sm}, obtaining a sizable enhancement of this rate requires either introducing extremely large new physics couplings, or a large number of coannihilation partners. Both scenarios will lead to a rich phenomenology, with a huge diversity of colored particles and interactions that can be investigated at colliders. The next section discusses the expected collider signatures for the three classes of models studied throughout this paper.

\section{Collider phenomenology}
\label{sec:collider:pheno}
In a previous work~\cite{ElHedri:2017nny} we studied the possible LHC signatures for simplified dark matter models with a single coannihilation partner. Neglecting the interaction between the dark matter and X, these models can be probed by two types of collider searches: multijet plus missing $E_T$ searches targeting the production of a prompt X in association with initial state radiation (ISR)~\cite{ATLAS-CONF-2016-078,CMS-PAS-SUS-15-005,CMS-PAS-SUS-16-014,CMS-PAS-SUS-16-016,Khachatryan:2016kdk,ATLAS:2017cjl,ATLAS:2017dnw, Sirunyan:2017cwe}, and, for scenarios where X and the dark matter are very close in mass, searches for long-lived strongly interacting particles~\cite{Khachatryan:2016sfv,Khachatryan:2015jha,Aaboud:2016dgf,Aaboud:2016uth,Aad:2013gva,Aad:2015rba}. Combining the associated LHC constraints with the relic density requirements, the multijet plus missing $E_T$ searches are expected to probe the $m_\mathrm{DM} \lesssim 1$~TeV and $\Delta \gtrsim 10$\% region. Searches for long-lived particles, on the other hand, typically probe the $\Delta \lesssim 1$\% region, that corresponds to multi-TeV dark matter masses~\cite{ElHedri:2017nny}. Exploring the remaining region --- multi-TeV dark sector particles with moderately compressed spectra --- requires both a more powerful collider and new types of detectors, specially designed to look for compressed topologies. In what follows, we describe how a future $100$~TeV collider would constrain the parameter space of single and multi-partner coannihilating models.

While the study detailed in~\cite{ElHedri:2017nny} focused on simplified models including only one coannihilation partner, we consider here models with multiple X$_i$. Adding coannihilation partners to a model can have multiple effects on its collider phenomenology. First, the new X$_i$ particles will be associated with their own set of constraints from both jets + $\slashed{E}_T$ and long-lived searches. Additionally, the introduction of new particles and vertices is going to modify the pair production cross-section of the X$_i$, possibly introducing new physics X$_j$-mediated processes in addition to the usual QCD processes. Finally, the new vertices associated with the mixed annihilation process will give rise to new production channels for the dark sector particles at colliders in addition to the X$_i$\,$\overline{\mathrm{X}}_i$ pair production. In particular, the mixed annihilation process X$_i$\,X$_j\,\to\,$SM\,SM can be reversed to lead to X$_i$\,X$_j$ production. In models where this process occurs through a new physics mediator, it is also possible to look for the signatures associated to this mediator. In what follows, we evaluate the influence of these different effects for a model with a SM singlet Majorana fermion dark matter DM$_\mathrm{M}$ and two coannihilation partners X$_\mathrm{C3}$ and X$_\mathrm{M8}$. We study the scenario where the mixed X$_\mathrm{C3}$\,X$_\mathrm{M8}$ annihilation occurs through an $s$-channel quark, keeping in mind that introducing a new physics mediator instead would lead to a richer phenomenology and hence stronger constraints. The dark sector particles in our model are similar to the squark, the gluino, and the bino in SUSY~\cite{Baer:1995nc,Baer:2007uz}. Here, however, we consider only one flavor of squark and model the interaction between the X$_i$ and the dark matter using an effective operator.

The parameter space of our two-partner model is spun by the dark matter mass $m_\mathrm{DM}$, the relative mass splittings $\Delta_1$ and $\Delta_2$ of the two coannihilation partners, the X$_\mathrm{C3}$\,X$_\mathrm{M8}$\,$q$ coupling $\alpha_\mathrm{NP}$, and the parameters describing the X$_i$\,DM interaction. Since, as shown in~\cite{ElHedri:2017nny}, the collider bounds on prompt X$_i$ pair-production from multijet + $\slashed{E}_T$ searches have a very weak dependence on $\Delta_i$, we set $\Delta_1 = \Delta_2 = \Delta$ for the rest of this study. As in~\cite{ElHedri:2017nny}, we treat the X$_i$\,DM$\,\to\,$SM\,SM coannihilation as a subdominant process for the determination of the dark matter relic density and we model it using an effective operator, with suppression scale $\Lambda$. The value of this scale, as well as the structure of the effective interaction will not affect the collider phenomenology of models where X$_i$ is prompt; they will, however, affect the lifetime of X$_i$, and hence the constraints from the long-lived particle searches. In what follows, since we will study the phenomenology of our models at a $100$~TeV collider, we set $\Lambda = 20$~TeV. We use the effective operators introduced in~\cite{ElHedri:2017nny} for the X$_\mathrm{C3}$\,DM\,SM\,SM and X$_\mathrm{M8}$\,DM\,SM\,SM interactions, choosing the SM particles to be either quarks or gluons. Note that, with this requirement, since the dark matter is a fermion, one of the products of the X$_\mathrm{C3}$\,DM coannihilation process will necessary be a gluon. The X$_\mathrm{C3}$\,DM\,SM\,SM effective operator will thus be at least of dimension six and one-loop suppressed. The decay width of X$_\mathrm{C3}$ should therefore be extremely suppressed, which will lead to particularly strong bounds from the long-lived particle searches.  

\begin{figure}
	\centering
	\includegraphics[width=0.7\linewidth]{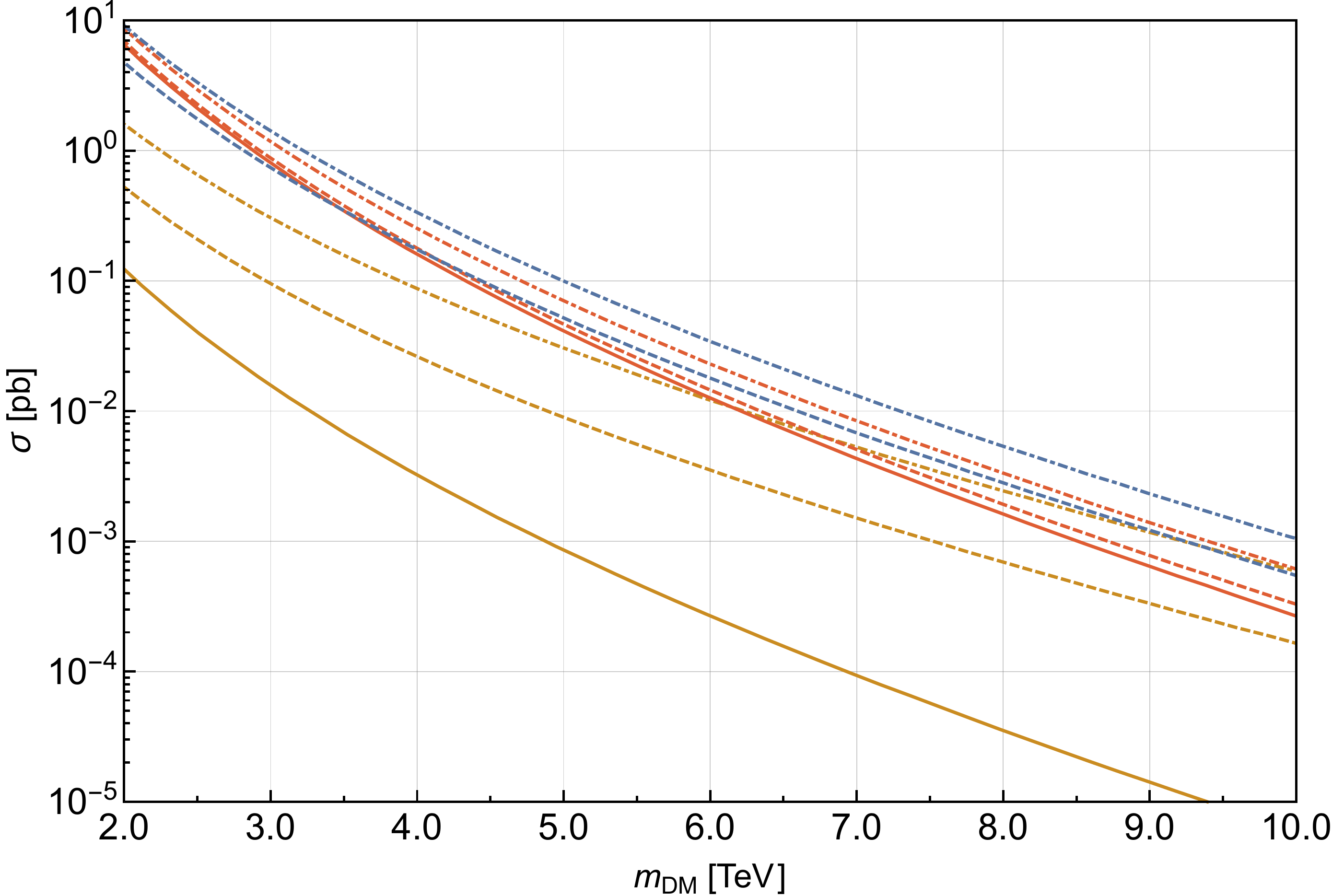}
	\caption{Production cross-sections for X$_\mathrm{C3}$\,$\overline{\mathrm{X}}_\mathrm{C3}$ (orange), X$_\mathrm{M8}\,$X$_\mathrm{M8}$ (red), and X$_\mathrm{C3}$\,X$_\mathrm{M8}$ (blue). For each process, we take the new physics coupling to be $\alpha_\mathrm{NP} = 0$ (solid), $7.0 \, \alpha_s$ (dashed), and $12.9 \, \alpha_s$ (dash-dotted).}
	\label{fig:collider:xsec:prod}
\end{figure}

We first evaluate the impact of the new X$_\mathrm{C3}$\,X$_\mathrm{M8}$\,$q$ vertex described in equation~\eqref{eq:lagrangians:mixed:sm} on the production rate of X$_\mathrm{C3}$ and X$_\mathrm{M8}$ at a $100$~TeV collider. To this end, we compute the X$_\mathrm{C3}$\,$\overline{\mathrm{X}}_\mathrm{C3}$, X$_\mathrm{M8}\,$X$_\mathrm{M8}$, and X$_\mathrm{C3}$\,X$_\mathrm{M8}$ production cross-sections in association with one ISR jet for different values of $\alpha_\mathrm{NP}$. We compute these cross-sections at leading order, imposing a mild $p_T$ cut of $100$~GeV on the ISR jet. We checked that increasing this cut does not affect our conclusions. We inform our choice of $\alpha_\mathrm{NP}$ by considering the ratio $R$ of the dark matter effective annihilation cross-section in the two-partner model over its value in a one-partner model for $\Delta = 0$, as defined in equation~\eqref{eq:sigma:mix:ratio}. In particular, we choose $\alpha_\mathrm{NP} = 0$, $7.0\,\alpha_s$, and $12.9\,\alpha_s$, that correspond to $R \approx 0.57$, $1$, and $2$ respectively, obtained by following the procedure discussed in section~\ref{sec:mixed:mediator:sm}. Note that $R_\mathrm{min} \approx 0.57$ is the minimal possible value for $R$. The production rates of the coannihilation partners for these different couplings are shown in figure~\ref{fig:collider:xsec:prod} as a function of the mass of the X$_i$. While introducing the new physics coupling always increases the pair-production rate of the X$_i$, this increase remains negligible for the pair-production rate of the X$_\mathrm{M8}$ ``gluino-like'' coannihilation partner. While the ``squark'' X$_\mathrm{C3}$ pair-production cross-section can be increased by up to two orders of magnitude by the introduction of the new couplings, it remains usually lower or similar to the X$_\mathrm{M8}$ pair-production cross-section. Finally, the ``mixed'' X$_\mathrm{C3}$\,X$_\mathrm{M8}$ production rate is comparable to the X$_\mathrm{M8}$ pair-production rate for all values of $m_\mathrm{X}$. For the multijet plus $\slashed{E}_T$ searches and the compressed topologies that we are studying, all three processes are expected to lead to extremely similar signatures. It is therefore possible to estimate the associated constraints by considering a classic gluino-neutralino simplified model where the gluino production cross-section is multiplied by a factor of a few compared to SUSY. This enhancement, however, would only correspond to an increase of at most a TeV in the mass reach of the multijet search. Hence, the multijet plus $\slashed{E}_T$ constraints on our two-partner model will be similar to the ones associated to the gluino-neutralino simplified model, that have already been computed in the literature~\cite{Cohen:2013xda}. 

\begin{figure}[!t]
	\centering
	\includegraphics[width=0.6\textwidth]{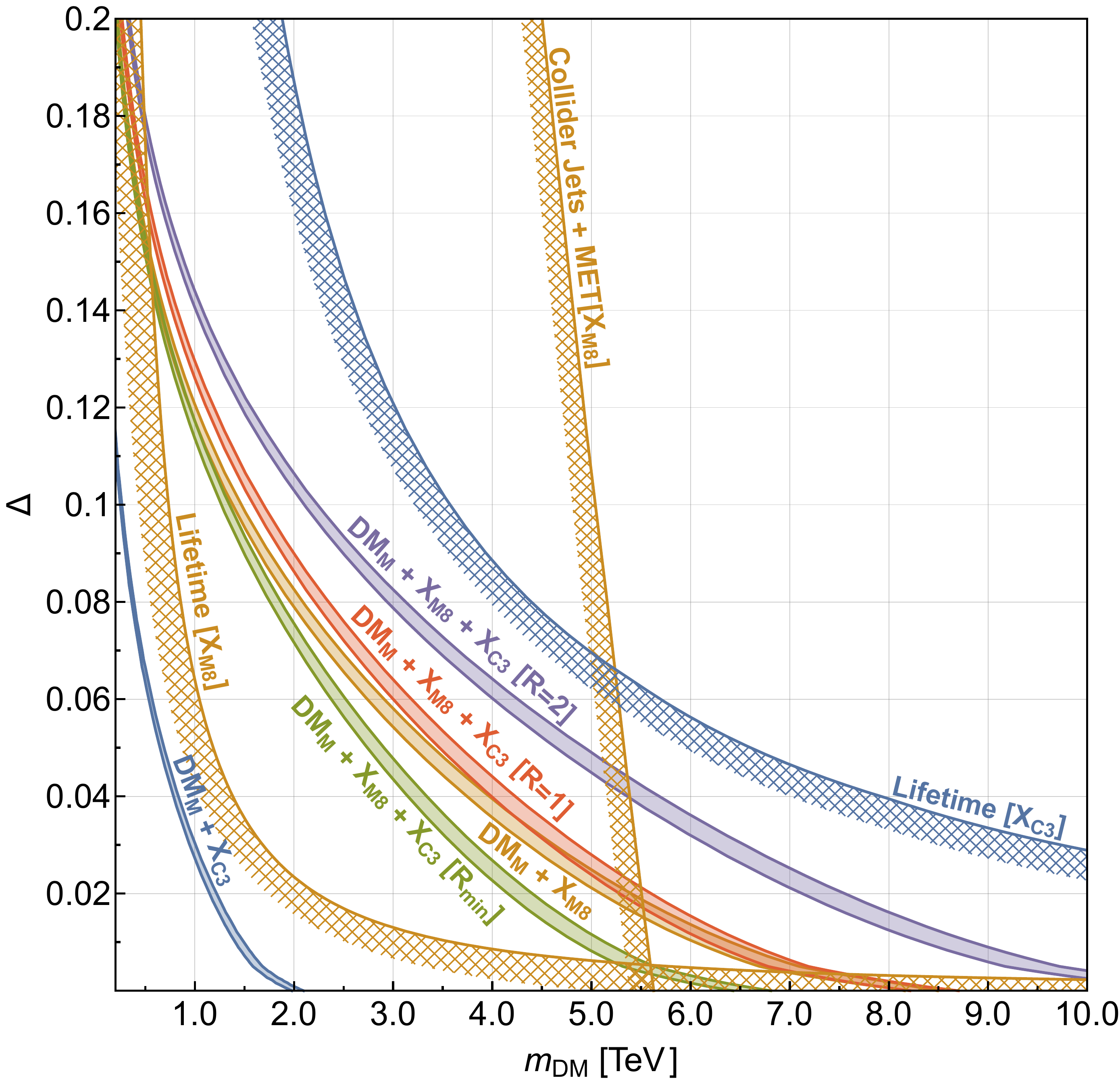}
	\caption{Example model for colored coannihilation which shows the general interplay among collider limits and relic abundance constraints. We compare two models with either a single X$_\mathrm{C3}$ (blue) or a single X$_\mathrm{M8}$ (orange) coannihilation partner to a model with both X$_\mathrm{C3}$ and X$_\mathrm{M8}$ at the same time. We show three scenarios for the two partner model: one where the mixed annihilation is absent (green) and two other scenarios with $R=1$ (red) and $R=2$ (purple). Imposed are the exclusion limits from a prospected $100$ TeV hadron collider for the single partner models.}
	\label{fig:collider:example}
\end{figure}

We use the result discussed above to compute the bounds on $m_\mathrm{X}$ from multijet plus $\slashed{E}_T$ searches for a $100$~TeV center of mass energy and $3$~ab$^{-1}$ luminosity using the results derived in~\cite{Cohen:2013xda} for a gluino-neutralino simplified model. In addition, we estimate the bounds associated to future long-lived particle searches using the procedure described in~\cite{ElHedri:2017nny}, also assuming a luminosity of $3$~ab$^{-1}$. These different constraints are shown along with the regions of parameter space allowed by the relic density requirement in figure~\ref{fig:collider:example}. One striking result for the X$_\mathrm{C3}$\,X$_\mathrm{M8}$ model is that the whole $R \leq 2$ parameter space allowed by the relic density constraints would be within the reach of long-lived particle searches since the decay rate of X$_\mathrm{C3}$ is particularly suppressed. Constraints from these searches can be alleviated if $\Delta_\mathrm{C3}$ is increased with respect to $\Delta_\mathrm{M8}$ but most of the parameter space of our model is still likely to remain excluded, as such an increase will also considerably tighten the relic density constraints on $m_\mathrm{DM}$ and $\Delta_\mathrm{M8}$. Note, however, that this result is valid only if X$_\mathrm{C3}$ decays via an effective operator. In a SUSY-like scenario where X$_\mathrm{C3}$ can decay directly to a quark and a neutralino, for example, constraints from long-lived particle searches will be considerably alleviated. The constraint from multijet plus $\slashed{E}_T$ searches on the other hand is particularly robust since it does not depend on the structure of the X$_i$\,DM interaction. For $R = 0.57$, this search will be extremely efficient and can probe the parameter space down to extremely compressed regions, with $\Delta \lesssim 1\%$. In these regions, it is expected that the X$_i$ will have suppressed decay rates independently from the structure of the X$_i$\,DM\,SM\,SM operator. For $R = 1$ and $2$, the direct searches will probe the parameter space down to $\Delta \approx 2$\% and $4$\% respectively. Although a large region remains unexplored, we have to remember that the multijet bounds shown in figure~\ref{fig:collider:example} have been computed assuming that the design of the detectors for the future $100$~TeV collider would be similar to the one of ATLAS and CMS. In fact, our result shows that designing detectors targeting compressed regions of parameter space would be a crucial step in the search for thermal dark matter~\cite{Mahbubani:2017gjh,Alpigiani:2017haj,Chou:2016lxi,Liu:2018wte,Gligorov:2017nwh}.

Finally, we would like to comment on other smoking gun signatures that could be observed in models with multiple coannihilation partners. First, as noted throughout this paper, in most regions of the parameter space, increasing the number of coannihilation partners in a given model leads to looser relic density bounds only for couplings larger than a few times $\alpha_s$. Such scenarios should therefore be associated with a strongly-coupled sector at energy scales close to $m_\mathrm{DM}$. This strongly-coupled sector will typically be associated with a large number of new particles, notably composite states and light Goldstone bosons, that could also be observed at colliders~\cite{Alanne:2017ymh,Balkin:2017yns,Balkin:2017aep,Ma:2017vzm}. Additionally, as discussed in section~\ref{sec:mixed:annihilation}, the mixed X$_i$\,X$_j$ annihilation process can be mediated by a new physics particle. This mediator does not need to be close in mass to the dark matter and will therefore give rise to characteristic collider signatures. In particular, for $s$-channel models, the mediator can be light and decays to two SM particles. It could therefore be a privileged target for resonance searches~\cite{Aaboud:2018tqo,Aaboud:2018mjh,Aaboud:2018fgi,Aaboud:2017nmi,Aaboud:2016qeg,CMS-PAS-EXO-16-035,CMS-PAS-EXO-16-034,CMS-PAS-EXO-16-031,CMS-PAS-EXO-16-056,CMS-PAS-EXO-17-003,CMS-PAS-EXO-17-022,CMS-PAS-EXO-17-029,CMS-PAS-EXO-18-006}, with signatures similar to the ones studied in~\cite{Baker:2015qna,Buschmann:2016hkc}. In the $t$-channel, the decay mode of the mediator will be similar to the one of the X$_i$ but since the mediator does not have to be close in mass to the dark matter, multijet plus $\slashed{E}_T$ searches should now be much more powerful.

\section{Conclusions}
\label{sec:conclusions}
Confronting experimental limits to relic density constraints has been a major tenet of the search for thermal dark matter during the last few years. In particular, for coannihilating dark matter scenarios, using the simplified model formalism has allowed to translate results from the LHC searches into limits on the mass of the dark matter~\cite{Abdallah:2014hon,Arina:2018zcq,Busoni:2014gta,Brennan:2016xjh,Pree:2016hwc,Jacques:2016dqz,Bauer:2016gys,Abercrombie:2015wmb,Boveia:2016mrp} and on its splitting with its coannihilation partner X. These results in turn have been used to motivate both building a $100$~TeV collider and develop new detector technologies, targeted towards models with particularly compressed spectra. In this study, we evaluated the robustness of these simplified model results against introducing new coannihilation partners. 

Throughout this paper we focused on models with colored coannihilation partners that are charged under QCD although our results can be straightforwardly extended to other models with unbroken gauge groups. We find that, in general, increasing the number of coannihilation partners in these models lead to tighter relic density constraints, and hence to lower dark matter masses, in most regions of the parameter space. The few regions in which adding new coannihilation partners allows for larger dark matter masses correspond to scenarios where the dark matter effective annihilation rate is dominated by ``mixed'' annihilation processes between different species of coannihilation partners. We showed that these different regions can all be characterized by focusing on models with only two coannihilation partners. 

For a representative subset of these models, we considered three possible classes of mixed annihilation processes: mediated via an SM quark, and $s$- and $t$-channel via a new physics mediator. In the first class, the mixed annihilation rates dominate only when the new physics coupling of the SM to the coannihilation partners is much larger than the strong coupling. In the second class, introducing a new coannihilation partner can lead to a significant increase of the dark matter annihilation rate near the resonant annihilation region, where the mediator mass is close to the sum of the masses of the coannihilation partners. This increase, however, only takes place when the mediator of the mixed annihilation process has a particularly narrow width. Finally, mixed annihilation processes happening in the $t$-channel can dominate over similar self-annihilation processes only if the latter are associated to particularly small rates, either being $p$-wave suppressed or involving color triplet particles with a small spin. Since such suppressed processes lead to relatively tight relic density bounds, of around a TeV, on the dark matter mass raising the dark matter annihilation rate for these models is unlikely to affect the current estimates of the upper bound on the energy scale of the coannihilating models, which lies around $10$~TeV. This upper bound can therefore be loosened only in very specific models involving resonant $s$-channel annihilation processes, or in models involving particularly large new physics couplings, that are likely to exhibit a non-perturbative dynamics.

We finally studied the possible signatures associated with models with multiple coannihilation partners at a future $100$~TeV collider. We focused on a simple SUSY-like scenario where a scalar color triplet and a Majorana color octet coannihilate with a Majorana fermion dark matter candidate. We showed that, for this scenario, the interaction between the triplet and the octet only leads to order one changes in the total pair-production rate of these new particles. We also showed that the different coannihilation particles in a given model can be associated to complementary bounds --- with one particle having a large pair-production rate and the other one being long-lived for example --- that would lead to a spectacular reach. Finally, we would like to emphasize that, even in regions where the dark matter annihilation rate is increased compared to the one-partner models, a $100$~TeV collider could probe mass splittings between the dark matter and its coannihilation partners down to a few percent. In order to explore the remaining region, it is essential to develop new detectors, experiments and analysis techniques, specially geared towards these small mass splittings.

We showed in this paper that, even in non-minimal scenarios, dark matter models with an arbitrary number of coannihilating partners generally cannot have dark matter particles heavier than about $10$~TeV. Exceeding this energy scale requires introducing either new particles and vertices or particularly large couplings, which would imply a particularly rich and diverse phenomenology at colliders. Exploring this phenomenology should therefore be a crucial element of the dark matter search program at a future 100~TeV machine. Finally, we would like to emphasize that the methodology detailed throughout this paper can be straightforwardly adapted to study other extensions of the current simplified dark matter models, such as scenarios with multiple dark matter candidates or new unbroken gauge groups. In a time where minimal scenarios are being increasingly cornered, our approach then provides an effective toolkit to comprehend complete models and thus guide the design of the next generations of colliders to allow them to say the final word on thermal dark matter.

\acknowledgments
We would like to thank Anna Kaminska and Jos\'e Zurita for valuable discussions. This research is supported by the Cluster of Excellence Precision Physics, Fundamental Interactions and Structure of Matter (PRISMA-EXC 1098), by the ERC Advanced Grant EFT4LHC of the European Research Council, and by the Mainz Institute for Theoretical Physics. SEH is supported by the NWO Vidi grant ``Self-interacting asymmetric dark matter''.

\appendix
\section{Condition on the effective annihilation cross-section}
\label{sec:general:case:calculation}
In this appendix we consider a general model with one dark matter candidate and $N$ coannihilation partners as described in section~\ref{sec:multiple:different:species}. Recall that the corresponding effective annihilation cross-section as defined in equation~\eqref{eq:general:coannihilation} is 
\begin{equation} \label{eq:general:coannihilation:explicit}
	\sigma^\mathrm{eff}_{\mathrm{X}_1 \cdots \mathrm{X}_N}  = \frac{1}{(g_\mathrm{DM} + \sum_{i = 1}^N g_{\mathrm{X}_i})^2} \left( \sum_{i = 1}^{N} g_{\mathrm{X}_i}^2 \sigma_{\mathrm{X}_i} + 2 \sum_{i = 1}^N \sum_{j = i + 1}^N g_{\mathrm{X}_i} g_{\mathrm{X}_j} \sigma^\mathrm{mix}_{\mathrm{X}_i \mathrm{X}_j} \right) ,
\end{equation}
where we now wrote the double sum explicitly. Here, we show that requiring all submodels of the form DM + X$_i$ + X$_j$ to satisfy~\eqref{eq:sigma:mix:limit:two:species} is a \emph{sufficient} condition for $\sigma^\mathrm{eff}_{\mathrm{X}_1 \cdots \mathrm{X}_N} $ to be smaller than $\mathrm{max} \left( \sigma^\mathrm{eff}_{\mathrm{X}_1}, \ldots, \sigma^\mathrm{eff}_{\mathrm{X}_N} \right)$. 

Without loss of generality, we assume that $\sigma^\mathrm{eff}_{\mathrm{X}_1} \geq \sigma^\mathrm{eff}_{\mathrm{X}_2} \geq \ldots \geq \sigma^\mathrm{eff}_{\mathrm{X}_N}$. We then saturate the constraint in equation~\eqref{eq:sigma:mix:limit:two:species} for each $\sigma^\mathrm{mix}_{\mathrm{X}_i \mathrm{X}_j}$ to obtain
\begin{equation}
	\begin{aligned}
	\sigma^\mathrm{eff}_{\mathrm{X}_1 \cdots \mathrm{X}_N} \! \leq & \frac{1}{(g_\mathrm{DM} \! + \! \sum_{i = 1}^N g_{\mathrm{X}_i})^2} \! \left( \! \sum_{i = 1}^N g_{\mathrm{X}_i}^2 \sigma_{\mathrm{X}_i} \! + \! \sum_{i = 1}^N \! \sum_{j = i + 1}^N \! g_{\mathrm{X}_j} (2 g_\mathrm{DM} \! + \! 2 g_{\mathrm{X}_i} \! + \! g_{\mathrm{X}_j}) \frac{g_{\mathrm{X}_i}^2}{(g_\mathrm{DM} \! + \! g_{\mathrm{X}_i})^2} \sigma_{\mathrm{X}_i} \right. \\
   & \left. - \sum_{i = 1}^N \! \sum_{j = i + 1}^N \! g_{\mathrm{X}_j}^2 \sigma_{\mathrm{X}_j} \! \right) \\
   \leq & \frac{1}{(g_\mathrm{DM} \! + \! \sum_{i = 1}^N g_{\mathrm{X}_i})^2} \sum_{i = 1}^N \sigma^\mathrm{eff}_{\mathrm{X}_i} \left( \! (g_\mathrm{DM} \! + \! g_{\mathrm{X}_i})^2 (2 \! - \! i) + \! \sum_{j = i + 1}^N \! g_{\mathrm{X}_j} (2g_\mathrm{DM} \! + \! 2 g_{\mathrm{X}_i} \! + \! g_{\mathrm{X}_j}) \! \right) .
	\end{aligned}
\end{equation}
    Denoting by $\sigma^\mathrm{eff}_{\mathrm{max}}$ the maximal effective annihilation cross-section for models with only one of the X$_i$ we can then write
\begin{equation}
	\begin{aligned}
	\sigma^\mathrm{eff}_{\mathrm{X}_1 \cdots \mathrm{X}_N}  \leq \, \sigma^\mathrm{eff}_{\mathrm{max}} \frac{\sum_{i = 1}^N \left[ (g_\mathrm{DM} + g_{\mathrm{X}_i})^2 (2 - i) +  \sum_{j = i + 1}^N g_{\mathrm{X}_j} (2 g_\mathrm{DM} + 2g_{\mathrm{X}_i} + g_{\mathrm{X}_j}) \right]}{(g_\mathrm{DM} + \sum_{i = 1}^N g_{\mathrm{X}_i})^2}.
	\end{aligned}
\end{equation}
In order for the effective annihilation cross-section for the complete model to be maximal $\sigma^\mathrm{eff}_{\mathrm{max}}$, we need the ratio
\begin{equation}
	\begin{aligned}
	R = \frac{\sum_{i = 1}^N \left[ (g_\mathrm{DM} + g_{\mathrm{X}_i})^2 (2 - i) + \sum_{j = i + 1}^N g_{\mathrm{X}_j} (2g_\mathrm{DM} + 2g_{\mathrm{X}_i} + g_{\mathrm{X}_j})\right]}{(g_\mathrm{DM} + \sum_{i = 1}^N g_{\mathrm{X}_i})^2} \equiv \frac{\mathcal{N}}{\mathcal{D}}, 
	\end{aligned}
\end{equation}
to be at most one. We note by explicit computation that
\begin{equation}
	\begin{aligned}
		\frac{\partial R}{\partial g_{\mathrm{X}_l}} = 2 \frac{\mathcal{D} - \mathcal{N}}{(g_\mathrm{DM} + \sum_{i = 1}^N g_{\mathrm{X}_i})^3} \quad \Rightarrow \quad \begin{cases} \frac{\partial R}{\partial g_{\mathrm{X}_l}} > 0 & \mathrm{if} \; R < 1 \\ \frac{\partial R}{\partial g_{\mathrm{X}_l}} = 0 & \mathrm{if} \; R = 1 \\ \frac{\partial R}{\partial g_{\mathrm{X}_l}} < 0 & \mathrm{if} \; R > 1 \end{cases} ,
	\end{aligned}
\end{equation}
which implies that $R$ has a stationary point at $R = 1$ that can be approached from either $R > 1$ or $R < 1$ but not both simultaneously. Now if we can show that one point in the $g_{\mathrm{X}_i}$ space has $R < 1$ we then know that for all points $R \leq 1$. In fact it can be shown that for the point $g_{\mathrm{X}_i} = g_\mathrm{DM}$ we always have $R < 1$ for $N > 1$. Hence, when all the DM  + X$_i$ + X$_j$ models verify condition~\ref{eq:sigma:mix:limit:two:species}, the dark matter effective annihilation rate is also decreased when all the coannihilation partners are present simultaneously.

\section{Sommerfeld corrections}
\label{sec:sommerfeld:corrections}
In our analysis, we take the Sommerfeld corrections into account when computing the annihilation cross-sections of the colored coannihilation partners. We base ourselves on~\cite{Cassel:2009wt} and on a previous work~\cite{ElHedri:2016onc} which focuses on the self-annihilation rates of colored particles into gluons and quarks through the strong interaction. Here we extend this work to include the Sommerfeld corrections for the mixed annihilation rates and the new physics contributions to the self-annihilation rates stemming from the interactions shown in the Lagrangians~\eqref{eq:lagrangians:mixed:sm},~\eqref{eq:lagrangians:mixed:schannel},~\eqref{eq:lagrangians:mixed:tchannel:scalarx} and~\eqref{eq:lagrangians:mixed:tchannel:fermionx} and depicted in figures~\ref{fig:mixed:channels:sm:mediator} and~\ref{fig:mixed:channels:np:mediator}. We adopt here the same strategy as in~\cite{ElHedri:2016onc}, decomposing the product of the $SU(3)$ representations of the initial state particles into a sum of color eigenstates, each of them associated to a Coulomb potential. This procedure allows to derive particularly simple expressions for the Sommerfeld corrections to the the self-annihilation rates of colored particles using CP conservation and the symmetry properties of their color representations. For mixed annihilation rates, which involve initial state particles of different types, the situation is more complicated and we need to consider each process individually. In the rest of this appendix we use the formalism and notation of~\cite{ElHedri:2016onc}. Moreover, we update the Mathematice package~\cite{ElHedri:2016pac} to compute Sommerfeld corrections for the processes discussed here.

Although non-perturbative effects like bound state formation~\cite{Liew:2016hqo} can occur alongside Sommerfeld corrections, we focus solely on the latter for this study. As shown in~\cite{Liew:2016hqo,ElHedri:2017nny}, the effect of bound state formation on the effective annihilation cross-section is generally milder than the Sommerfeld corrections. When taking ratios of cross-sections as for the mixed condition derived in section~\ref{sec:relic:density:annihilation} these non-perturbative effects partially factor out between the one and two-partner models. We will therefore not consider bound state formation throughout this work.

The sizes of the non-perturbative corrections to the mixed annihilation rates heavily depend on the quantum numbers of the coannihilation partners involved as well as the structure of the annihilation process. In this section we adopt the strategy detailed in~\cite{ElHedri:2016onc}, computing the Sommerfeld corrections separately for each product of two diagrams that enters the total cross-section, that is, squared and interference terms. As discussed in section~\ref{sec:mixed:annihilation} mixed annihilation can proceed through either a SM quark, an $s$-channel mediator or a $t$-channel mediator. Each scenario involves a different set of diagrams, each of them associated to a specific color structure.

We first notice that there is a wide range of possibile configurations for the mixed annihilation diagrams since the annihilating particles can transform as triplets, sextets, or octets under the strong gauge group. We thus need to calculate the QCD decomposition of all the possible pairs of colored initial states that can annihilate into colored SM particles. All the relevant initial state color combinations for either self-annihilation or mixed annihilation processes are
\begin{equation} \label{eq:decomposition:annihilations}
	\begin{aligned}
		\irrep{3} \otimes \irrep{3} & = \overline{\irrep{3}}_\textbf{A} + \irrep{6}_\textbf{S} \\
		\irrep{3} \otimes \overline{\irrep{3}} & = \irrep{1} \oplus \irrep{8} \\
		\irrep{3} \otimes \irrep{6} & = \irrep{8} \oplus \irrep{10} \\
		\irrep{3} \otimes \overline{\irrep{6}} & = \overline{\irrep{3}} \oplus \overline{\irrep{15}} \\
		\irrep{3} \otimes \irrep{8} & = \irrep{3} \oplus \overline{\irrep{6}} \oplus \irrep{15} \\
		\irrep{6} \otimes \irrep{6} & = \overline{\irrep{6}}_\textbf{S} \oplus \irrep{15}_\textbf{A} \oplus \irrep{15}_\textbf{S}^\prime \\
		\irrep{6} \otimes \overline{\irrep{6}} & = \irrep{1} \oplus \irrep{8} \oplus \irrep{27} \\
		\irrep{6} \otimes \irrep{8} & = \overline{\irrep{3}} \oplus \irrep{6} \oplus \overline{\irrep{15}} \oplus \irrep{24} \\
		\irrep{8} \otimes \irrep{8} & = \irrep{1}_\textbf{S} \oplus \irrep{8}_\textbf{A} \oplus \irrep{8}_\textbf{S} \oplus \irrep{10}_\textbf{A} \oplus \overline{\irrep{10}}_\textbf{A} \oplus \irrep{27}_\textbf{S} ,
	\end{aligned}
\end{equation}
as well as the conjugate expressions. Analogous to~\cite{ElHedri:2016onc,deSimone:2014pda} the QCD potential for an initial state whose particles have representations $\irrep{R}$ and $\irrep{R}^\prime$ is decomposed into channels of definite color as
\begin{equation}
	V_{\irrep{R} \otimes \irrep{R}^\prime} = \frac{\alpha_s (\hat{\mu})}{2r} \sum_{\irrep{Q}} \Big[ C_2(\irrep{Q}) \mathbbm{1}_{\irrep{Q}} - C_2(\irrep{R}) \mathbbm{1} - C_2(\irrep{R}^\prime) \mathbbm{1} \Big] =  \frac{\alpha_s (\hat{\mu})}{r} \sum_{\irrep{Q}} \alpha_{\irrep{Q}} \mathbbm{1}_{\irrep{Q}} .
\end{equation}
The coefficient $\alpha_{\irrep{Q}}$ can be calculated for all the decompositions in equation~\eqref{eq:decomposition:annihilations} and is given in the following table.
\begin{equation} \label{eq:decomposition:potential:mixed}
	\begin{tabular}{c | x{6.9mm} | x{6.8mm} | x{6.8mm} | x{6.8mm} | x{6.8mm} | x{6.9mm} | x{6.8mm} | x{6.8mm} | x{6.8mm} | x{6.8mm} | x{6.8mm} | x{6.8mm} | x{6.8mm}}
		\irrep{Q} & $\irrep{1}$ & $\irrep{3}$ & $\overline{\irrep{3}}$ & $\irrep{6}$ & $\overline{\irrep{6}}$ & $\irrep{8}$ & $\irrep{10}$ & $\overline{\irrep{10}}$ & $\irrep{15}^\prime$ & $\irrep{15}$ & $\overline{\irrep{15}}$ & $\irrep{24}$ & $\irrep{27}$ \tn
		\hline
		$V_{\irrep{3} \otimes \irrep{3}}$ & & & $-\frac{2}{3}$ & $\frac{1}{3}$ & & & & & & & & & \tn
		$V_{\irrep{3} \otimes \overline{\irrep{3}}}$ & $-\frac{4}{3}$ & & & & & $\frac{1}{6}$ & & & & & & \tn
		$V_{\irrep{3} \otimes \irrep{6}}$ & & & & & & $-\frac{5}{6}$ & $\frac{2}{3}$ & & & & & & \tn
		$V_{\irrep{3} \otimes \overline{\irrep{6}}}$ & & & $-\frac{5}{3}$ & & & & & & & & $\frac{1}{3}$ & & \tn
		$V_{\irrep{3} \otimes \irrep{8}}$ & & $-\frac{3}{2}$ & & & $-\frac{1}{2}$ & & & & & $\frac{1}{2}$ & & & \tn
		$V_{\irrep{6} \otimes \irrep{6}}$ & & & & & $-\frac{5}{3}$ & & & & $\frac{4}{3}$ & $-\frac{2}{3}$ & & & \tn
		$V_{\irrep{6} \otimes \overline{\irrep{6}}}$ & $-\frac{10}{3}$ & & & & & $-\frac{11}{6}$ & & & & & & & $\frac{2}{3}$ \tn
		$V_{\irrep{6} \otimes \irrep{8}}$ & & & $-\frac{5}{2}$ & $-\frac{3}{2}$ & & & & & & & $-\frac{1}{2}$ & $1$ & \tn
		$V_{\irrep{8} \otimes \irrep{8}}$ & $-3$ & & & & & $-\frac{3}{2}$ & 0 & 0 & & & & & $1$ \tn
	\end{tabular}
\end{equation}
In this table a blank entry indicates that the color decomposition of the representation product $\irrep{R} \otimes \irrep{R}^\prime$ does not contain $\irrep{Q}$. We note that the values of $\alpha_{\irrep{Q}}$ are independent of the considered annihilation processes and their diagrammatic structures.

The remaining necessary step to calculate the Sommerfeld corrections is to compute the size of the contributions of the different $\irrep{Q}$ initial states to the total annihilation cross-section. As mentioned earlier in this section we need to treat each product of two diagrams entering in the cross-section separately. In section~\ref{sec:mixed:models} we detailed which diagrams are involved in the mixed annihilation processes. The vertices involved in these diagrams are associated with the following color factors
\begin{equation} \label{eq:color:structures}
	\begin{aligned}
	C_{\irrep{3} \irrep{3} \irrep{3}} & = \epsilon^{ijk}, \quad C_{\irrep{3} \irrep{3} \irrep{6}} = K^u_{ij}, \quad C_{\irrep{3} \irrep{3} \irrep{8}} = T^a_{ij}, \quad C_{\irrep{6} \irrep{6} \irrep{8}} = (T_6)^a_{uv}, \\
	C_{\irrep{8} \irrep{8} \irrep{8}} & = (T_8)^a_{bc} = - i f^{abc}, \quad C_{\irrep{3} \irrep{6} \irrep{8}} = (T^a)_j^l \, \epsilon^{ijk} K^u_{kl}  
	\end{aligned}
\end{equation}
where the indices give the color representations of the particles involved. Since Sommerfeld corrections for self-annihilation cross-sections have been computed in~\cite{ElHedri:2016onc}, we will focus here on the mixed annihilation rates. However, in models involving a SM mediator or a $t$-channel NP mediator, allowing for mixed annihilation also opens new channels for the self-annihilation of the X$_i$ into SM quarks. The corresponding diagrams interfere with the existing QCD $s$-channel annihilation process and we need to compute the size of the Sommerfeld corrections for the associated new contributions. 

For each annihilation process, we write the total cross-section as the sum over all the color eigenstates $\irrep{Q} \in \irrep{R}\otimes\irrep{R}'$
\begin{equation}
	\sigma_{\irrep{R} \otimes \irrep{R}^\prime} = \sum_{\irrep{Q}} \beta_{\irrep{Q}} \, \sigma_{\irrep{Q}} .
\end{equation}
This allows us to write the Sommerfeld corrected cross-section in the notation of~\cite{ElHedri:2016onc} as
\begin{equation} \label{eq:sommerfeld:corrections:decomposed}
	\sigma_{\irrep{R} \otimes \irrep{R}^\prime}^{(S)} = \sum_{\irrep{Q}} \beta_{\irrep{Q}} \, \sigma^{(S)}_C \left[ - \alpha_{\irrep{Q}} \, \alpha_s \right] .
\end{equation}
Note that if $\alpha_{\irrep{Q}}$ is negative we have Sommerfeld enhancement and if it is positive we have a Sommerfeld reduction of the annihilation rate in the specific color channel $\irrep{Q}$.

We first look consider processes occuring via a $s$-channel NP mediator. These cases are particularly simple to treat as the color representation of the initial state has to be the same as the one of the mediator, and we therefore obtain $\beta_{\irrep{Q}} = 1$ for a mediator with charge $\irrep{Q}$. Then the single Sommerfeld correction factor $\alpha_{\irrep{Q}}$ can be read of table~\eqref{eq:decomposition:potential:mixed}. The $s$-channel NP models do not induce self-annihilation processes for the coannihilation partners so we do not need to compute any additional corrections for these models. A similar simplification occurs for the interference of the new physics self-annihilation processes in SM and $t$-channel models with the QCD self-annihilation into SM quarks. Since the latter proceeds through an $s$-channel gluon these terms are always associated with $\beta_{\irrep{8}} = 1$, and non-octet initial states do not contribute. Finally, the SM mediated mixed annihilation of X$_1$ and X$_2$ into a SM quark and a gluon involves one $s$-channel and two $t$-channel diagrams. For the $s$-channel amplitude squared and any of its interferences with the other diagrams the previous argument applies and we have $\beta_{\irrep{3}, \overline{\irrep{3}}} = 1$ depending on whether the mediator is a quark or an antiquark.

\begin{figure}[!t]
	\centering
	\begin{tikzpicture}[line width=1.4pt, scale=1]
	\draw[fermionbar] (0.8,0.8)--(0,0.5);
	\draw[fermionbar] (0.8,-0.8)--(0,-0.5);
	\draw[fermion] (-0.8,0.8)--(0,0.5);
	\draw[fermion] (-0.8,-0.8)--(0,-0.5);
	\draw[fermion] (0,0.5)--(0,-0.5);
	
	\node at (-1.1,0.8) {$\irrep{R}_1$};
	\node at (-1.1,-0.8) {$\irrep{R}_2$};
	\node at (-0.35,0) {$\irrep{M}$};
	\node at (1.05,0.8) {$\irrep{S}_1$};
	\node at (1.05,-0.8) {$\irrep{S}_2$};
\end{tikzpicture}
	\caption{The color flow in the $t$-channel diagrams that contribute to the mixed annihilation cross-sections. The arrows denote the direction in which the color charges of the respective particles flow.}
	\label{fig:color:flow}
\end{figure}
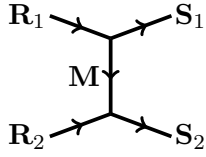

The only remaining contributions to the annihilation cross-sections are the terms involving only products of $t$ and $u$-channel amplitudes, that appear in the SM and $t$-channel NP models. The corresponding processes can be uniquely represented by $\irrep{R}_1 \, \irrep{R}_2 \to \irrep{S}_1 \, \irrep{S}_2$ where $\irrep{R}_i$ and $\irrep{S}_i$ are the color representations of the initial and final states respectively and we denote the color of the mediator with $\irrep{M}$. The color flow for these types of diagrams is defined in figure~\ref{fig:color:flow}. We denote the color structure of the diagram as $\mathcal{C}_{\irrep{R}_1 \irrep{R}_2, \irrep{M}, \irrep{S}_1 \irrep{S}_2}$, keeping in mind that $|\mathcal{C}_{\irrep{R}_1 \irrep{R}_2, \irrep{M}, \irrep{S}_1 \irrep{S}_2}|^2 = |\mathcal{C}_{\irrep{R}_2 \irrep{R}_1, \overline{\irrep{M}}, \irrep{S}_2 \irrep{S}_1}|^2$. For the initial state representations and their color decompositions described in equation~\eqref{eq:decomposition:annihilations} there are many possible combinations for $\mathcal{C}$. In the next tables we list the cross-section decomposition coefficients $\beta_{\irrep{Q}}$ for the models introduced in section~\ref{sec:mixed:annihilation}. For the SM-mediated models discussed in section~\ref{sec:mixed:model:sm}, the squared and mixed products of the three leftmost diagrams in figure~\ref{fig:mixed:channels:sm:mediator} are associated with the following coefficients
\begin{equation} \label{eq:decomposition:xsec:sm:mediated}
	\begin{tabular}{c | x{6.8mm} | x{6.9mm} | x{6.8mm} | x{6.8mm} | x{6.8mm} | x{6.8mm}}
		\irrep{Q} & $\irrep{3}$ & $\overline{\irrep{3}}$ & $\irrep{6}$ & $\overline{\irrep{6}}$ & $\irrep{15}$ & $\overline{\irrep{15}}$ \tn
		\hline
		$|\mathcal{C}_{\irrep{3} \irrep{3}, \irrep{3}, \irrep{8} \overline{\irrep{3}}}|^2$ & & $\frac{1}{4}$ & $\frac{3}{4}$ & & & \tn
		$|\mathcal{C}_{\irrep{3} \irrep{3}, \irrep{3}, \irrep{8} \overline{\irrep{3}}} \, \mathcal{C}_{\irrep{3} \irrep{3}, \overline{\irrep{3}}, \overline{\irrep{3}} \irrep{8}}|$ & & $-\frac{1}{2}$ & $\frac{3}{2}$ & & & \tn
		\hline
		$|\mathcal{C}_{\irrep{3} \overline{\irrep{6}}, \irrep{3}, \irrep{8} \overline{\irrep{3}}}|^2$ & & $\frac{1}{16}$ & & & & $\frac{15}{16}$ \tn
		$|\mathcal{C}_{\overline{\irrep{6}} \irrep{3}, \overline{\irrep{6}}, \irrep{8} \overline{\irrep{3}}}|^2$ & & $\frac{5}{8}$ & & & & $\frac{3}{8}$ \tn
		$|\mathcal{C}_{\irrep{3} \overline{\irrep{6}}, \irrep{3}, \irrep{8} \overline{\irrep{3}}} \, \mathcal{C}_{\overline{\irrep{6}} \irrep{3}, \overline{\irrep{6}}, \irrep{8} \overline{\irrep{3}}}|$ & & $\frac{1}{4}$ & & & & $\frac{3}{4}$ \tn
		\hline
		$|\mathcal{C}_{\irrep{3} \irrep{8}, \irrep{3}, \irrep{8} \irrep{3}}|^2$ & $\frac{1}{64}$ & & & $\frac{9}{32}$ & $\frac{45}{64}$ & \tn
		$|\mathcal{C}_{\irrep{8} \irrep{3}, \irrep{8}, \irrep{8} \irrep{3}}|^2$ & $\frac{9}{16}$ & & & $\frac{1}{8}$ & $\frac{5}{16}$ & \tn
		$|\mathcal{C}_{\irrep{3} \irrep{8}, \irrep{3}, \irrep{8} \irrep{3}} \, \mathcal{C}_{\irrep{8} \irrep{3}, \irrep{8}, \irrep{8} \irrep{3}}|$ & $\frac{1}{8}$ & & & $\frac{1}{4}$ & $\frac{5}{8}$ & \tn
		\hline
		$|\mathcal{C}_{\irrep{6} \irrep{8}, \irrep{6}, \irrep{8} \overline{\irrep{3}}}|^2$ & & $\frac{5}{32}$ & $\frac{9}{16}$ & & & $\frac{9}{32}$ \tn
		$|\mathcal{C}_{\irrep{8} \irrep{6}, \irrep{8}, \irrep{8} \overline{\irrep{3}}}|^2$ & & $\frac{1}{16}$ & $\frac{5}{8}$ & & & $\frac{5}{16}$ \tn
		$|\mathcal{C}_{\irrep{6} \irrep{8}, \irrep{6}, \irrep{8} \overline{\irrep{3}}} \, \mathcal{C}_{\irrep{8} \irrep{6}, \irrep{8}, \irrep{8} \overline{\irrep{3}}}|$ & & $-\frac{1}{8}$ & $\frac{3}{4}$ & & & $\frac{3}{8}$ \tn
	\end{tabular}
\end{equation}
For the $t$-channel mediated processes, represented by the last diagram in figure~\ref{fig:mixed:channels:sm:mediator} and the last two diagrams in figure~\ref{fig:mixed:channels:np:mediator}, there are two combinations that decompose onto a single channel. This is due to the fact the channels, the color representations of the initial and final states must be the same, i.e.~$\irrep{Q} \in \irrep{R}_1 \otimes \irrep{R}_2 \cap \irrep{S}_1 \otimes \irrep{S}_2$ . Using equation~\eqref{eq:decomposition:annihilations} we thus find that $|\mathcal{C}_{\irrep{3} \overline{\irrep{6}}, \irrep{8}, \irrep{3} \irrep{3}}|^2$ and $|\mathcal{C}_{\irrep{6} \irrep{6}, \irrep{8}, \overline{\irrep{3}} \overline{\irrep{3}}}|^2$ are non-zero only for initial pairs of particles in the $\overline{\irrep{3}}$ and $\irrep{6}$ representation respectively. The results for the remaining combinations are
\begin{equation} \label{eq:decomposition:xsec:tchannel:mediated}
	\begin{tabular}{c | x{6.8mm} | x{6.9mm} | x{6.8mm} | x{6.8mm} | x{6.8mm} | x{6.8mm}}
		\irrep{Q} & $\irrep{1}$ & $\irrep{3}$ & $\overline{\irrep{3}}$ & $\irrep{6}$ & $\overline{\irrep{6}}$ & $\irrep{8}$ \tn
		\hline
		$|\mathcal{C}_{\irrep{3} \overline{\irrep{3}}, \overline{\irrep{3}}, \overline{\irrep{3}} \irrep{3}}|^2$ & $\frac{1}{3}$ & & & & &  $\frac{2}{3}$ \tn
		$|\mathcal{C}_{\irrep{3} \overline{\irrep{3}}, \irrep{6}, \overline{\irrep{3}} \irrep{3}}|^2$ & $\frac{2}{3}$ & & & & & $\frac{1}{3}$ \tn
		$|\mathcal{C}_{\irrep{3} \overline{\irrep{3}}, \irrep{8}, \irrep{3} \overline{\irrep{3}}}|^2$ & $\frac{1}{18}$ & & & & & $\frac{17}{18}$ \tn
		\hline
		$|\mathcal{C}_{\irrep{3} \irrep{3}, \irrep{8}, \irrep{3} \irrep{3}}|^2$ & & & $\frac{2}{3}$ & $\frac{1}{3}$ & & \tn
		$|\mathcal{C}_{\irrep{3} \irrep{3}, \irrep{8}, \irrep{3} \irrep{3}}^t \, \mathcal{C}_{\irrep{3} \irrep{3}, \irrep{8}, \irrep{3} \irrep{3}}^u|$ & & & $2$ & $-1$ & & \tn
		\hline
		$|\mathcal{C}_{\irrep{3} \irrep{8}, \overline{\irrep{3}}, \overline{\irrep{3}} \overline{\irrep{3}}}|^2$ & & $\frac{1}{4}$ & & & $\frac{3}{4}$ & \tn
		$|\mathcal{C}_{\irrep{3} \irrep{8}, \overline{\irrep{3}}, \overline{\irrep{3}} \overline{\irrep{3}}}^t \, \mathcal{C}_{\irrep{3} \irrep{8}, \overline{\irrep{3}}, \overline{\irrep{3}} \overline{\irrep{3}}}^u|$ & & $-\frac{1}{2}$ & & & $\frac{3}{2}$ & \tn
		\hline
		$|\mathcal{C}_{\irrep{6} \overline{\irrep{6}}, \irrep{3}, \irrep{3} \overline{\irrep{3}}}|^2$ & $\frac{1}{6}$ & & & & & $\frac{5}{6}$ \tn
		$|\mathcal{C}_{\irrep{6} \overline{\irrep{6}}, \irrep{8}, \overline{\irrep{3}} \irrep{3}}|^2$ & $\frac{4}{9}$ & & & & & $\frac{5}{9}$ \tn
		\hline
		$|\mathcal{C}_{\irrep{6} \irrep{8}, \irrep{3}, \irrep{3} \irrep{3}}|^2$ & & $\frac{3}{8}$ & $\frac{5}{8}$ & & & \tn
		$|\mathcal{C}_{\irrep{6} \irrep{8}, \irrep{3}, \irrep{3} \irrep{3}}^t \, \mathcal{C}_{\irrep{6} \irrep{8}, \irrep{3}, \irrep{3} \irrep{3}}^u|$ & & $-\frac{3}{2}$ & $-\frac{5}{2}$ & & & \tn
		\hline
		$|\mathcal{C}_{\irrep{8} \irrep{8}, \overline{\irrep{3}}, \irrep{3} \overline{\irrep{3}}}|^2$ & $\frac{1}{8}$ & & & & & $\frac{7}{8}$ \tn
		$|\mathcal{C}_{\irrep{8} \irrep{8}, \overline{\irrep{3}}, \irrep{3} \overline{\irrep{3}}}^t \, \mathcal{C}_{\irrep{8} \irrep{8}, \irrep{3}, \overline{\irrep{3}} \irrep{3}}^u|$ & $-1$ & & & & & $2$ \tn
		$|\mathcal{C}_{\irrep{8} \irrep{8}, \irrep{6}, \irrep{3} \overline{\irrep{3}}}|^2$ & $\frac{1}{4}$ & & & & & $\frac{3}{4}$ \tn
		$|\mathcal{C}_{\irrep{8} \irrep{8}, \irrep{6}, \irrep{3} \overline{\irrep{3}}}^t \, \mathcal{C}_{\irrep{8} \irrep{8}, \overline{\irrep{6}}, \overline{\irrep{3}} \irrep{3}}^u|$ & $\frac{1}{3}$ & & & & & $\frac{2}{3}$ \tn
	\end{tabular}
\end{equation}
In these tables some of the representations $\irrep{Q}$ appearing in the decomposition~\eqref{eq:decomposition:annihilations} are not present since they do not appear in the color decompositions of the SM final states and therefore cannot contribute to the annihilation processes. Since the diagram in figure~\ref{fig:color:flow} also has $u$-channel variants we use the superscripts $t$ and $u$ to indicate the respective channel whenever the color flow of the interference term is ambiguous. Now, using equation~\eqref{eq:sommerfeld:corrections:decomposed} and by reading of the values for $\alpha_{\irrep{Q}}$ in table~\eqref{eq:decomposition:potential:mixed} and for $\beta_{\irrep{Q}}$ in tables~\eqref{eq:decomposition:xsec:sm:mediated} and~\eqref{eq:decomposition:xsec:tchannel:mediated}, we can calculate the Sommerfeld corrections for all processes considered in this work.

\bibliographystyle{JHEP}
\bibliography{mcds_v1}

\end{document}